\documentclass[10pt]{article}
\pdfoutput=1
\usepackage[utf8]{inputenc}
\usepackage{cite}
\usepackage{amsmath,amssymb,amsbsy,amstext,amsthm,simplewick,amsfonts}
\usepackage{graphicx}
\usepackage{subcaption}
\usepackage{lscape}
\usepackage{cancel}
\usepackage{wrapfig}
\usepackage{upgreek}
\usepackage{bm} 
\usepackage{framed}
\usepackage{bbm}
\usepackage{textcomp}
\usepackage{multirow}

\usepackage{tikz}
\tikzset{every picture/.style={line width=0.75pt}} 
\newcommand{\fig}[1]{ \vcenter{\hbox{\includegraphics{#1.pdf}}} }
\newcommand{\discn}[1]{ \underset{#1}{\text{disc}} } 
\usepackage{pifont}
\usetikzlibrary{matrix,shapes,fit,tikzmark,calc}
\usepackage{adjustbox}
\usepackage{makecell}
\usepackage{tcolorbox}
\usepackage{physics}
\usepackage{empheq}
\usepackage[normalem]{ulem}
\usepackage{enumitem}
\usepackage{braket} 
\usepackage{array}
\usepackage{dsfont}
\usepackage{ulem}

\usepackage{titlesec}
\titleformat{\section}{\normalfont\fontsize{12}{16}\bfseries}{\thesection}{1em}{}

\numberwithin{equation}{section}

\def\be{\begin{equation}}
\def\ee{\end{equation}}

\def\ba{\begin{eqnarray}}
\def\ea{\end{eqnarray}}

\def\t{\tau}

\def\bfx{\textbf{x}}

\def\bfk{\textbf{k}}

\def\bfx{\textbf{x}}
\def\bfq{\textbf{q}}

\newcommand{\Disc}[1]{ \underset{#1}{\text{Disc}} } 
\newcommand{\DiscT}[1]{ \underset{#1}{ \widetilde{\text{Disc}}} } 
\newcommand{\disc}[1]{ \underset{#1}{\text{disc}} }

\usepackage[framemethod=TikZ]{mdframed}
\mdfdefinestyle{boxed}{%
linecolor=black,
roundcorner=5pt,
innerleftmargin=0,%
innerrightmargin=0,
backgroundcolor=gray!10,%
}

\definecolor{blue3}{RGB}{31,119,180}
\definecolor{red3}{RGB}{214,39,40}
\definecolor{orange3}{RGB}{255,127,14}
\definecolor{green3}{RGB}{44,160,44}

\usepackage{colortbl}
\definecolor{lightgreen}{cmyk}{0.2, 0, 0.2, 0.2}
\definecolor{lightgray}{cmyk}{0.1,0.2,0,0.1}
\definecolor{lightgray2}{cmyk}{0.1,0.1,0,0.1}

\usepackage{hyperref}
\hypersetup{colorlinks=true,linkcolor=teal,citecolor=orange3,urlcolor=green3,pdfencoding=auto}

\makeatletter
\newlength{\apb@width}
\newcommand{\autoparbox}[2][c]{\settowidth{\apb@width}{#2}\parbox[#1]{\apb@width}{#2}}

\makeatother


\def\t{{\bm t}}

\def\bfp{\textbf{p}}

\def\beq{\begin{equation}}
\def\eeq{\end{equation}}

\allowdisplaybreaks[1]
\begin{document}


\begin{titlepage}
\setcounter{page}{1} \baselineskip=15.5pt

\thispagestyle{empty}

\renewcommand*{\thefootnote}{\fnsymbol{footnote}}

\begin{center}

{\fontsize
{20}{20} \bf The Cosmological Tree Theorem} \\
\end{center}

\vskip 18pt
\begin{center}
\noindent
{\fontsize{12}{18}\selectfont Santiago Ag\"{u}\'{i} Salcedo$\,{}^{a}$ and Scott Melville$\,{}^{b, a}$}
\end{center}

\begin{center}
\vskip 8pt
${}^{a}$\textit{DAMTP, University of Cambridge, Wilberforce Road, Cambridge, CB3 0WA, UK}  \\
${}^{b}$\textit{Astronomy Unit, Queen Mary University of London, Mile End Road, London, E1 4NS, UK} 
\end{center}


\vspace{1.4cm}

		\noindent A number of diagrammatic ``cutting rules'' have recently been developed for the wavefunction of the Universe which determines cosmological correlation functions. 
These leverage perturbative unitarity to relate particular ``discontinuities'' in Feynman-Witten diagrams (with cosmological boundary conditions) to simpler diagrams, in much the same way that the Cutkosky rules relate different scattering amplitudes. 
In this work, we make use of a further causality condition to derive new cutting rules for Feynman-Witten diagrams.
These lead to the cosmological analogue of Feynman's tree theorem for amplitudes, which can be used to systematically expand any loop diagram in terms of (momentum integrals of) tree-level diagrams. 
As an application of these new rules, we show that certain singularities in the wavefunction cannot appear in equal-time correlators due to a cancellation between ``real'' and ``virtual'' contributions that closely parallels the KLN theorem. 
Finally, when combined with the Bunch-Davies condition that certain unphysical singularities are absent, these cutting rules completely determine any tree-level exchange diagram in terms of simpler contact diagrams.
Altogether, these results remove the need to ever perform nested time integrals when computing cosmological correlators.  
		

	\end{titlepage}

	
	\setcounter{tocdepth}{3}
	{
		\hypersetup{linkcolor=black}
		\tableofcontents
	}

	\renewcommand*{\thefootnote}{\arabic{footnote}}
	\setcounter{footnote}{0} 
	
	\newpage
	
	\section{Introduction}
		
	\noindent Loops are hard. 
	It is an unhappy fact of life: to make precise predictions in most quantum field theories, one must compute Feynman diagrams that contain loops. 
	Even on Minkowski spacetime, this is difficult due to momentum integrals that often diverge. 
	On a cosmological (time-dependent) spacetime background, things are even worse: loop diagrams contain potentially divergent integrals both over momenta and over time. 
	Since loop corrections can play an important role in a wide variety of cosmological settings---including precise calculations of the primordial power spectrum~\cite{Weinberg:2005vy,Boyanovsky:2005px,Boyanovsky:2005sh, Weinberg:2006ac,Sloth:2006az,Sloth:2006nu,Bilandzic:2007nb,Seery:2007wf,Seery:2007we,vanderMeulen:2007ah, Adshead:2008gk,Adshead:2009cb, Senatore:2009cf,Chen:2016nrs, Ota:2022xni, Inomata:2022yte, Firouzjahi:2023aum, Fumagalli:2023loc} and non-Gaussianity~\cite{Riotto:2008mv, Assassi:2012et, Gorbenko:2019rza, Green:2020txs, Cohen:2021fzf, Premkumar:2021mlz, Cohen:2021jbo,  Wang:2021qez, Xianyu:2022jwk, Heckelbacher:2022hbq, Motohashi:2023syh,Iacconi:2023slv, Lee:2023jby}; inflationary features that may seed primordial black holes \cite{Sasaki:2018dmp,Kristiano:2022maq,Riotto:2023hoz,Kristiano:2023scm,Riotto:2023gpm,Choudhury:2023jlt, Franciolini:2023lgy, Choudhury:2023rks} or produce a stochastic gravitational wave background \cite{Ananda:2006af, Baumann:2007zm, Fumagalli:2020nvq, Fumagalli:2021mpc, Aragam:2023adu}; de Sitter holography \cite{Strominger:2001gp, Heckelbacher:2020nue, Penin:2021sry, Fichet:2021xfn} and the IR stability of de Sitter spacetime \cite{Ford:1984hs, Antoniadis:1985pj, Starobinsky:1994bd, Dolgov:2005se, Marolf:2010zp,Burgess:2010dd}---it is important to develop tools for evaluating (and better understanding) these loop diagrams. 
	Our goal in this work is to take a step in this direction, and in particular address the issue of the time integrals. The main result is a \emph{cosmological tree theorem} which can be used to expand any loop diagram in terms of (momentum integrals of) tree-level diagrams, which are comparatively much easier to evaluate. 
	This theorem applies on any time-dependent spacetime background, and therefore can play a useful role in future studies of the cosmological implications of loop corrections.

	\paragraph{The cosmological wavefunction.}
	The quantum-mechanical object that we focus on computing is the wavefunction of the Universe \cite{Hartle:1983ai}.	
	This wavefunction characterises the state of the quantum fields at a given time, and has recently played an increasingly central role in inflationary cosmology. 
	Any equal-time correlation function can be calculated from it using the Born rule, and these cosmological correlators encode valuable information about the new high-energy fields that can be excited during inflation \cite{Arkani-Hamed:2015bza}.
    But beyond efficiently encoding all correlators, use of the wavefunction has been driven by its many parallels with scattering amplitudes.
    The $S$-matrix programme of the 1960's developed a suite of constraints and calculational tools for amplitudes using the fundamental axioms of unitarity, causality and locality \cite{Eden:1966dnq}, and the modern incarnation of these ideas includes powerful bootstrap techniques (see e.g. \cite{Kruczenski:2022lot, Bern:2022jnl, Baumgart:2022yty, deRham:2022hpx} for recent reviews).
	Proceeding along similar lines has led to a variety of ``bootstrap'' approaches for efficiently computing the cosmological wavefunction.
	
\paragraph{Cosmological bootstrap(s).}
	One influential approach uses the de Sitter isometries \cite{Arkani-Hamed:2018kmz,Baumann:2019oyu,Baumann:2020dch, Baumann:2021fxj}, building on earlier work \cite{McFadden:2009fg, McFadden:2010vh, Bzowski:2013sza, Bzowski:2015pba} 
	by introducing the requirement that certain unphysical singularities must vanish for a Bunch-Davies initial state (see \cite{Baumann:2022jpr} for a recent review).
	Another symmetry-based approach is the cosmological scattering equation \cite{Gomez:2021qfd, Gomez:2021ujt}, or the analytic continuation from (Euclidean) AdS where harmonic analysis is possible \cite{Anninos:2014lwa,Konstantinidis:2016nio,Baumann:2019ghk, Sleight:2019mgd, Sleight:2019hfp, Sleight:2020obc, Sleight:2021plv, DiPietro:2021sjt}, including recent progress towards a non-perturbative Kallen-Lehmann representation of the two-point function~\cite{Hogervorst:2021uvp, Penedones:2023uqc, Loparco:2023rug}.
	Other bootstrap approaches take a general ansatz with the desired analytic structure and apply various known limits and consistency relations~\cite{Pajer:2020wnj,Pajer:2020wxk, Jazayeri:2021fvk, Bonifacio:2021azc}, without the need for full de Sitter symmetry, or build the desired coefficient by gluing together simpler seed functions~\cite{Meltzer:2021zin, Hillman:2021bnk, Qin:2022fbv, Tong:2021wai, Qin:2023bjk}. 
Altogether, these important advances represent an ``NLO revolution'' for cosmological correlators: they have enabled the computation of primordial non-Gaussianity and other observables in a range of different models beyond the leading-order contact diagrams \cite{Cabass:2021fnw, Jazayeri:2022kjy,Cabass:2022rhr,Pimentel:2022fsc, Bonifacio:2022vwa, Ghosh:2023agt}, much like the NNLO revolution currently taking place for scattering amplitudes (where on-shell unitarity methods and factorisation have enabled an explosion of new two-loop computations, see e.g. \cite{Duhr:2016nrb}).

	\paragraph{Unitarity.}
		One aspect of many of these approaches is the use of perturbative unitarity.
		Much like for amplitudes, unitary time evolution is encoded in perturbation theory\footnote{
		Beyond the diagrammatic expansion, these unitarity relations can be understood as an infinite number of conserved charges which follow from the total probability (the norm of the wavefunction) remaining constant in time \cite{Cespedes:2020xqq}.
		} as a set of ``cutting rules'' which can relate any Feynman-Witten diagram for the wavefunction to simpler diagrams with fewer internal lines \cite{Goodhew:2020hob, Cespedes:2020xqq, Melville:2021lst, Goodhew:2021oqg} (see also \cite{Meltzer:2020qbr} for recent cutting rules in AdS)\footnote{
		A more geometric formulation of the wavefunction is provided by the cosmological polytope \cite{Arkani-Hamed:2017fdk, Arkani-Hamed:2018bjr,Benincasa:2018ssx, Benincasa:2019vqr, Benincasa:2020aoj, Benincasa:2021qcb}, and \cite{Albayrak:2023hie} recently derived a number of new cutting rules in that language.
		}. 
		In the context of amplitudes, pairing unitarity and the optical theorem with causality and analyticity leads to powerful UV/IR relations that connect the coefficients in any low-energy Effective Field Theory with properties of the underlying high-energy completion (see e.g. \cite{Adams:2006sv, Bellazzini:2016xrt,deRham:2017avq, deRham:2017zjm,Tolley:2020gtv, Bellazzini:2020cot,Davighi:2021osh,Li:2021cjv} for an incomplete list).
	Despite some progress in this direction~\cite{Baumann:2015nta, Baumann:2019ghk, Grall:2021xxm, Creminelli:2022onn, Salcedo:2022aal}, the application of analogous UV/IR relations in cosmology has mostly been held back by our understanding of causality/analyticity on time-dependent spacetime backgrounds.

	\paragraph{Causality.}
	Relativistic causality is the central pillar on which the Minkowski $S$-matrix programme is built. 
	This is the postulate that a local observer can only access information from inside their past lightcone and affect events inside their future lightcone---or stated mathematically: all local operators must commute at space-like separations. At the level of the $S$-matrix, this property implies\footnote{
    The connection between causality and analyticity is fairly rigorous in simple theories, e.g. a real scalar field with a mass gap, although in a completely general setting one often postulates analyticity in place of causality since it can be precisely implemented at the level of the $S$-matrix.
    } a particular analytic structure for $2 \to 2$ scattering in the complex $s$-plane at fixed $t$, where $s$ and $t$ are the usual Mandelstam variables.
    On a curved spacetime background, the implications of relativistic causality are less clear due to the different local lightcones (mathematically, there is no analogue of $s$ and $t$), although see \cite{Chen:2021bvg, deRham:2021bll,Arkani-Hamed:2021ajd, CarrilloGonzalez:2022fwg, Bellazzini:2021shn, Bellazzini:2022wzv, Caron-Huot:2022ugt, Caron-Huot:2022jli} for recent progress in implementing causality as a constraint in gravitational field theories. 

    An alternative way forward is to consider the weaker condition of \emph{non-relativistic} causality. This is the requirement that that any source which is local-in-time can only affect its future, and is the property that underpins the classic Kramers-Kronig relations for the dispersion of light in a medium. 
    In a cosmological context, this non-relativistic causality was recently used in \cite{Salcedo:2022aal} to find analogous dispersion relations for wavefunction coefficients.
    Here, we are going to explore another consequence of non-relativistic causality: namely that it forbids loops (closed time-like curves).

\paragraph{Feynman tree theorem.}
In the context of amplitudes, this aspect of non-relativistic causality was exploited by Feynman long ago \cite{Feynman:1963ax, DeWitt:1967ub} to derive what is often dubbed ``Feynman's tree theorem''. 
This theorem provides a systematic way to cut open any closed loop in a Feynman diagram and replace it by a sum over simpler diagrams with fewer loops. 
It makes manifest the fact that, while drawing loops is often helpful to visualise certain quantum corrections, since no physical signal could ever travel on such a trajectory they must be expendable. 
Feynman's theorem has since been developed in various ways \cite{Vaman:2005dt, Brandhuber:2005kd, Caron-Huot:2010fvq}, and is now widely used in the literature, often together with on-shell unitarity and recursion relations (see e.g. \cite{Bern:1994zx, Bern:1994cg} and \cite{Britto:2004nc,Anastasiou:2006gt,Forde:2007mi,Bern:2007dw,Berger:2009zb,Badger:2008cm} for some early work in that direction). 
One prominent incarnation of Feynman's theorem is the ``loop-tree duality'' of \cite{Catani:2008xa,Rodrigo:2008fp} (see also \cite{Bierenbaum:2010cy,Bierenbaum:2010xg, Bierenbaum:2012th,Buchta:2014dfa, Buchta:2015xda}), which can be used to efficiently implement IR subtractions numerically \cite{Buchta:2015wna,Hernandez-Pinto:2015ysa,Sborlini:2016gbr,Sborlini:2016hat,Driencourt-Mangin:2017gop}.
A version of Feynman's tree theorem has even been applied to the two-point function on a quasi-de Sitter background in order to determine the signs of higher-loop diagrams without explicitly computing them \cite{Tsamis:1996qq}. 
In short, Feynman's remarkably simple relation has had a remarkable impact on how we compute and understand loop corrections.

\paragraph{Main results.}
	In this work, we consider the constraints imposed by non-relativistic causality
	for the cosmological wavefunction.  
Concretely, our main results are to:
\begin{itemize}

\item[(i)] derive a cosmological analogue of the Feynman tree theorem, which can be used to express a Feynman(-Witten) diagram with any number of loops in terms of purely tree-level diagrams, 

\item[(ii)] use this tree theorem to determine, from purely tree-level data, simple expressions for the one-loop wavefunction coefficients and the corresponding one-, two- and three-point cosmological correlators, 

\item[(iii)] demonstrate a cosmological analogue of the KLN theorem, namely that particular singularities in the wavefunction cancel out in cosmological correlators,

\item[(iv)] describe a novel bootstrap procedure which combines unitarity cutting rules, the absence of unphysical singularities and our causality condition in order to determine any tree-level exchange diagram from simpler cut diagrams with no internal lines. 

\end{itemize}
	As a result, \emph{any} Feynman-Witten diagram (with an arbitrary number of edges and loops, and on an arbitrary time-dependent spacetime background) can now be expressed in terms of the tree-level single-vertex diagrams of the theory.

As an example of the utility of our tree theorem, we consider the problem of determining from the wavefunction the equal-time correlation functions relevant for cosmology.  
We find that many (otherwise mysterious) cancellations between loop- and tree-level contributions can be explained by our Cosmological Tree Theorem, which provides a systematic way of simplifying the map from wavefunction to correlators. 
For example, for the one-loop power spectrum (or one-loop bispectrum), the standard approach via the Born rule generates $7$ (or $13$) separate Feynman-Witten wavefunction diagrams. 
Applying the Cosmological Tree Theorem reduces this to just $2$ (or $5$) tree-level diagrams for massless fields. 
The cancellations which take place between loop- and tree-level coefficients are analogous to the cancellation of IR divergences that takes place when computing an observable cross-section from a scattering amplitude: a phenomenon often called the ``KLN theorem'' \cite{Kinoshita:1962ur,Lee:1964is} (though see the recent discussion in \cite{Frye:2018xjj,Hannesdottir:2019opa} for several subtleties in this cancellation).
The Cosmological Tree Theorem~\eqref{eqn:CTT} for wavefunction coefficients and the resulting Cosmological KLN theorem (together with the explicit results \eqref{eqn:PowerspectrumCTTdimreg} and \eqref{eqn:B_to_I} for the one-loop power spectrum and bispectrum) are the main results of this work.

\paragraph{Synopsis.}
We end this introduction with a description of our notation and a technical summary of the above results (i-iv). 
Then in section~\ref{sec:tree} we begin in earnest by deriving simple consequences of causality for tree-level diagrams and describing a bootstrap procedure for determining an arbitrary exchange diagram from its cut contact diagrams. 
In section~\ref{sec:loop} we move on to loop diagrams and illustrate our new Cosmological Tree Theorem with several simple examples, before providing a general proof for an arbitrary diagram.
In section~\ref{sec:corr}, we show how the Cosmological Tree Theorem can be used to identify the analytic structure of both the wavefunction and the corresponding in-in correlators, and give a simplified mapping from the wavefunction coefficients to the one-loop power spectrum (the longer expression for other correlators can be found in Appendix~\ref{app:bispectrum}).
Finally, we conclude in section~\ref{sec:disc} with a discussion of the future directions that this opens up. 

\subsection{Notation and conventions}
\label{sec:conventions}

While the majority of our notation is fairly standard, we collect it here for ease of reference. 
A reader already familiar with recent literature on the wavefunction of the Universe may wish to simply note the definition~\eqref{eqn:Disc_def} of our $\disc{}$/$\Disc{}$ operations and their graphical representation~\eqref{eqn:Disc_graph} and then proceed to our main results in section~\ref{sec:summary}.

\paragraph{Time and momenta.}
We will consider quantum field theories on an isotropic time-dependent spacetime background, which can be written as $ds^2 = a^2 (t) \left( - d t^2 + d \bfx^2 \right)$ in conformal coordinates.
We label each field $\phi_{\bfk} (t)$ by its temporal location $t$ and spatial momentum $\bfk$ using the spatial Fourier transformation, 
\begin{align}
	f(\bfx) &= \int \frac{d^3 \bfk}{ (2 \pi)^3} \; f_\bfk\; \exp(i\bfk \cdot \bfx ) \,,
\end{align}
which commutes with time derivatives and integrals.
We adopt the following shorthands for time and momenta integrals, and frequently absorb factors of $(2\pi)^3$ into the Dirac delta function,
\begin{align}
		\int_t &\equiv \int_{-\infty}^{0} d t
  \;, \quad \int_{\bfp} \equiv \int \frac{d^3 \bfk}{ (2 \pi)^3 }  \; , \quad
	\tilde{\delta}^3 \left( \bfk \right) \equiv ( 2 \pi )^3 \delta^3 \left( \bfk \right) \; . 
\end{align}
Bold type always refers to spatial vectors, and we write their magnitude as $k\equiv | \bfk | \equiv +\sqrt{ \delta^{ij} \bfk_i \bfk_j } \equiv \sqrt{\bfk \cdot \bfk}$.

\paragraph{Wavefunction coefficients.}
The central object of our study is the wavefunction of the Universe, $\Psi$, which describes how an initial Bunch-Davies vacuum\footnote{
When discussing a general conformally flat spacetime, we refer to the vacuum which coincides with the instantaneous ground state in the far past as the Bunch-Davies vacuum.
} state $|\Omega \rangle$ evolves in time from the far past to conformal time $t =0$. This time evolution is implemented by the unitary operator $\hat{U}$, and we project the resulting state onto the field eigenstate $| \phi \rangle$ at $t =0$, which gives the wavefunction,
\begin{align}
	\Psi [ \phi  ] \equiv  \langle \phi  | \hat{U} | \Omega \rangle &\equiv \exp \left[+ 
	\sum_{n}^{\infty} \int_{\bfk_1,..,\bfk_n}\frac{1}{n!} \psi_{ \bfk_1 ... \bfk_n }  \phi_{\bfk_1} ... \phi_{\bfk_n}  \right]\;  .
	\label{eqn:psi_def}
\end{align}
The $\psi_{\bfk_1 ... \bfk_n}$ appearing in the exponential parameterisation of the wavefunction are the \emph{wavefunction coefficients}. 
We occasionally write $\psi_n$ as shorthand for $\psi_{\bfk_1 ... \bfk_n}$.

\paragraph{Propagators.}
To develop a perturbative expansion for these coefficients, we define the usual \emph{bulk-to-bulk} and \emph{bulk-to-boundary} propagators,
\begin{align}
	\langle \phi = 0 | i \hat{\Pi}_{\bfk'} \hat{\phi}_{\bfk} (t) | \Omega \rangle &\equiv K_k (t) \, \delta^3 \left( \bfk + \bfk' \right) \nonumber \\ 
	\langle \phi = 0 | T \hat{\phi}_{\bfk} (t_1) \hat{\phi}_{\bfk'} (t_2) | \Omega \rangle &\equiv G_k (t_1, t_2 ) \, \delta^3 \left( \bfk + \bfk' \right)
\end{align} 
where we have written the field and its conjugate momentum in the Heisenberg picture, and when no time argument is given these are to be evaluated at $t = 0$. $T$ denotes time-ordering. 
In the free theory \eqref{eqn:Lfree}, these are related by\footnote{
Note that $G_k (t_1, t_2)$ differs from the usual Feynman propagator by the final term in \eqref{eqn:G_to_K}, which is a result of the different bra boundary condition (which is the zero-field eigenstate rather than the vacuum).
}, 
\begin{align}
 G_k ( t_1, t_2 ) = i P_k \left[   
 \Theta \left( t_1 - t_2 \right) K_{k}^* (t_1) K_k (t_2)
 +  \Theta \left( t_2 - t_1 \right) K_{k} (t_1) K^*_k (t_2)
 - K_k (t_1) K_k (t_2)
 \right]
 \label{eqn:G_to_K}
\end{align} 
where $P_k$ is the free-theory power spectrum, related to the wavefunction by $ 2 \text{Re} \, \psi_{\bfk \bfk'}  \equiv - \delta^d \left( \bfk + \bfk' \right)/ P_k$. 
To prevent a proliferation of $\delta$ functions, it will often be convenient to write the power spectrum as,
\begin{align}
 P_{\bfk \bfk'} \equiv P_k \;  \tilde{\delta}^d \left( \bfk + \bfk' \right) \; . 
 \label{eqn:P_def_intro}
\end{align}
For concreteness, when we give examples below for a massless scalar field on Minkowki or de Sitter, the corresponding power spectrum and propagators are given by,
\begin{align}
&\text{Minkowski:} \;\; &a(t) &= 1 \; ,   &P_k &= \frac{1}{2k} \; , &K_k (t) &= e^{i k t} \nonumber \\ 
&\text{de Sitter:} \;\; &a(t) &= \frac{- 1}{H t} \; ,  &P_k &= \frac{H^2}{2 k^3} \; , &K_k (t) &= e^{i k t} \left( 1 - i k t \right) 
\label{eqn:G_eg_intro}
\end{align}
together with \eqref{eqn:G_to_K} for $G_k (t_1, t_2)$. 

\paragraph{Perturbation theory.}
We consider theories with a (generally time-dependent) Hamiltonian that can be separated into a solvable quadratic part plus an interacting part which can be treated perturbatively.
For the quadratic part, a scalar field would have the Lagrangian,
\begin{align}
	\mathcal{L}_{\rm free} (t) = \int d^3 \bfx \,  \sqrt{-g} \left(  \tfrac{1}{2} g^{\mu\nu}  \nabla_\mu \phi \nabla_\nu \phi - \tfrac{1}{2} m^2 \phi^2 \right) 
	\label{eqn:Lfree}
\end{align}
where $g_{\mu \nu} (t)$ is the time-dependent background metric\footnote{
We focus on the evolution of quantised fluctuations on a fixed time-dependent background (i.e. a decoupling limit $M_P \to \infty$ in which backreaction can be neglected).
}.	
Our only requirement on the interactions is that the Hamiltonian $\mathcal{H}_{\rm int} (t)$ be Hermitian (i.e. time evolution remains unitary in the interacting theory). 

Since each wavefunction coefficient in \eqref{eqn:psi_def} corresponds to the connected part of,
\begin{align}
	\frac{  \langle  \phi = 0 | i \hat{\Pi}_{\bfk_1}  ...  i \hat{\Pi}_{\bfk_n} \hat{U} | \Omega \rangle  }{ \langle \phi = 0 | \hat{U} | \Omega \rangle } = \frac{ 1 }{\Psi [ 0 ]} \frac{\delta^n \Psi [\phi ]}{ \delta \phi_{\bfk_1}  ... \delta \phi_{\bfk_n}  } \bigg|_{\phi  = 0} \; , 
	\label{eqn:Psi_matrix}
\end{align}
we can represent this matrix element as a diagrammatic series over Feynman(-Witten) diagrams in which,
\begin{itemize}

	\item Each ``external'' line (which connects an interaction vertex at time $t$ to the boundary) carrying momentum $\bfk$ represents a factor of the free-theory $K_k (t)$, 
	
	\item Each ``internal'' line (which connects two interaction vertices at times $t_1$ and $t_2$) carrying momentum $\bfp$ represents a factor of the free-theory $G_p (t_1, t_2)$,
	
	\item Each vertex which connects $n$ lines of momenta\footnote{
We use a condensed notation in which $\bfk$ represents the momentum \emph{and also} any other relevant quantum numbers which distinguish the fields, e.g. their species (if more than one is present), their helicity (if spin is non-zero), etc. The $\delta/\delta \phi_{\bfk}$ then corresponds in the obvious way to taking a functional derivative with respect to whatever field is identified by $\bfk$. 	
	} $\{ \bfk_1 , ... , \bfk_n \}$ at time $t$ corresponds to a factor of $\delta^n \mathcal{L}_{\rm int} (t) / \delta \phi_{\bfk_1} ... \phi_{\bfk_n} $,
	
	\item All vertex times and internal momenta are then integrated over, i.e. perform $\int_{-\infty}^0 dt$ for every vertex and $\int_{\bfp}$ for every internal line, 
	
	\item Finally, our normalisations are such that we include a factor of $+i$ for every vertex and $-i$ for every propagator: this leads to an overall $i^{1-L}$, where $L$ is the number of loops in the diagram \cite{Melville:2021lst}.

\end{itemize}
For instance, a theory in which $\mathcal{L}_{\rm int}$ contains the interaction $\tfrac{1}{3!} \lambda (t) \phi^3$ would also contain the diagram, 
\begin{align}
    \fig{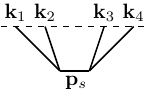} &\equiv i \int_{t_1 t_2} \;  \lambda (t_1) \lambda (t_2) \; K_{k_{1}}(t_{1})K_{k_{2}}(t_{1})G_{p_{s}}(t_{1},t_{2})K_{k_{3}}(t_{2})K_{k_{4}}(t_{2}) \, \; \tilde{\delta} \left( \sum_{b=1}^4 \bfk_b  \right)\label{eqn:psi4exc}
\end{align}
where $\bfp_s = \bfk_1 + \bfk_2$ is fixed by momentum conservation.
Expanding \eqref{eqn:Psi_matrix} order by order in $\mathcal{H}_{\rm int}$ then gives each wavefunction coefficient as an expansion in Feynman diagrams with increasing numbers of loops, 
\begin{align}
 \psi_{n} =  \psi_{n}^{\rm tree} +  \psi_{n}^{1\text{-loop}} + \psi_{n}^{2\text{-loop}} + ...
 \label{eqn:loop_exp} 
\end{align}
Explicitly, in the main text we make use of,
\begin{align}
 \psi_{\bfk_1 \bfk_2 \bfk_3}^{\rm tree} &=  \fig{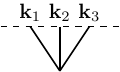}  \; ,   \label{eqn:psi_intro_diagrams} \\ 
 \psi_{\bfk_1 \bfk_2 \bfk_3 \bfk_4}^{\rm tree} &= \fig{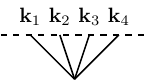} + \fig{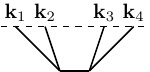} + \fig{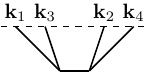} + \fig{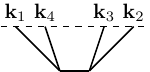}  \nonumber \\ 
 \psi_{\bfk_1 \bfk_2}^{\rm 1-loop} &=  \frac{1}{2} \; \fig{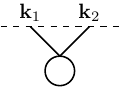} +\frac{1}{2} \; \fig{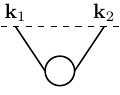}+\frac{1}{2} \; \fig{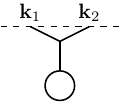}      \; , \;\; \psi_{\bfk}^{\rm 1-loop} = \frac{1}{2} \fig{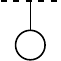}   \nonumber
\end{align}
where the factors of $1/2$ for the loop diagrams are the usual symmetry factors which appear when expanding a matrix element of the form~\eqref{eqn:Psi_matrix}\footnote{
These symmetry factors appear in precisely the same way for scattering amplitudes, and we discuss them further in Appendix~\ref{app:symm}. 
}. 
In Appendix~\ref{app:bispectrum} we give the analogous expansions for $\psi_5^{\rm tree}$, $\psi_3^{\rm 1-loop}$ and $\psi_1^{\rm 2-loop}$ as further examples.

\paragraph{Discontinuities.}
The wavefunction coefficients will, in general, be non-analytic functions of the spatial momenta due to the dependence of $K_k (t)$ on the ``energy'' $\omega_k$ of each external line.
On Minkowski, $\omega_k = +\sqrt{k^2 + m^2}$ is given by the usual on-shell condition for free propagation. On a general time-dependent spacetime, we identify $\omega_k$ with the phase of $K_k (t)$ in the far past, $K_k (t) \sim e^{+ i \omega_k t}$: for quasi-de Sitter spacetimes, this gives $\omega_k = k$ for any finite\footnote{
Note that in the flat space limit, $H \to 0$ with $m, k$ fixed, $K_k (t)$ reduces to the Minkowski propagator and $\omega_k$ becomes $\sqrt{k^2 + m^2}$. 
} $m^2/H^2$.
As discussed in \cite{Goodhew:2020hob, Cespedes:2020xqq, Melville:2021lst, Goodhew:2021oqg}, we can exploit this particular non-analyticity to project any $K_k$ or $G_p$ within a diagram onto its real or imaginary part.
The essential idea is to analytically continue a particular energy (or subset of energies) to negative values and then exploit the Hermitian analyticity of the propagators,  
\begin{align}
	K_{k} (t) \big|_{\omega_k \to - \omega_k} &= K^*_k (t) \;,    &G_{p} (t_1, t_2) \big|_{\omega_p \to - \omega_p} &= G_p^* (t_1, t_2) \;. 
	\label{eqn:hermitian_analyticity}
\end{align}
This is usually achieved by continuing $k \to - k$ in order to cross a branch cut (e.g. sending $+\sqrt{k^2 + m^2} \to - \sqrt{k^2 + m^2}$). 

In this work, we view each wavefunction coefficient or Feynman diagram as a function of both the momenta $\{ \bfk \}$ and their energies $\{ \omega_k \}$, and make use of the following two ``discontinuity'' operations,
\begin{align}
	\disc{k_1, ..., k_j} 
	\left[  \; F \left(  \omega_{k_1} , ... , \omega_{k_n} ; \{ \bfk \}  \right) \; \right] &\equiv F \left( \omega_{k_1}, ... , \omega_{k_n} ; \{ \bfk \}  \right)  -   F \left(  - \omega_{k_1} , ... , - \omega_{k_j} , \omega_{k_{j+1}} , ...,  \omega_{k_n} ; \{ \bfk \}   \right)   \; , \nonumber \\ 
	\Disc{k_1, ..., k_j} 
\left[  \; F \left(  \omega_{k_1} , ... , \omega_{k_n} ; \{ \bfk \}  \right) \; \right] &\equiv F \left( \omega_{k_1}, ... , \omega_{k_n} ; \{ \bfk \}  \right)  -   F^* \left(  \omega_{k_1}, ... ,  \omega_{k_j} , -\omega_{k_{j+1}} , ... , - \omega_{k_n} ; \{ -\bfk \}   \right)   \; . 
 \label{eqn:Disc_def}
\end{align}
In words: $\disc{}$ corresponds to analytically continuing all indicated energies and is used to extract the imaginary part of external lines.
On the other hand, $\Disc{}$ corresponds to analytically continuing all energies \emph{except} those indicated, and is used to extract the imaginary part of internal lines.
For example,
\begin{align}
\disc{k_1} \left[  G_{p_s} K_{k_1} K_{k_2}  \right] &=  2i G_{p_s} K_{k_2}  \, \text{Im} \, K_{k_1}  \; ,   
 &\Disc{k_2} \left[   G_{p_s} K_{k_1} K_{k_2} \right] &=   2 i K_{k_1} \, \text{Im} \left(  G_{p_s} K_{k_2} \right)  \; . 
\end{align}
Note that for contact Feynman diagrams with no internal lines, $\disc{k} \left[  i \psi \right] = \Disc{k} \left[  i \psi \right]$. 

In contrast to previous works\footnote{
In particular, while our $\Disc{}$ coincides with the $\Disc{}$ of~\cite{Melville:2021lst} and the $\DiscT{}$ of~\cite{Baumann:2021fxj}, the use of $\disc{}$ appears novel.
}, we make use of these two different operations in order to avoid ever analytically continuing an internal energy. As far as we are concerned, each $\psi_n$ depends only on the $\{ \bfk \}$ and $\{ \omega_k \}$ of the external momenta, and we may only analytically continue those (i.e. internal energies like $p_s = | \bfk_1 + \bfk_2|$ are treated as a function of $\bfk_1$ and $\bfk_2$ due to momentum conservation)\footnote{
This was recently made precise in \cite{Salcedo:2022aal} through a particular ``off-shell'' extension of the wavefunction on Minkowski. An analogous construction for general time-dependent spacetime backgrounds will be reported elsewhere \cite{MelvillePimentel}. 
}.

To represent the discontinuities diagrammatically, we introduce graphs with highlighted lines to denote taking a single discontinuity with respect to those momenta. 
If only external lines are highlighted, it is the $\disc{}$ operation, while if any internal line is highlighted it is the $\Disc{}$ operation. For instance\footnote{
The factors of $i$ for highlighted internal lines are such that the $\Disc{}$ selects $\text{Im} \, G_p$ at tree-level and $\text{Re} \, G_p$ for a one-loop diagram (given the factor of $i^{1-L}$ in our Feynman rules). 
},
 \begin{align}
\fig{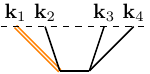} &\equiv \disc{k_1} \left[ \fig{exch_s} \right] \; , 
 &\fig{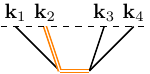} &\equiv -i \, \Disc{k_2} \left[ 
i \;  \fig{exch_s}
 \right] \; . 
 \label{eqn:Disc_graph}
 \end{align}
If no lines are highlighted, no discontinuity is to be taken.
It will also be convenient to use a dotted external line to denote the analytically continued propagator $K_{k}^* (t)$, since then the $\disc{}$ corresponds to,
\begin{align}
\fig{disc_eg_1}  \; &\equiv \;  \fig{exch_s}  \; - \; \fig{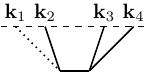} \; . 
\end{align}

Finally, a comment about the terminology ``discontinuity''. The name is inspired by the analogy with scattering amplitudes, where the discontinuity across e.g. the $s$-channel branch cut can be written in two equivalent ways,
\begin{align}
\lim_{\epsilon \to 0} \left[ \mathcal{A}_{12 \to 34} ( s + i \epsilon , t ) - \mathcal{A}_{12 \to 34} (s - i \epsilon , t ) \right] =  \mathcal{A}_{12 \to 34} ( s  , t ) - \mathcal{A}^*_{34 \to 12} (s , t ) \; . 
\end{align}
Morally, the left-hand-side corresponds to our $\disc{}$ operation (since $s \pm i\epsilon$ gives an energy $\omega_s = \pm \sqrt{s}$) and the right-hand-side corresponds to our $\Disc{}$ operation (since by CPT the time-reversed process can be viewed as flipping the signs of all momenta). 
To make the analogy more concrete, we showed in \cite{Salcedo:2022aal} that the Minkowski wavefunction in the complex $\omega^2_1 = k_1^2 + m_1^2$ plane has a branch cut only along the positive real axis, and $\disc{\omega_1}$ is precisely the discontinuity across this cut.

\subsection{Summary of main results}
\label{sec:summary}

To complement our more thorough and pedagogical presentation in sections~\ref{sec:wvfn} and~\ref{sec:corr}, here we provide a self-contained summary of the main narrative.

\paragraph{Review of cutting rules from unitarity.}
Existing cutting rules leverage the \emph{unitarity} of time evolution in the interacting theory (together with the Hermitian analyticity of the free-theory propagators~\eqref{eqn:hermitian_analyticity}) in order to reduce the number of internal lines in a Feynman diagram.
For instance, the simplest such rule is \cite{Melville:2021lst, Goodhew:2021oqg}, 
\begin{align}
\text{Unitarity} \;\; \Rightarrow \;\; 	\fig{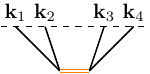}
	=
	- \int_{\bfq \bfq'} P_{\bfq \bfq'} \left(\fig{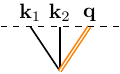}\right) \left(\fig{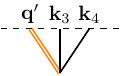}\right) \; , \label{eqn:unit_eg}
\end{align}
which says that the $\Disc{}$ operation (depicted by a highlighted line) effectively ``cuts'' an internal line, brings the two resulting half-edges to the boundary and multiplies by a factor of the boundary power spectrum $P_{\bfq \bfq'}$.\footnote{
For later convenience, we have chosen to express these cuts using an integral over the resulting external momenta of the cut line. In tree-level examples such as this one, these momenta integrals are trivial thanks to the $\delta$ functions inside $P_{\bfq \bfq'}$ and each wavefunction coefficient. 
} 
This identity can then be used to fix the discontinuity of $\psi_4$ in terms of its cuts into $\psi_3 \times \psi_3$, 
\begin{align}
- i \,  \Disc{} \left[ i\,  \psi^{\rm tree}_{\bfk_1 \bfk_2 \bfk_3 \bfk_4} \right] = \sum_{\rm perm.}^3 \int_{\bfq \bfq'} P_{\bfq \bfq'} \,  \disc{q} \left[ \psi^{\rm tree}_{\bfk_1 \bfk_2 \bfq} \right] \disc{q'} \left[  \psi^{\rm tree}_{\bfk_3 \bfk_4 \bfq'}  \right] \; .
\end{align}
in direct analogy with the usual Cutkosky cutting rules for scattering amplitudes.

\paragraph{New cutting rules from causality.}
In this work, we leverage \emph{causality} of the free theory to further constrain the perturbative wavefunction coefficients. 
For instance, the entire exchange diagram above can be written as,
\begin{align}
\fig{exch_s}
	=
- \int_{\bfq \bfq'} P_{\bfq \bfq'} \left(\fig{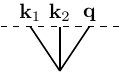}\right) \left(\fig{con3_R_C}\right) + \fig{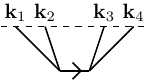} 
\label{eqn:exch_eg}
\end{align}
where a directed arrow from $t_1$ to $t_2$ represents the retarded propagator,
\begin{align}
 G^R_p (t_1, t_2 ) \equiv 2 P_p\text{Im} \left[ K_p (t_1) K_p^* (t_2)  \right] \Theta (t_1 - t_2 )  
 = G_{p}(t_{1},t_{2}) - 2P_{p}K_{p}(t_{1})\Im \left[ K_{p}(t_{2}) \right]  \; .
 \label{eqn:GR_def_intro}
\end{align}
Since $G_p^R$ is real its discontinuity vanishes, so taking $\Disc{}$ of \eqref{eqn:exch_eg} immediately reproduces \eqref{eqn:unit_eg}. 
However, while the final term in \eqref{eqn:exch_eg} cannot be constrained by unitarity alone, it can be constrained by causality. 
In particular, we show that,
\begin{align}
\text{Causality} \;\; \Rightarrow \;\; \fig{exch_R}  - \fig{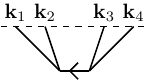}  =  - i P_{p_s}  \disc{q , q'} \left[  \fig{con3_L}  \, \fig{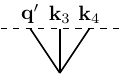} \right]
\label{eqn:caus_cut_intro}
\end{align}
as a consequence of the fact that $G_p^R (t_1, t_2 ) - G_p^R (t_2, t_1)$ is a smooth function of $t_1$ and $t_2$ and so can be written in terms of $K_p$'s without any step functions.   
This represents a qualitatively new way to cut diagrams, and demonstrates how causality can complement unitarity in fixing wavefunction coefficients.
Furthermore, we find that demanding the absence of certain unphysical (``folded'') singularities can be used to completely fix the remaining freedom in the retarded exchange diagram,
\begin{align}
\begin{gathered} 
    \text{Analyticity} \\  \text{\small (no folded} \\ \text{\small singularities)}
    \end{gathered}
 \;\; \Rightarrow \;\; \fig{exch_R}   \;\; \text{is fixed by} \;\;   \fig{con3_L}  \; ,
\end{align}
which is ultimately a re-writing of the bootstrap approach developed in \cite{Jazayeri:2021fvk}---the main difference\footnote{
There is another, more subtle, difference: the bootstrap procedure we describe here based on \eqref{eqn:exch_eg} and $G_p^R$ does not require $\mathcal{H}_{\rm int}$ to be unitarity, it relies only on unitarity/causality of the free theory.
} is that our retarded exchange is guaranteed to be an even function of the exchanged momenta, which can simplify the bootstrap procedure.
Altogether, the combination of unitarity, causality and a particular analytic structure are enough to completely determine this (and indeed \emph{any} tree-level) exchange diagram in terms of its cut diagrams with fewer internal lines.

\paragraph{Causality at loop level.}
The consequences of causality are even more striking for loop diagrams. 
While unitarity alone can determine the $\Disc{}$ of an arbitrary loop diagram in terms of diagrams with fewer internal lines \cite{Melville:2021lst}, the combination of unitarity and causality immediately fixes the entire diagram (both $\Disc{}$ and non-$\Disc{}$ parts). 
The simplest example of this phenomenon is the following one-loop diagram,
\begin{align}
	 \fig{psi2_a} &= - \int_{\bfq \bfq'} P_{\bfq \bfq'} \left( \fig{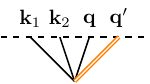} \right) + \fig{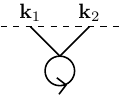} \; , 
	 \label{eqn:psi2_eg}
\end{align}
which has been re-written in terms of $G_p^R$ using \eqref{eqn:GR_def_intro}. 
Taking the $\Disc{}$ again removes the final term and reproduces the unitarity cutting rule of \cite{Melville:2021lst}. The new observation that we exploit here is that,
\begin{align}
\text{Causality} \;\; \Rightarrow \;\;  \fig{loop_R} = 0 \; . 
\label{eqn:no_loops_intro}
\end{align} 
Causality forbids loops (closed time-like curves), and hence once any loop diagram is expanded in terms of retarded $G_p^R$ propagators \emph{all of the loops must vanish}. 
Consequently, expressions like \eqref{eqn:psi2_eg} can express any loop diagram in terms of (momentum integrals of) tree-level diagrams.

\paragraph{Cosmological tree theorem.}
Our main result is to do this systematically for any closed loop within a diagram $D$, and hence prove the \emph{cosmological tree theorem} \eqref{eqn:CTT}. Schematically, this takes the form, 
\begin{align}
 - D  \; = \;\; \sum_{ \substack{\text{cuts} \\ C }} \; \;\; \int_{\substack{ \text{cut line} \\ \text{momenta} }}  \;\; \; \prod_{ \substack{ \text{subdiagrams} \\ n}} \; \disc{  }  \left[ D^{(n)}_C \right]
 \label{eqn:CTT_schematic}
\end{align}
where the sum is over all possible ways of cutting one or more internal lines in the loop, and as a result of the cuts $C$ the diagram may split into disconnected components which we label $D_C^{(n)}$. 
As a result of the cuts, every term on the right-hand-side has at least one fewer loops than the original diagram. Applied recursively to every closed loop in the diagram, this can be used to replace any arbitrary loop-diagram with (momentum integrals of) tree-level diagrams.
As a simple example, the $\psi_2^{\rm 1-loop}$ given in \eqref{eqn:psi_intro_diagrams} can be written as,
\begin{align}
2 \psi_{\bfk_1 \bfk_2}^{\text{1-loop}} 
 =
 \int_{\bfq \bfq'} P_{\bfq \bfq'} \, \disc{ q'} \left[ \psi_{\bfk_1 \bfk_2 \bfq \bfq'}^{\rm tree}  \right] + \int_{ \substack{ \bfq_1 \bfq_1' \\ \bfq_2 \bfq_2' } } P_{\bfq_1 \bfq_1'} P_{\bfq_2 \bfq_2'} \, \disc{ q_2' } \left[ \psi_{\bfk_1 \bfq_1 \bfq_2'} \right] \disc{ q_1'} \left[ \psi_{\bfk_2 \bfq_2 \bfq_1'} \right] \; . 
 \label{eqn:psi2_intro}
\end{align}
While the momenta integrals may still pose a challenge, this has achieved the following important simplification: \emph{there are no longer any time integrals}. In practice, this means that once the tree-level wavefunction coefficients have been determined to a sufficiently high number of external legs, 
then the loop momenta integrands are completely fixed by the above tree theorem. This is an important step towards a general Landau analysis of the singularities in $\psi_n$, since from $\psi_n^{\rm tree}$ alone we can now determine all possible poles in the loop integrand.

\paragraph{Cosmological KLN theorem.}
Finally, as an application of our new relations, we consider how causality constrains cosmological correlators. 
While the standard Born rule mapping from wavefunction to correlator introduces many terms at loop-level, our tree theorem can be used to considerably simplify this map.
In particular, we show that the power spectrum of massless fields at one-loop order can be written as (a momentum integral of) just two tree-level diagrams,
\begin{align}
&\frac{ \langle \Omega | \hat{\phi}_{\bfk_1} \hat{\phi}_{\bfk_2} | \Omega \rangle }{ P_{k_1} P_{k_2} } = \frac{ P_{\bfk_1 \bfk_2} }{ P_{k_1} P_{k_2} }  +  \int_{\bfq \bfq'} P_{\bfq \bfq'} \, \text{Re} \left[ 
  \fig{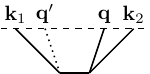} + \fig{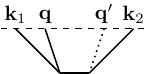}
 \right]  \label{eqn:P_CTT_intro} \\
 &+  \int_{\substack{\bfq_1 \bfq_1' \\ \bfq_2 \bfq_2'}} P_{\bfq_1 \bfq_1'} P_{\bfq_2 \bfq_2'} \, \left( 
2  \text{Re} \left[  \fig{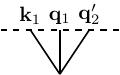} \right]\text{Re} \left[  \fig{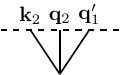} \right] - \text{Re} \left[  \fig{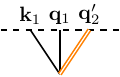}   \, \fig{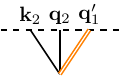} \right]
 \right) \nonumber 
\end{align}
where the dotted line represents the analytic continuation to negative energy $\omega_{q'}$ and $P_k$ is the free-theory power spectrum. 
The analogous expression for the one-loop bispectrum is given in (\ref{eqn:B_to_I}, \ref{eqn:B_CTT_1}, \ref{eqn:B_CTT_2}). 
Interestingly, we find that the pattern of analytic continuations and cuts is always such that no vertex (or connected set of vertices) can have a total energy which depends on both the total external energy and the loop momenta\footnote{
For instance in \eqref{eqn:P_CTT_intro}, there is no vertex (or set of vertices) which depends on both $\omega_{k_1} + \omega_{k_2}$ and $\bfq = - \bfq'$. 
}. 
In the wavefunction coefficients, such vertices generically do appear and lead to branch cuts in the total external energy once the loop integration is performed. These branch cuts do not appear in equal-time correlators, and our tree theorem makes their cancellation manifest. 
We demonstrate that this cancellation is not confined to the power spectrum or bispectrum, but in fact takes place very generally. Given the close analogy with the cancellation of IR divergences in amplitudes, we refer to this result as the \emph{cosmological KLN theorem}.

\section{Wavefunction coefficients}
\label{sec:wvfn}

In this section, we describe how the causal properties of the retarded propagator $G_p^R$ can be used to constrain the wavefunction coefficients defined in~\eqref{eqn:psi_def}. 
We begin with a discussion of tree-level diagrams in subsection~\ref{sec:tree}, and then move on to loops in subsection~\ref{sec:loop}.

\subsection{Causality and a tree-level bootstrap}
\label{sec:tree}

Here we consider the constraints that causality can place on a general tree-level exchange diagram, which we represent diagrammatically as,
\begin{align}
\fig{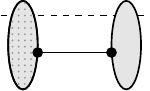} \; . 
\label{eqn:blobs}
\end{align}
Each gray blob represents a particular subdiagram with arbitrarily many external and internal lines (i.e. a general function of $K_k$ and $G_p$ and their conjugates), and they are connected by the single internal line shown. 
We have added a dotted pattern to the left blob to indicate that it need not be the same subdiagram as the right blob.

\paragraph{Introducing directed edges.}
From the definition \eqref{eqn:GR_def_intro} of the retarded propagator and the definition \eqref{eqn:Disc_graph} of the discontinuity, we can trade any internal line (factor of $G_p$) for a directed line (factor of $G_p^R$) as follows,
\begin{align}
\fig{blobs}
	\;\; =  \;\; \fig{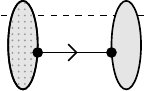}
- \int_{\bfq \bfq'} P_{\bfq \bfq'} \, \fig{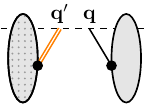}   \; .
\label{eqn:blobs_1}
\end{align}
Note that since $G_R$ is real, 
\begin{align}
 \Disc{} \left[ i G_p^R (t_1, t_2) \right]  = 0  \; ,
\end{align}
i.e. a directed internal line will vanish when highlighted. 
Taking the $\Disc{}$ of \eqref{eqn:blobs_1} therefore produces,
\begin{align}
\fig{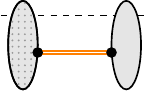}  = - \int_{\bfq \bfq'}  P_{\bfq \bfq'}  \fig{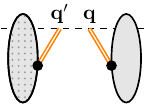}
\label{eqn:unitarity_cut}
\end{align}
which are the tree-level unitarity cutting rules of \cite{Melville:2021lst, Goodhew:2021oqg}. 
Note that while \eqref{eqn:unitarity_cut} requires Hermiticity of the interaction Hamiltonian (i.e. real coupling constants so the discontinuity on the left-hand-side implements $G_{p} \to 2i \, \text{Im} \, G_{p}$), the causal representation~\eqref{eqn:blobs_1} does not (since $\disc{}$ does not require any complex conjugation, there is no reality restriction on the couplings)\footnote{
As described in Appendix~\ref{sec:overview}, the definition \eqref{eqn:GR_def_intro} which underpins \eqref{eqn:blobs_1} follows from the unitarity and causality of the free theory~\eqref{eqn:Lfree} only. 
}.

\paragraph{Cutting rules from causality.}
But now notice that since $G_p (t_1, t_2)$ is symmetric in $t_1$ and $t_2$, we could equally have written,
\begin{align}
\fig{blobs}
	\;\; =  \;\; \fig{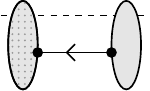}
- \int_{\bfq \bfq'} P_{\bfq \bfq'} \, \fig{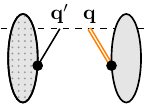}   \; ,
\end{align}
using $G_p^R (t_2, t_1)$, which is essentially the advanced propagator from $t_1$ to $t_2$.  
Additional cutting rules therefore follow from comparing these two different expressions: for instance their difference is,
\begin{align}
\fig{blobs_R}  \;\; -  \;\; \fig{blobs_L}
	=
 \int_{\bfq \bfq'} P_{\bfq \bfq'} \, \disc{q, q'} \left[  \fig{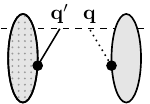}   \right]  \;. 
 \label{eqn:caus_cut_0}
\end{align}
where the dotted external line indicates the analytically continued propagator $K_q^*(t)$. 
So by combining retarded and advanced propagation (or equivalently $\psi ( \{ \bfk_L \} , \{ \bfk_R \} ) - \psi ( \{ \bfk_R \} , \{ \bfk_L \} )$, where $\{ \bfk_L \}$ and $\{ \bfk_R \}$ are the external momenta of the left- and right-hand blobs) we can effectively cut the internal line and produce two disconnected subdiagrams. 
This cutting rule is tied to causality, since it can be written as the propagator identity,
\begin{align}
i G_p^R (t_1, t_2) - i G_p^R (t_2, t_1 ) =  2 i P_p \text{Im} \left[  K_p (t_1 ) K_p^* (t_2) \right] \left( \Theta (t_1 - t_2 ) + \Theta ( t_2- t_1 ) \right) =  P_p \; \disc{p} \left[  K_p (t_1 ) K_p^* (t_2) \right]
\end{align} 
where the right-hand-side can only be written in terms of $K_p$ alone since $G_p^R$ is proportional to the Heaviside step function $\Theta (t_1 - t_2)$.
In an acausal theory, with no such retarded propagator (no way of separating sources in the future from responses in the past), it would not be possible to construct a cutting rule of the form~\eqref{eqn:caus_cut_0}\footnote{
Note that analogous cutting rules exist for $S$-matrix elements, in which case the internal line of the Feynman diagram would be the usual Feynman propagator and the highlighted line would correspond to simply $K_p^*$ rather than $\text{Im} K_p$. 
}. 
As mentioned in the introduction, these causal cutting rules complement previous unitarity cutting rules in that they constrain the part of the diagram with vanishing $\Disc{}$. 
What we have shown is that unitarity and causality together can fix the diagram up to a residual freedom which corresponds to adding,
\begin{align}
\fig{blobs_R}  \;\; +  \;\; \fig{blobs_L} \; .
\label{eqn:blobs+}
\end{align}
We give an overview and explicit comparison of the different unitarity and causality cutting rules in appendix~\ref{sec:overview}, where we also make precise the connection with unitarity / causality of the two-point function.

\paragraph{Causal bootstrap.}
The cutting rule~\eqref{eqn:caus_cut_0} is written in terms of graphs with directed edges, but ideally we would phrase such constraints in terms of the original diagram~\eqref{eqn:blobs_1} with undirected edges since this is what determines the wavefunction coefficients / cosmological correlators.
To do this requires the symmetric part \eqref{eqn:blobs+} of the retarded propagator. 
Remarkably, we have found that in practice it is not necessary to compute this object explicitly, since it can inferred from the unitarity/causality cuts by demanding the absence of certain unphysical singularities. 
This provides yet another way to ``bootstrap'' these wavefunction coefficients (i.e. determine them without actually computing nested time integrals like~\eqref{eqn:psi4exc}). 
Concretely, we input the following three ingredients:
\begin{itemize}

\item[(i)] \emph{Causality of the free theory}, so that each bulk-to-bulk $G_p (t_1, t_2)$ may be expanded in terms of a retarded $G_p^R (t_1, t_2) = i \Delta^S_p (t_1, t_2) \Theta (t_1 - t_2 )$ for some $\Delta_p^S$, 

\item[(ii)] \emph{Unitarity of the free theory}, which fixes $\Delta_p^S (t_1, t_2) = 2 P_p \text{Im} \left[ K_p (t_1) K_p^* (t_2) \right]$ in terms of the bulk-to-boundary propagator,

\item[(iii)] \emph{Bunch-Davies initial state}, so that the only singularities are at kinematics for which the total energy entering one or more vertex vanishes,

\end{itemize}
and the output is a procedure for determining any tree-level exchange diagram from its cut diagrams with fewer internal lines. 
It was previously shown in \cite{Jazayeri:2021fvk} that the absence of certain unphysical (``folded'') singularities can be combined with unitarity of the interacting theory to bootstrap an exchange diagram from its unitarity cuts. 
The approach described here is essentially the same, but we apply this bootstrap to determine the retarded part of the exchange~\eqref{eqn:blobs+} rather than the full diagram: this means that we only require unitarity of the free theory (not necessarily of the interacting theory) and in practice can exploit the fact that retarded propagator is a real, even function of $p$ to simplify the computations and remove the need for any further conditions (locality, etc.) to completely fix the exchange diagram up to contact terms.

\paragraph{An example.}
Before describing the general procedure, let us give a simple example. 
Suppose we have determined the contact $\psi_3^{\rm tree}$ diagram in \eqref{eqn:psi_intro_diagrams}, and wish to determine from it the exchange diagrams in $\psi_4^{\rm tree}$.
Replacing the internal line with a retarded propagator using \eqref{eqn:GR_def_intro} produces \eqref{eqn:exch_eg}, in which the first term is determined by $\psi_3^{\rm tree}$ but the second term,
\begin{align}
\fig{exch_R}   \equiv  P_{p_s} \disc{p_s} \left[
 \int_{t_1 t_2} K_{k_1} (t_1) K_{k_2} (t_1) K_{k_3} (t_2) K_{k_4} (t_2) \,   K_{p_s} (t_1) K_{p_s}^* (t_2)  \Theta\left( t_1 - t_2 \right) 
 \right]
 \label{eqn:DR_eg}
\end{align}
seems to require a nested time integral. 
Rather than compute this explicitly, we can notice that since it takes the form $P_{p_s} \, \disc{p_s} \left[ f (p_s) \right]$, it must be an \emph{even} function of $p_s$.\footnote{
This follows from the fact that $\disc{p_s}$ is clearly odd by definition~\eqref{eqn:Disc_def}, and the power spectrum $P_{p_s}$ is either odd or can be made odd by a suitable renormalisation (since any even $p_s^{2n}$ term can be absorbed into the local counterterm $(\partial^n \phi)^2$).
} 
We can therefore play the following trick. 
On general grounds, the Bunch-Davies wavefunction coefficient may only have singularities when the total energy flowing into one or more vertices vanishes. In this example, that corresponds to the total energy $\omega_T = \omega_{k_1} + \omega_{k_2} + \omega_{k_3} + \omega_{k_4}$ and the partial energies $\omega_L = \omega_{k_1} + \omega_{k_2} + \omega_{p_s}$ and $\omega_R = \omega_{k_3} + \omega_{k_4} + \omega_{p_s}$. 
Since the $\psi_3^{\rm tree}$ discontinuity in the first term of \eqref{eqn:exch_eg} introduce unphysical singularities at $\omega_{p_s} = \omega_{k_1} + \omega_{k_2}$ and $\omega_{k_3} + \omega_{k_4}$, we can demand that these are cancelled by the even function \eqref{eqn:DR_eg}. 
This effectively fixes \eqref{eqn:DR_eg} up to terms which are analytic in $p_s^2$ (and such terms can be absorbed into contact diagrams). 

For example, for a $\tfrac{\lambda}{3!} \phi^3$ interaction on a Minkowski spacetime background that produces,
\begin{align}
\fig{contactkkk} \; &= \; \frac{ \lambda \tilde{\delta}^3 \left( \bfk_1 + \bfk_2 + \bfk_3  \right) }{ \omega_{k_{1}} + \omega_{k_{2}} + \omega_{k_{3}} }, 
\label{eqn:psi3_Mink}
\end{align}
the equation \eqref{eqn:exch_eg} becomes,
\begin{align}
\fig{exch_s}
	\; = \;
& \frac{ \lambda \tilde{\delta}^3 \left( \sum_{b=1}^4 \bfk_b \right)  }{  \omega_L \omega_R \left(  \omega_{k_3} \; +\;  \omega_{k_4}  - \omega_{p_s} \right)  } 
+
\fig{exch_R} 
 \label{eqn:exch_Mink_eg}
\end{align}
The only even function of $p_s$ which cancels the unphysical singularity (and does not introduce any further unphysical singularities) is,
\begin{align}
 \fig{exch_R} 
 =
 \frac{ \lambda \tilde{\delta}^3 \left( \sum_{b=1}^4 \bfk_b \right)  }{  \omega_T  }  \;  \frac{ 1 }{   \omega_{p_s}^2 - ( \omega_{k_3} + \omega_{k_4}  )^2  } 
\end{align}
up to an analytic function of $p_s^2$. 
Substituting this back into \eqref{eqn:exch_Mink_eg}, we arrive at the simple answer,
\begin{align}
\fig{exch_s} = \frac{\lambda \tilde{\delta}^3 \left( \sum_{b=1}^4 \bfk_b \right) }{ \omega_T \omega_L \omega_R } \; . 
\label{eqn:exch_Mink_eg2}
\end{align}
Notice that we have therefore determined this exchange diagram \emph{without doing the nested time integrals} in \eqref{eqn:psi4exc}.
Of course, in this simple Minkowski example the time integrals are fairly straightforward (and indeed produce \eqref{eqn:exch_Mink_eg2}). 
But this same approach can be applied more generally to \emph{any} tree-level diagram on \emph{any} time-dependent spacetime background.

\paragraph{General procedure.}
For any tree-level diagram $D$, we can split it into a retarded and a cut part as shown in \eqref{eqn:blobs_1}: we will write this as $D = D_R + D_C$. 
Now the bootstrap proceeds as follows. 
Label the partial energies of the original diagram $D$ which depend on the momentum of the internal line's energy $\omega_p$ as $\{ E_1 (\omega_p) , ... , E_n (\omega_p) \}$. $D$ will generically contain singularities when any $E_j (\omega_p) = 0$. 
Due to the analytic continuation, the cut $D_C$ terms will contain additional singularities at $E_j (- \omega_p ) = 0$.
These unphysical singularities are not present in $D$, and so they must exactly cancel with singularities in $D_R$.
However, since $D_R$ is proportional to $P_p \, \disc{p}$ it must be an even function of $p$, and so if it contains a singularity at $E_j (-\omega_p)$ it must also contain one at $E_j (+\omega_p )$ with the same residue. 
Proceeding in this way, one can determine all of the singularities at $E_j ( \pm \omega_p )$ and their residues, which fixes $D_R$ up to a remainder that is analytic in $\omega_p^2$.
This remainder, if non-zero, can be absorbed into the local interaction which resembles $D$ but with the internal line collapsed into a single vertex.
This represents the freedom to add any contact diagram to $D$---this will not affect the cut diagrams $D_C$ and is ultimately related to our freedom to perform field redefinitions\footnote{
Unlike $S$-matrix elements, wavefunction coefficients and equal-time correlators are not invariant under field redefinitions.
}.

\paragraph{Another Minkowski Example.}
As a second simple example of this procedure, consider the following tree-level diagram,
\begin{align}
 D = \fig{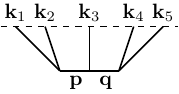}
\end{align}
The partial energies flowing into each vertex or collection of vertices are,
\begin{align}
 E_1 (\omega_p) &= \omega_{k_1} + \omega_{k_2} + \omega_p \; , \;\; &E_2 (\omega_p) &= \omega_{k_3} + \omega_p + \omega_q \; , \;\; &E_3  &=  \omega_{k_4} + \omega_{k_5} + \omega_q \; , \;\; \nonumber \\
 E_4 (\omega_p) &= \omega_{k_3} + \omega_{k_4} + \omega_{k_5} + \omega_p \; , \;\; &E_5 &= \omega_{k_1} + \omega_{k_2} + \omega_{k_3} + \omega_q \; , \;\; &\omega_T &= \omega_{k_1} + \omega_{k_2} + \omega_{k_3} + \omega_{k_4} + \omega_{k_5}
\end{align}
where we have highlighted the $p$ dependence of $E_1$, $E_2$ and $E_3$. 
Replacing this internal line with retarded propagators as in \eqref{eqn:blobs_1} produces the following cut contribution\footnote{
Note that we could have cut $\bfq$ and taken the $\disc{}$ of a single line in the resulting $\psi_3^{\rm tree}$: since this would exactly mimic the previous example, here we have chosen to cut $\bfp$ and take the $\disc{}$ of the resulting 4-point exchange diagram.
},
\begin{align}
D_C = - \frac{1}{2 \omega_p E_3} \left[
 \frac{1}{  E_1 (\omega_p) E_2 (\omega_p) E_4 (\omega_p) } 
 -  \frac{1}{ E_1 (\omega_p) E_2 (-\omega_p) E_4 (-\omega_p) }
 \right]
\; ,  
\end{align}
for the cubic interaction on Minkowski~\eqref{eqn:psi3_Mink} (and the 4-point exchange we bootstrapped from it in \eqref{eqn:exch_Mink_eg2}). 
$D_C$ has unphysical singularities at both $E_2 (-\omega_p)$ and $E_4 (-\omega_p) = 0$, which cannot appear in $D$. 
Consequently, we can construct $D_R$ using the ansatz $\sum_{j} Z_j/(E_j ( \omega_p) E_j (-\omega_p) )$ and fix each $Z_j$ so that these folded singularities cancel. 
The result is, 
\begin{align}
D_R =  \frac{1}{ E_3} \Bigg[&  
\frac{ 1}{  E_2 ( \omega_p) E_2 (-\omega_p)  E_1 ( \omega_{k_3} + \omega_q)  E_4 ( - \omega_{k_3} - \omega_q)   }  \nonumber \\
&+
\frac{ 1}{  E_4 ( \omega_p) E_4 (-\omega_p)  E_1 (  \omega_{k_3} + \omega_{k_4} + \omega_{k_5} )  E_2 ( - \omega_{k_3} - \omega_{k_4} - \omega_{k_5} )   }
\Bigg]
\label{eqn:DR_eg_2}
\end{align}
so that the total $D_C + D_R$ is simply,
\begin{align}
 D = \frac{1}{ \omega_T E_1 (\omega_p) E_2 ( \omega_p ) E_3 } \left( \frac{1}{ E_4 (\omega_p) } + \frac{1}{ E_5 }  \right) \; .
 \label{eqn:exch5_eg}
\end{align}
Again, notice how there was no need to ever perform a nested time integral: the retarded part $D_R$ was fixed entirely by the singularities of $D_C$. 

In this simple Minkowski setting, it is straightforward to verify \eqref{eqn:exch5_eg} by explicitly performing the time integrals. 
The virtue of fixing $D$ using causality is that we remove the need to perform any time integrals, once the basic $\psi_3^{\rm tree}$ vertex is known.  
This comes at the cost of introducing unphysical singularities in the intermediate steps which are ultimately absent in the final result \eqref{eqn:exch5_eg}. 
The simple structure of \eqref{eqn:exch5_eg}, and indeed of all Minkowski wavefunction coefficients, can be made manifest using the ``old-fashioned perturbation theory'' based on the Lippmann-Schwinger equation (see e.g. \cite{Benincasa:2019vqr,Hillman:2021bnk, Salcedo:2022aal} for recent reviews). This approach also removes the need for time integrals, and introduces only the physical singularities in the partial energies. 
However, the utility of OFPT is largely limited to Minkowski spacetime.
So the second virtue of fixing $D$ using causality as described above, is that this method can be applied to any time-dependent background.

\paragraph{de Sitter Example.}
To demonstrate this, consider the same exchange diagram \eqref{eqn:exch_Mink_eg} but this time for the interaction $\mathcal{L}_{\rm int} =  \frac{\lambda}{3!} \dot{\pi}^3 a(t)$ for a massless scalar field $\pi$ on de Sitter.
The contact diagram is readily determined by doing a single time integral of the $K_k$ propagator in \eqref{eqn:G_eg_intro},
\begin{align}
   \fig{contactkkk} &= -\frac{2 \lambda k_{1}^2 k_{2}^2 k_{3}^2}{H ( k_1 + k_2 + k_3 )^3} \; ,
   \label{eqn:psi3_dS}
\end{align}
where for now we will omit the $\tilde{\delta}^3$ functions.
The cut contributions to the exchange diagram are then,
\begin{align}
D_C &= - \frac{  2 \lambda^2 k_1^2 k_2^2 k_3^2 k_4^2 p }{ (k_{12} + p_s )^3 } \left[ \frac{1}{  (k_{34} + p_s )^3 } -
 \frac{ 1 }{   (k_{34} - p_s )^3 }
 \right]
\end{align}
where $k_{ij} \equiv k_i + k_j$. 
Proceeding as above by making a general ansatz $D_R = \sum_{i,j} Z_j/ ( E_j (p) E_j (-p) )^i$, we can immediately determine the retarded contributions,
\begin{align}
 D_R &=  - \frac{ 4 \lambda^2 k_1^2 k_2^2 k_3^2 k_4^2 }{k_T^5} \left[ 
   \frac{6 k_{34}^2 }{  \left( p_s^2 - k_{34}^2 \right)}  
- \frac{ 6 p_s^2 k_T  k_{34} }{   ( p_s^2 - k_{34}^2 )^2 }   
   +  \frac{ p_s^2 k_T^2  \left( 3 k_{34}^2 + p_s^2 \right) }{ \left( p_s^2 - k_{34}^2 \right)^3 } 
     \right] \; , 
\end{align}
again up to an analytic function of $p^2_s$. 
Summing $D_C$ and $D_R$ gives,
\begin{align}
   \fig{exchkkkk} &=  - \frac{4 \lambda^{2} k_{1}^2 k_{2}^2 k_{3}^2 k_{4}^2 }{k_T^5} \bigg(
    \frac{ 4 p_s^4 k_T^2 }{s^3}  + \frac{ 6 p_s^3 ( k_T^3 - s k_T ) }{s^3}  + \frac{3 p_s^2}{s^3} (2 s^2 - s k_T^2 + k_T^4) + 6  
    \bigg) \; ,
    \label{eqn:dS_eg_1}
\end{align}
for two $\dot{\pi}^3$ interactions on de Sitter, where we have used the variables,
\begin{equation}
k_{T} \equiv k_{1}+k_{2}+k_{3}+k_{4} \;\;\;\; \text{and} \;\;\;\; s \equiv - (k_1 + k_2 + p_s ) (k_3 + k_4 + p_s) \; . 
\end{equation}
In our opinion, this bootstrap route is far simpler than performing the nested time integration in~\eqref{eqn:psi4exc}.
We verified \eqref{eqn:dS_eg_1} by carrying out the time integration explicitly, and also by checking that it correctly factorises in limits\footnote{
Note that in the amplitude limit, $k_T \to 0$, the variable $s \to (k_1 + k_2 )^2 - p_s^2$ and becomes the usual Mandelstam invariant. 
In fact, in that limit,
\begin{align}
    \eqref{eqn:dS_eg_1} \sim - \frac{24 \lambda^2 k_1^2 k_2^2 k_3^2 k_4^2}{k_T^5} \; \frac{ (k_1 + k_2 )^2 }{s}  \propto \frac{\mathcal{A} }{ k_T^5 \prod_{a=1}^4 k_a^2 P_{k_a} } 
    \label{eqn:sunset_dS_amplitude_limit}
\end{align}
and indeed coincides with the expected amplitude $\mathcal{A}$ for an $\dot{\pi}^3 \times \dot{\pi}^3$ interaction (see e.g. \cite{Goodhew:2020hob, Pajer:2020wxk} for a description of the proportionality constant in \eqref{eqn:sunset_dS_amplitude_limit}).
} as $k_T \to 0$, $E_L \to 0$ or $E_R \to 0$. 
It also matches the result given in \cite[(6.47)]{Jazayeri:2021fvk}. 

~\\
While the three examples given in this section are fairly simple diagrams for which the nested time integration can be performed explicitly, the point that we wish to stress is that this procedure can be applied to \emph{any} tree-level diagram on \emph{any} time-dependent spacetime, including those for which the time integrals become arduous or intractable. 
This removes the need to ever perform a nested time integral in cosmology at tree-level. 
By repeatedly applying \eqref{eqn:blobs_1} and fixing the $D_R$ part by matching against the unphysical singularities in $D_C$, all internal lines can be removed and any $D$ can be expressed as a product of simple contact diagrams (which require only a single time integral) and their discontinuities.

\paragraph{Further identities.}
Finally, we remark that \eqref{eqn:caus_cut_0} is by no means the only constraint which follows from the properties of $G_p^R$.
Unlike the bulk-to-bulk $G_p$, the retarded $G_p^R$ is simply proportional to $\Theta (t_1 - t_2)$.
As a result, we have propagator identities of the form,
\begin{align}
  \text{Im} \left[ K_{p_1} (t_1) K_{p_2}^* (t_2) \right] G_{p_2}^R (t_1, t_2) =   G_{p_1}^R (t_1, t_2) \, \text{Im} \left[ K_{p_2} (t_1) K_{p_1}^* (t_2) \right] 
\end{align}
which corresponds to the diagrammatic identity,
\begin{align}
  P_{\bfq_1 \bfq'} \;\; \disc{q_1, q_1'} \left[  \fig{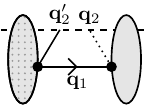} \right]  = P_{\bfq_2 \bfq_2'} \;\; \disc{q_2, q_2'} \left[  \fig{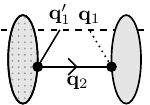}\right] \; . 
  \label{eqn:Ctype_2}
\end{align}
In words, these identities say that in any diagram with \emph{colinear} momenta (i.e. a pair of external legs with $\bfq + \bfq' = 0$), the $\disc{}$ with respect to these momenta can be used to \emph{exchange} the momenta of external and internal lines. 
This is a further constraint which all Feynman-Witten diagrams must satisfy due to the causal properties of the free theory.
Unlike the above causal cutting rules, these identities only apply to particular colinear configurations of momenta. At present, we have not found any particularly useful application for them (except that they guarantee the consistency of different loop-level cutting rules discussed in the next section).
There may be some connection to the factorisation in non-local soft limits recently studied in \cite{Qin:2023bjk}. It would be interesting to explore these relations further in future.

\subsection{Loops and the cosmological tree theorem}
\label{sec:loop}

Now we turn our attention to diagrams containing loops. 
In particular, we introduce and then prove our Cosmological Tree Theorem, which can replace any loop diagram with a sum over (momentum integrals of) tree diagrams. 
To build some intuition for how this works, we begin by outlining several simple examples in subsection~\ref{sec:CTT:subsec:examples}, postponing the general statement of the theorem to subsection~\ref{sec:CTT:subsec:General}. 
While we initially focus on scalar fields, we explain the straightforward generalisation to spinning fields at the end of this section.

\subsubsection{Some examples}
\label{sec:CTT:subsec:examples}

First we show, simply by inspection, that the one-loop correction to the wavefunction coefficients which determine the power spectrum and the bispectrum can be written in terms of simpler tree-level diagrams.

\paragraph{One vertex.}
The simplest one-loop diagram contains only a single vertex. 
For instance, a $\tfrac{\lambda}{4!} \phi^4$ interaction contributes to the quadratic wavefunction coefficient through the diagram,
\begin{align}
\fig{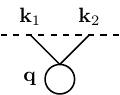}
=  \int_{\bfq}  \int_t \, \lambda \,  K_{k_1} (t) K_{k_2} (t) G_{q} (t,t) \;  \tilde{\delta} \left( \bfk_1 + \bfk_2 \right)  \; . 
\label{eqn:fish_OFPT}
\end{align}
The coupling $\lambda$ can depend on both time and, in the case of derivative interactions, all three momenta entering the vertex. 
Notice that since the bulk-to-bulk propagator in this diagram is evaluated at coincident times, it no longer contains any $\Theta( t_1 - t_2 )$ step functions. 
It can be written simply as,
\begin{equation}\label{eqn:CTT:sametime}
    G_{q}(t,t) = 2P_{q} K_{q}(t)\text{Im}(K_{q}(t)).
\end{equation}
This allows us to express the momentum integrand in \eqref{eqn:fish_OFPT} in terms of a tree-level diagram without any internal lines, using the $\disc{}$ operation \eqref{eqn:Disc_def} to extract the required imaginary part of an external propagator. 
In this example, the identity \eqref{eqn:CTT:sametime} implies, 
\begin{align}
- \fig{loopkk_q} 
=
\int_{\bfq \bfq'} P_{\bfq \bfq'} \left( \fig{contactkkqqC} \right)
\label{eqn:fish_CTT}
\end{align}
where the highlighted line corresponds to taking a $\disc{}$, as defined in \eqref{eqn:Disc_graph}. 
The identity~\eqref{eqn:fish_CTT} is the simplest example of our tree theorem.
It shows that the time integration in \eqref{eqn:fish_OFPT} can be replaced by cutting open the loop and taking a discontinuity.
This turns out to be a very general property of loop diagrams and is connected to the causal structure of the propagator (anticipated in~\eqref{eqn:no_loops_intro}).

Before moving on to our next example, let us make a few remarks about~\eqref{eqn:fish_OFPT}.
\begin{itemize}

    \item[(i)] For massless fields on a Minkowski spacetime background, these diagrams are particularly simple, 
    \begin{align}
    \fig{contactkkkk} &= \frac{  \lambda \, \tilde{\delta}^4 \left( \sum_{b=1}^4 \bfk_b  \right) }{ \omega_{k_{1}}+\omega_{k_{2}}+\omega_{k_{3}}+\omega_{k_{4}}} \; ,  
  &  \fig{loopkk_q} &=  \frac{ \lambda \tilde{\delta}^4 \left( \bfk_1 + \bfk_2  \right) }{\omega_{k_1} + \omega_{k_2}} \int_{\bfq} \frac{ 1 }{ \omega_{k_1} + \omega_{k_2} + 2 \omega_q} ,
\label{eqn:fish_mink}
    \end{align}
    and it is straightforward to confirm that \eqref{eqn:fish_CTT} holds, essentially as a consequence of the partial fraction identity,
    \begin{align}
     \frac{ 1 }{ ( \omega_{k_1} + \omega_{k_2} ) (  \omega_{k_1} + \omega_{k_2} + 2 \omega_q )} = - \frac{1}{2\omega_q} \left( \frac{1}{ \omega_{k_1} + \omega_{k_2} + \omega_q + \omega_{q'} } - \frac{1}{ \omega_{k_1} + \omega_{k_2} + \omega_q - \omega_{q'} } \right) \Big|_{q' = q} \; . 
     \label{eqn:partial_frac}
    \end{align}
    The simplicity of the integrand in \eqref{eqn:fish_mink} can again be understood from old-fashioned perturbation theory and its resulting recursion relation. One way to think of~\eqref{eqn:fish_CTT} is that it gives up some of that simplicity (by splitting the single rational integrand \eqref{eqn:fish_mink} into the two separate terms in \eqref{eqn:partial_frac}) in order to buy a greater generality: \eqref{eqn:fish_CTT} can be immediately applied to fields of any mass on any time-dependent spacetime background\footnote{
 We will return to the comparison with OFPT at the end of this section.}.   

    \item[(ii)] 
    As an example of this generality, consider the contact diagram given by the interaction $\frac{\lambda}{4!} \dot{\pi}^{4}$ on a fixed de Sitter spacetime. In that case, performing the explicit time integration leads to, 
     \begin{align}
    \fig{contactkkkk} &=  \lambda \tilde{\delta}^3 \left( \sum_{b=1}^4 \bfk_b \right)  \frac{ 24k_{1}^{2}k_{2}^{2}k_{3}^{2}k_{4}^{2} }{(k_{1}+k_{2}+k_{3}+k_{4})^{5}}  , \nonumber \\ 
    \fig{loopkk_q} &=  - \lambda \tilde{\delta}^3 \left( \bfk_1 + \bfk_2  \right)  \int_{\bfq}\left(
 \frac{ 24 k_1^2 k_2^2 q^4 }{ ( k_1 + k_2 + 2 q)^5}
 -
 \frac{ 24 k_1^2 k_2^2 q^4}{ ( k_1 + k_2 )^5 }  \right) ,
\label{eqn:fish_dS}
    \end{align}
    which again satisfy~\eqref{eqn:fish_CTT}. 

    \item[(iii)] In many ways the cutting rule~\eqref{eqn:fish_CTT} is a completion of those developed in~\cite{Melville:2021lst}, since it expresses both the real \emph{and} imaginary part of the loop diagram in terms of tree-level cuts. It is straightforward to recover the results of \cite{Melville:2021lst} by taking a further discontinuity of \eqref{eqn:fish_CTT},
    \begin{align}
    - \fig{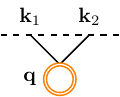}  
        =  \int_{\bfq \bfq'} P_{\bfq \bfq'} \, \left( \fig{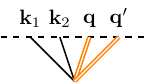} \right) 
        \label{eqn:1loop_unit_cut}
    \end{align}
    where we have used that $\Disc{q} \left[ \disc{q'} \left[ K_q (t) K_{q'} (t) \right] \right] = \disc{q, q'} \left[ K_q (t) K_{q'} (t) \right]$. 
Note that the unitarity cutting rule~\eqref{eqn:1loop_unit_cut} requires unitarity time evolution in the fully interacting theory (so that $\lambda$ commutes with the $\Disc{}$ operation defined in~\eqref{eqn:Disc_def}), whereas the tree theorem~\eqref{eqn:fish_CTT} uses only the unitarity/causality of the free theory\footnote{
Since $\partial_t \disc{p} \left[ K_p (t) \right] = \disc{p} \left[ \partial_t K_p (t) \right]$, $\lambda$ automatically commutes with $\disc{}$ for interactions with time-derivatives.
}. 


\end{itemize}

\paragraph{Two vertices.}
The next-simplest one-loop diagrams contains two vertices.
For instance, a $\tfrac{\lambda}{3!} \phi^3$ interaction contributes to the quadratic wavefunction coefficient through the diagram,
\begin{align}
    \fig{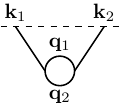} = \int_{\bfq_1 \bfq_2} \int_{t_1, t_2}  \lambda (t_1) \lambda (t_2) K_{k_1}(t_{1}) G_{q_{1}}(t_{1},t_{2}) G_{q_{2}} (t_{2},t_{1}) K_{k}(t_{2}) \tilde{\delta}^3 (\bfk_1 + \bfq_1 - \bfq_2) \tilde{\delta}^3 (\bfk_2 - \bfq_1 + \bfq_2) \; . 
    \label{eqn:sunset_OFPT}
\end{align}
Ordinarily, a product of two bulk-to-bulk propagators would contain two Heaviside $\Theta$ functions. However, when evaluated at cyclic arguments as in \eqref{eqn:sunset_OFPT}, this product only contains a single $\Theta$ function.
It can therefore be written in terms of a single bulk-to-bulk propagator,
\begin{align}\label{eqn:G1G2}
   G_{p_{1}}(t_{1},t_{2})G_{p_{2}}(t_{2},t_{1}) =&
   2 P_{p_{1}}  \Im(K_{p_{1}}(t_{2})) G_{p_{2}}(t_{2},t_{1}) K_{p_{1}}(t_{1})  
   +2 P_{p_{2}} \Im(K_{p_{2}}(t_{1})) G_{p_{1}}(t_{1},t_{2}) K_{p_{2}}(t_{2})   \nonumber \\
   & -4P_{p_{1}}P_{p_{2}} K_{p_{1}}(t_{1})  \Im(K_{p_{2}}(t_{1})) K_{p_{2}}(t_{2}) \Im(K_{p_{1}}(t_{2})).
\end{align}
This suggests that it should be possible to express the momentum integrand of \eqref{eqn:sunset_OFPT} using diagrams with at most one internal line (and therefore tree-level). 
In fact, 
it is straightforward to verify that the three terms of \eqref{eqn:G1G2} can be written as,
\begin{align}
    - \fig{loopkk_qq} &= \int_{\bfq_1 \bfq_1'} P_{\bfq_1 \bfq_1'} \left( \fig{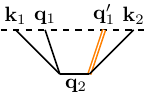} \right) + \int_{\bfq_2 \bfq_2'} P_{\bfq_2 \bfq_2'} \left( \fig{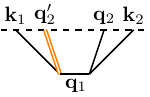} \right)   \nonumber \\ 
    &+ \int_{\substack{\bfq_1 ,\bfq_2 \\ \bfq_1' ,\bfq_2'}} \, P_{\bfq_1 \bfq_1'} P_{\bfq_2 \bfq_2'}  \left(  \fig{contactkqqC_1}  \right)   \left(  \fig{contactkqqC_2}  \right) 
    \label{eqn:sunset_CTT}
\end{align}
where again we have used coloured lines to denote the imaginary part of a propagator. 

The identity~\eqref{eqn:sunset_CTT} is another example of our tree theorem.
It shows, as in the previous example, that the time integrals appearing in the conventional expression for this loop diagram~\eqref{eqn:sunset_OFPT} can be exchanged for suitable discontinuities of tree-level diagrams. 
The qualitative difference with~\eqref{eqn:fish_CTT} is that now the loop consists of two lines, and we must sum over cutting either/both of these lines. 
This need to sum over all possible cuts was also found in the cutting rules previously developed in~\cite{Melville:2021lst, Goodhew:2021oqg}. 

The other distinct feature of a loop with more than one edge is that it can be oriented in either of two ways (i.e. loop momenta can flow clockwise or counterclockwise around the loop).
This manifests as a second identity for the bulk-to-bulk propagator,
\begin{align}
   G_{p_{1}}(t_{1},t_{2})G_{p_{2}}(t_{2},t_{1}) =&
   2 P_{p_{1}} K_{p_{1}}(t_{2}) G_{p_{2}}(t_{2},t_{1})  \Im(K_{p_{1}}(t_{1}))
   +2 P_{p_{2}} K_{p_{2}}(t_{1}) G_{p_{1}}(t_{1},t_{2}) \Im(K_{p_{2}}(t_{2}))  \nonumber \\
   & -4P_{p_{1}}P_{p_{2}} \Im(K_{p_{1}}(t_{1})) K_{p_{2}}(t_{1}) \Im(K_{p_{2}}(t_{2})) K_{p_{1}}(t_{2})  .
   \label{eqn:G1G2_2}
\end{align}
which corresponds to cutting open the loop with the opposite orientation, producing an identity which differs from \eqref{eqn:sunset_CTT} only in the colouring of the lines\footnote{
If it seems that \eqref{eqn:sunset_CTT_2} is a trivial relabelling of \eqref{eqn:sunset_CTT}, note that the two internal lines could correspond to different fields.
},
\begin{align}
    -\fig{loopkk_qq} &= \int_{\bfq_1 \bfq_1'} P_{\bfq_1 \bfq_1'} \left( \fig{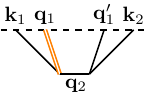} \right) + \int_{\bfq_2 \bfq_2'} P_{\bfq_2 \bfq_2'} \left( \fig{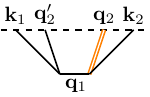} \right)   \nonumber \\ 
    &+ \int_{\substack{\bfq_1 ,\bfq_2 \\ \bfq_1' ,\bfq_2'}} \, P_{\bfq_1 \bfq_1'} P_{\bfq_2 \bfq_2'}  \left(  \fig{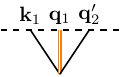}  \right)   \left(  \fig{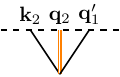}  \right) \; . 
    \label{eqn:sunset_CTT_2}
\end{align}
That there are two distinct ways of cutting open a loop into trees will turn out to be quite general, and we show in subsection~\ref{sec:CTT:subsec:General} below that this is tied to the existence of two causal propagators (i.e. retarded/advanced). 

Some comments about~\eqref{eqn:sunset_CTT} and~\eqref{eqn:sunset_CTT_2}: 
\begin{itemize}

    \item[(i)] It is instructive to compare with the simple expressions for a massless scalar on Minkowski.
    The relevant tree-level diagram $\psi_3^{\rm tree}$ was given in \eqref{eqn:psi3_Mink}, and from it we constructed the $\psi_4$ exchange diagram \eqref{eqn:exch_Mink_eg2} using the tree-level cutting rules.
    Now, we can use both of those expressions as input for \eqref{eqn:sunset_CTT} or \eqref{eqn:sunset_CTT_2}, which determines the loop integrand to be, 
\begin{align}
\fig{loopkk_qq} &= \frac{\lambda}{ \omega_{k_1} + \omega_{k_2} } \int_{\bfq_{1}, \bfq_{2}} 
\left(  
 \frac{1}{ \omega_{k_1} + \omega_{k_2} + 2 \omega_{q_1} } + \frac{1}{ \omega_{k_1} + \omega_{k_2} + 2 \omega_{q_2} } \right) 
\frac{ \tilde{\delta}^3 ( \bfk_1 + \bfq_1 - \bfq_2  ) \tilde{\delta} \left( \bfk_2 - \bfq_1 + \bfq_2 \right) }{ ( \omega_{k_1} + \omega_{q_1} +\omega_{q_2} ) ( \omega_{k_2} + \omega_{q_1} + \omega_{q_2} ) } \; . 
\end{align}
This is consistent with explicitly performing the nested time integrals in \eqref{eqn:sunset_OFPT}, but notice that using causality we required only $\psi_3^{\rm tree}$ as an input (i.e. a single time integral over only $K_k$'s).

    \item[(ii)] To illustrate this result on a non-trivial background (and for a derivative interaction), consider the interaction $\frac{\lambda}{ 3! } \dot{\pi}^3 a(t)$ on a fixed de Sitter background. This is typically the dominant source of primordial non-Gaussianity in the EFT of inflation \cite{Cheung:2007st}. 
    At tree-level, the seed wavefunction coefficient $\psi_3^{\rm tree}$ is given by \eqref{eqn:psi3_dS}, and we used this to construct the four-point exchange diagram in \eqref{eqn:dS_eg_1}.
  Using these as the input for \eqref{eqn:sunset_CTT} or \eqref{eqn:sunset_CTT_2} gives,
\begin{align}
 -    \fig{loopkk_qq} = \frac{ \lambda^{2} }{8 k}\int_{\bfq} &\bigg( 
 F(q_1, q_2 , q_{1}+q_{2} ) -  F(q_1, q_2 , -q_{1}+q_{2} ) -  F(q_1, q_2 , q_{1}-q_{2} ) +  F(q_1, q_2 , -q_{1}-q_{2} )   \nonumber \\
 &+ \frac{3 q_{1}^2 (5 k^4 + 10 k^3 q_{1} + 10 k^2 q_{1}^2 + 5 k q_{1}^3 + q_{1}^4)}{
(k + q_{1})^5}+ \left( q_{1}\leftrightarrow q_{2} \right) \bigg)
\label{eqn:sunset_dS}
\end{align}
where,
\begin{equation}
    F(q_1, q_2, p)= 2k^4q_{1}q_{2}  \frac{5 (k+p) (k+q_{1}+q_{2})+(k+p)^2+10 (k+q_{1}+q_{2})^2}{(k+q_{1}+q_{2})^3 (2 k+p+q_{1}+q_{2})^5}
\end{equation}
and we have omitted the overall $\tilde{\delta}^3 \left( \bfk_1 + \bfk_2 \right)$ and enforced $|\bfk_1| = | \bfk_2 | = k$ and $\bfq_2 = \bfq_1 + \bfk$.
It is remarkable that this result can be obtained immediately from the simple $\psi_3^{\rm tree}$ in~\eqref{eqn:psi3_dS}, without performing any further time integration\footnote{
We did, of course, check that the explicit time integration of \eqref{eqn:sunset_OFPT} matches \eqref{eqn:sunset_dS}. 
Note that this differs from \cite[(C.4)]{Melville:2021lst} by the additional contact terms in the second line.
}. 

    \item[(iii)] Note that while the bulk-to-bulk propagators are symmetric in their arguments, $G_p (t_1, t_2) = G_p (t_2, t_1)$, this symmetry is not manifest in the identities \eqref{eqn:G1G2} or \eqref{eqn:G1G2_2}. 
    We can make this explicit by defining the quantity,
    \begin{align}\label{eqn:L2}
   \mathcal{L}_{p_1 p_2} (t_1, t_2) \equiv& +  G_{p_{1}}(t_{1},t_{2})G_{p_{2}}(t_{2},t_{1})  + 4P_{p_{1}}P_{p_{2}} K_{p_{1}}(t_{1})  \Im(K_{p_{2}}(t_{1})) K_{p_{2}}(t_{2}) \Im(K_{p_{1}}(t_{2}))  \nonumber \\
   &-
   2 P_{p_{1}}  \Im(K_{p_{1}}(t_{2})) G_{p_{2}}(t_{2},t_{1}) K_{p_{1}}(t_{1})  
   - 2 P_{p_{2}} \Im(K_{p_{2}}(t_{1})) G_{p_{1}}(t_{1},t_{2}) K_{p_{2}}(t_{2})   
\end{align}
    so that $\mathcal{L}_{p_1 p_2} (t_1, t_2) = 0$ corresponds to \eqref{eqn:G1G2} and $\mathcal{L}_{p_1 p_2} (t_2, t_1) = 0$ corresponds to \eqref{eqn:G1G2_2}. 
    If we instead consider the difference $\mathcal{L}_{p_1 p_2} (t_1, t_2) - \mathcal{L}_{p_1 p_2} (t_2, t_1)$,
    we find that it only vanishes thanks to the non-trivial tree-level identity \eqref{eqn:Ctype_2}. 
	The consistency of different one-loop cutting rules is therefore guaranteed by the tree-level identities.
	This turns out to be a general pattern: the consistency of cutting higher-loop diagrams is guaranteed by lower-loop identities. 


\end{itemize}

Finally, note that at this order in perturbation theory there is also a 1PI-reducible diagram that contributes to $\psi_2$,
\begin{align}\label{eqn:tadpolepsi2}
    \fig{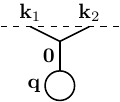} = \int_{\bfq} \int_{t_1, t_2}  \lambda_1 \lambda_2 K_{k_1}(t_{1}) K_{k_2} (t_1) G_{0}(t_{1},t_{2}) G_{q} (t_{2},t_{2}) \tilde{\delta}^3 (\bfk_1 + \bfk_2 ) \; . 
\end{align}
This can be expressed in terms of a tree-level diagram as in the one-vertex example using  \eqref{eqn:CTT:sametime},
\begin{align}
 -   \fig{loopkk_0_q} =  \int_{\bfq \bfq'} P_{\bfq \bfq'} \left(  \fig{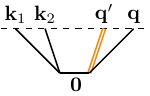}   \right) \; . 
\end{align}
In particular, note that the total one-loop correction sourced by this cubic interaction can be written as,
\begin{align}
    - \fig{loopkk_qq} - \fig{loopkk_0_q} 
    &=
     \int_{\bfq \bfq'} P_{\bfq \bfq'} \left( 
    \fig{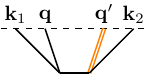} + \fig{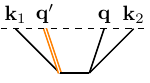} + \fig{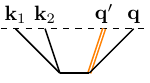}  
    \right)  \nonumber \\ 
    &+ \int_{\substack{\bfq_1 ,\bfq_2 \\ \bfq_1' ,\bfq_2'}} \, P_{\bfq_1 \bfq_1'} P_{\bfq_2 \bfq_2'}  \left(  \fig{contactkqqC_1}  \right)   \left(  \fig{contactkqqC_2}  \right)  \; .
\end{align}
where the first line now contains all three exchange channels of $\psi_4^{\rm tree}$. 
Together with \eqref{eqn:fish_CTT}, this establishes the relation~\eqref{eqn:psi2_intro} given in the introduction, which expresses the full $\psi_2^{\rm 1-loop}$ coefficient in terms of $\psi_4^{\rm tree}$ and $\psi_3^{\rm tree}$. 
So although we are deriving these identities by looking at particular diagrams, they naturally apply to the entire wavefunction coefficient (at a fixed order in perturbation theory).

\paragraph{Three-vertex loop.}
As a final example, we consider the one-loop diagram from three $\tfrac{\lambda}{3!} \phi^3$ vertices, 
\begin{align}\label{eqn:psi3phi3}
  \fig{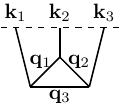}   \begin{gathered} = \int_{\bfq_{1}\bfq_{2}\bfq_{3}}\tilde{\delta}(\bfk_{1}+\bfq_{3}-\bfq_{1})\tilde{\delta}(\bfk_{2}+\bfq_{1}-\bfq_{2})\tilde{\delta}(\bfk_{1}+\bfq_{1}-\bfq_{2})\\
    \qquad\qquad \times \int_{\substack{t_{1}t_{2}\\t_{3}}} \lambda^{3}K_{k_{1}}(t_{1})G_{q_{1}}(t_{1},t_{2})K_{k_{2}}(t_{2})G_{q_{2}}(t_{2},t_{3})K_{k_{3}}(t_{3})G_{q_{3}}(t_{3},t_{1}) \, .
    \end{gathered}
\end{align}
While a product of three $G_p$ propagators generally introduces three $\Theta$ functions in the integrand, for this particular cyclic identification of the arguments we find that only two $\Theta$ functions are necessary. 
As in the above examples, this means that it is possible to expand this product in terms of fewer bulk-to-bulk propagators. For instance,
\begin{align}\label{eqn:G1G2G3prod}
G_{q_{1}}(t_{1},t_{2}) G_{q_{2}}(t_{2},t_{3})G_{q_{3}}(t_{3},t_{1}) &=
2P_{q_{1}}K_{q_{1}}(t_{1})\text{Im}(K_{q_{1}}(t_{2}))G_{q_{2}}(t_{2},t_{3})G_{q_{3}}(t_{3},t_{1})+\text{2 perm.}  \\
&-4P_{q_{1}}P_{q_{2}}K_{q_{1}}(t_{1})K_{q_{2}}(t_{2})\text{Im}(K_{q_{1}}(t_{2}))\text{Im}(K_{q_{2}}(t_{3}))G_{q_{3}}(t_{3},t_{1})+\text{2 perm.} \nonumber \\
 &+ 8P_{q_{1}}P_{q_{2}}P_{q_{3}}K_{q_{1}}(t_{1})K_{q_{2}}(t_{2})K_{q_{3}}(t_{3})\text{Im}(K_{q_{1}}(t_{2}))\text{Im}(K_{q_{2}}(t_{3}))\text{Im}(K_{q_{3}}(t_{1})),   \nonumber 
\end{align}
It is straightforward to verify that each of these seven terms corresponds to a particular discontinuity of a tree-level Feynman-Witten diagram, 
and consequently \eqref{eqn:psi3phi3} can be written as, 
\begin{align} 
    -\fig{TriangleGraph} &= \int_{\bfq_1 \bfq_1'} P_{\bfq_1 \bfq_1'} \left( \fig{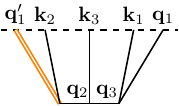} \right) + \text{2 perm.}  \nonumber \\
    &+\int_{\substack{\bfq_1 ,\bfq_2 \\ \bfq_1' ,\bfq_2'}} \, P_{\bfq_1 \bfq_1'} P_{\bfq_2 \bfq_2'}  \left(  \fig{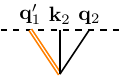}  \right) \left(  \fig{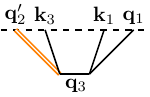}  \right) + \text{2 perm.}   \label{eqn:triangle_CTT} \\
    &+\int_{\substack{\bfq_1 ,\bfq_2,\bfq_3 \\ \bfq_1' ,\bfq_2',\bfq_3'}}P_{\bfq_1 \bfq_1'} P_{\bfq_2 \bfq_2'}P_{\bfq_3 \bfq_3'}\left(  \fig{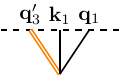}  \right)\left(  \fig{Psi3Phi3CTTq1con}  \right)\left(  \fig{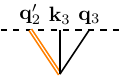}  \right)\nonumber 
\end{align}
where the permutations indicated correspond to the other ways of cutting the internal lines of the loop. Note that if we fix the orientation of the loop as above, the $\disc{}$ is always taken of the clockwise-most half-edge after the cut. 
Similarly to the two-vertex loop, the triangle graph \eqref{eqn:psi3phi3} loop can be oriented either clockwise or anticlockwise and there is therefore a second cutting rule 
which corresponds to summing over all ways of cutting the loop but instead taking the $\disc{}$ of the other (counter-clockwise) half-edges after the cut. 
The difference between these two cutting rules again vanishes thanks to a collinear tree-level identity which generalises \eqref{eqn:Ctype_2} to two internal edges. 

Returning again to the example of $\tfrac{\lambda}{3!} \phi^3$ on Minkowski---for which $\psi_3^{\rm tree}$ was given in \eqref{eqn:psi3_Mink} and used to bootstrap both the four- and five-point exchange diagrams, \eqref{eqn:exch_Mink_eg2} and \eqref{eqn:exch5_eg}---we now see that this input can be used in \eqref{eqn:triangle_CTT} to completely determine the loop momentum integrand for the triangle graph. For that example, \eqref{eqn:triangle_CTT} evaluates to,
\begin{align} 
  \fig{TriangleGraph} \;\;  \begin{gathered} = \lambda^{3}\int_{\bfq_{1}\bfq_{2}\bfq_{3}}\tilde{\delta}(\bfk_{1}+\bfq_{3}-\bfq_{1})\tilde{\delta}(\bfk_{2}+\bfq_{1}-\bfq_{2})\tilde{\delta}(\bfk_{1}+\bfq_{1}-\bfq_{2})   \qquad \\
\qquad\qquad \times \left( \frac{1}{ \omega_{T} e_{1} e_{2} e_{3}}\left(\frac{1}{ E_{3} (k_{T}+2 q_{3}) }+\frac{1}{ E_{1} (k_{T}+2 q_{3}) }\right)    +\text{2 perm.} \right)
\end{gathered}
 \label{eqn:triangle_polytope}
\end{align}
where $\omega_T = \omega_{k_1} + \omega_{k_2} + \omega_{k_3}$ and we have introduced the partial energies,
\begin{align}
e_{1}&= \omega_{k_{1}}+\omega_{q_{3}}+\omega_{q_{1}}\;,\;&e_{2}&= \omega_{k_{2}}+\omega_{q_{1}}+\omega_{q_{2}}\;,\; &e_{3}&=\omega_{k_{3}}+\omega_{q_{2}}+\omega_{q_{3}} \; , \nonumber \\
E_{1}&= \omega_{k_{2}} + \omega_{k_{3}} + \omega_{q_{1}} + \omega_{q_{3}}\;,\;&E_{2}&= \omega_{k_{3}} + \omega_{k_{1}} + \omega_{q_{2}} + \omega_{q_{1}}\;,\;&E_{3}&= \omega_{k_{1}} + \omega_{k_{2}} + \omega_{q_{3}} + \omega_{q_{2}} .\label{eqn:auxeEvariables} 
\end{align}
This agrees with the result from old-fashioned perturbation theory \cite{Arkani-Hamed:2017fdk, Salcedo:2022aal}, or from doing the three nested time integrals in \eqref{eqn:psi3phi3}.

\subsubsection{Proof of the general theorem at one loop}
\label{sec:CTT:subsec:General}

Having shown how a suitable expansion of bulk-to-bulk propagator products can be used to express the momentum integrand of some simple one-loop diagrams in terms of cut tree-level diagrams, the natural question is: how general is this phenomenon? 
In this subsection, we show that in fact \emph{any} one-loop diagram can be decomposed in terms of trees, and we give an explicit expression for the cutting rule that achieves this.

\paragraph{Closed time-like curves.}
The general proof of our tree theorem rests on causality: namely the free propagation described by the classical $G_p^R$ does not allow for closed time-like curves. 
Specifically, $G_p^R$ is defined as the Green function for the classical equation motion of the free theory with causal boundary conditions $G^R_p (t_1 , t_2 ) = 0$ if $t_2 > t_1$. 
For a quadratic Lagrangian like~\eqref{eqn:Lfree}, such a propagator can be explicitly constructed from the mode functions and is given by~\eqref{eqn:GR_def_intro}. We reproduce this equation here for convenience,
\begin{align}
 G^R_p (t_1, t_2 ) = 2 P_p\text{Im} \left( K_p (t_1) K_p^* (t_2)  \right) \Theta (t_1 - t_2 ) \; .
 \label{eqn:GR_def}
\end{align}
The key observation that translates this causal propagator into cutting rules for loop diagrams comes originally from Feynman~\cite{Feynman:1963ax}, who pointed out that a product of $N$ retarded propagators obeys,
\begin{equation}
    \prod_{a=1}^{N} G_{p_{a}}^R (t_{a},t_{a+1})  = 0 \;\; \text{when} \;\; t_{N+1} = t_1 \; .
    \label{eqn:FTT_GR}
\end{equation}
This is simply a consequence of being unable to order $t_1 > t_2 > ... > t_{N+1}$ if the vertices form a closed loop.

\paragraph{Propagator identities.}
Since we can write the bulk-to-bulk propagator~\eqref{eqn:G_to_K} in terms of the retarded propagator~\eqref{eqn:GR_def_intro}, Feynman's identity \eqref{eqn:FTT_GR} implies that the combinations,
\begin{align}
        \mathcal{L}_{p_1 ... p_N} ( t_1, t_2 , ... , t_N ) \equiv \prod_{a=1}^{N} \left. \Bigg(
        G_{p_{a}}(t_{a},t_{a+1})-2P_{p_{a}}K_{p_{a}}(t_{a})\Im(K_{p_{a}}(t_{a+1}))
        \Bigg) \right|_{t_{N+1} = t_1}
        \label{eqn:LN_def}
\end{align}
all vanish identically. 
As in \cite{Melville:2021lst,Goodhew:2021oqg}, and described in appendix~\ref{sec:overview}, for every propagator identity that relates different powers of $G_p$ and $K_p$ there is a cutting rule which relates Feynman diagrams with different numbers of internal lines.   

\paragraph{Cosmological Tree Theorem.}
To prove the tree theorem at one loop, consider the general one-loop diagram composed of $N$ interactions,
\begin{equation}\label{eqn:D_def}
        D =  \tilde\delta \left( \sum_{a=1}^N \bfk_a \right) \int_{\bfq_1} \left[ \prod_{a=1}^{N}\int_{t_{a}}  \lambda_a K_{k_{a}}(t_{a}) G_{q_a} (t_a, t_{a+1} )  \right]
\end{equation}
where $\bfk_a$ is the total momentum flowing into each vertex from the boundary, momentum conservation at each vertex requires that $\bfq_a - \bfq_{a+1} = \bfk_b$, and again we identify $t_{N+1} = t_1$  and $\bfq_{N+1} = \bfq_1$ for notational convenience. $\lambda_a$ is a vertex factor that can depend on each momenta flowing into the vertex and also the time $t_a$. 

We can use the fact that~\eqref{eqn:LN_def} vanishes to write this diagram $D$ as a sum over cuts.
More precisely, we ``cut'' a line by making the replacement,
\begin{align}
    G_{q_a} (t_a , t_{a+1} ) \to  \int_{\bfq_a'} P_{\bfq_a \bfq_a'} \;  K_{q_a} (t_a ) K_{q_a'} ( t_{a+1} )
    \label{eqn:cut_def}
\end{align}
in~\eqref{eqn:D_def}, which corresponds diagrammatically to replacing an internal line with two external half-edges.
To formalise the sum over cuts, we will denote the set of internal lines that make up the loop by $I$, and the set of lines which have been cut by $C$ (with sizes $|I|$ and $|C|$ respectively). 
Cutting the lines $C$ in diagram $D$ produces a new diagram $D_C$, which may no longer be connected. We use $D_C^{(n)}$ to refer to the connected subdiagrams of $D_C$. 
With that notation, the causal identity,
\begin{align}
   \left[  \prod_{a=1}^N \int_{t_a} K_{k_a} (t_a ) \right] \mathcal{L}_{p_1 ... p_N} (t_1, ... , t_N ) = 0
   \label{eqn:LN_dressed}
\end{align}
from~\eqref{eqn:LN_def} becomes the Cosmological Tree Theorem,
\begin{mdframed}[style=boxed]
\begin{align}
   -D  &=  \sum_{ \substack{ C \subseteq I \\ C \neq \{ \} } }^{2^{|I|}-1} \, 
      \left[ \prod_{a \in C}^{|C|}  \int_{\bfq_a \, \bfq_a'}  P_{\bfq_a \bfq_a'}  \right] \prod_{n} \, \disc{ \{ q_a' \} } \left[  D_C^{(n)}   \right]
    \; .
    \label{eqn:CTT}
\end{align}
\end{mdframed}
\eqref{eqn:CTT} is a more precise version of the schematic~\eqref{eqn:CTT_schematic}, and is the central result of this work.

\paragraph{An example.}
The previous examples in subsection~\ref{sec:CTT:subsec:examples} follow immediately from the general theorem~\eqref{eqn:CTT} when $|I| = 1, 2$ and $3$.
For instance, for the two-vertex loop, we have two internal lines, labelled by $I = \{ 1, 2\}$, and therefore 3 cut diagrams,
\begin{align}
    D_{ \{ 1 \}} &= \fig{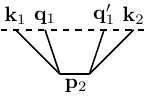} \; , \;\; 
    &D_{ \{ 2 \}} &= \fig{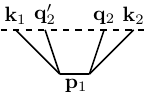} \; , \;\;
    &D_{ \{ 1 ,2 \}} &=  \underbrace{\fig{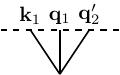}}_{D_{\{1,2\}}^{(1)}} \, \underbrace{\fig{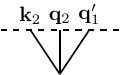}}_{ D_{\{ 1,2 \}}^{(2)} } \; . \nonumber 
\end{align}
Note that $D_{\{1\}}$ and $D_{\{ 2 \} }$ are connected, while $D_{\{1 , 2 \}}$ contains the two connected subdiagrams indicated.
Taking the discontinuities specified in~\eqref{eqn:CTT} then reproduces the coloured diagrams on the right-hand-side of~\eqref{eqn:sunset_CTT}.

\paragraph{Further identities.}
What about the other identities discussed in subsection~\ref{sec:CTT:subsec:examples}?
Well, note that since \eqref{eqn:CTT} uses discontinuities with respect to the $q_a'$ only, it is sensitive to how we have oriented the loop. 
Had we instead written each internal line as $G_{p_a} ( t_{a+1} , t_a )$, then applying the cuts as in~\eqref{eqn:cut_def} would have resulted in \eqref{eqn:CTT} with $\disc{ \{ q'_a \} }$ replaced by $\disc{ \{ q_a \} }$ (where the convention is that $\bfq_a'$ is the clockwise-most half-edge),
\begin{align}
   -D  &=  \sum_{ \substack{ C \subseteq I \\ C \neq \{ \} } }^{2^{|I|}-1} \, 
      \left[ \prod_{a \in C}^{|C|}  \int_{\bfq_a \, \bfq_a'}  P_{\bfq_a \bfq_a'}  \right] \prod_{n} \, \disc{ \{ q_a \} } \left[  D_C^{(n)}   \right]
    \; .
    \label{eqn:CTT_2}
\end{align}
At first sight, this may seem like a trivial re-labelling of \eqref{eqn:CTT}, however it is a non-trivial consequence of the fact that $\mathcal{L}_{p_1 ... p_N} ( t_N, t_1, ... , t_{N-1}  ) $ also vanishes, which follows from Feynman's identity~\eqref{eqn:FTT_GR} for the advanced propagator. 
So the Cosmological Tree Theorem for one-loop graphs is always, in effect, a \emph{pair} of relations, which correspond to cutting the loop open with either orientation. 

In fact, comparing the two tree theorems, we see that,
\begin{align}
    \sum_{ \substack{ C \subseteq I \\ C \neq \{ \} } }^{2^{|I|}-1} \, 
      \left[ \prod_{a \in C}^{|C|}  \int_{\bfq_a \, \bfq_a'}  P_{\bfq_a \bfq_a'}  \right] \left(  \prod_{n}  \disc{ \{ q_a \} } \left[  D_C^{(n)}   \right] -  \prod_{n}  \disc{ \{ q_a' \} } \left[  D_C^{(n)}   \right] \right) 
      = 0
      \label{eqn:CTT_3}
\end{align}
These integrands correspond to the tree-level colinear identities discussed at the end of subsection~\ref{sec:tree}. 

The fact that the tree-level causality conditions guarantee consistency of the different one-loop cutting rules turns out to be a general trend. 
Although we have presented a general argument for one-loop graphs, it is straightforward to replace $\prod_{a} K_{k_a} (t_a)$ above with any desired function of bulk-to-boundary and bulk-to-bulk propagators. \eqref{eqn:CTT} is therefore a general result which can be applied to any closed loop within a diagram. 
Since each loop can be oriented either clockwise or counterwise, there are a total of four different ways to cut a two-loop diagram into tree diagrams. 
The difference between these identities corresponds precisely to the one-loop identities discussed in subsection~\ref{sec:CTT:subsec:examples} above. 
The general pattern is that the $L$-loop identities guarantee the consistency of the different ways of cutting open an $L+1$-loop diagram.

\paragraph{Including spin.}
So far we have focused on the application of the Cosmological Tree Theorem to scalar fields. 
However, the underlying conditions (unitarity/causality of free propagators) that we have used do not specify the mass or the spin of the field. 
Indeed, the hermitian analyticity of the bulk-to-bulk propagators of spinning fields was proven in \cite{Goodhew:2021oqg}. 
Our Cosmological Tree Theorem can therefore be applied more widely to fields of any spin. 

To extend the Cosmological Optical Theorem from scalar to any other field content, all that is required is to decorate the internal lines to account for all of the internal quantum numbers which characterise each field. 
For instance, the tree theorem representation of the two-point wavefunction coefficient becomes:
\begin{align}
 - 2 \psi_{\bfk_1 \bfk_2}^{\alpha_{1}\alpha_{2}} &=\sum_{\lambda_1 \lambda_1'}   \int_{\bfq_{1}\bfq'_{1}}P_{\bfq_{1}\bfq'_{1}}^{\lambda \lambda'} \; \disc{q'_{1}} \left[ \psi_{\bfk_1 \bfq_1 \bfq_1' \bfk_2}^{\alpha_{1}\lambda_1 \lambda_1'\alpha_{2}} \right]
    + \sum_{\lambda_2 \lambda_2'}  \int_{\bfq_{2}\bfq'_{2}}P^{\lambda_2\lambda_2'}_{\bfq_{2}\bfq'_{2}} \; \disc{q'_{2}} \left[ \psi_{\bfk_1 \bfq_2' \bfq_2 \bfk_2}^{\alpha_{1}\lambda_2\lambda_2'\alpha_{2}} \right]  \nonumber \\
    &+ \sum_{\substack{ \lambda_1 \lambda_1' \\ \lambda_2 \lambda_2' } } \int_{\substack{\bfq_{1}\bfq'_{1} \\ \bfq_{2}\bfq'_{2}}}P_{\bfq_{1}\bfq'_{1}}^{\lambda_1 \lambda_1'}P^{\lambda_2\lambda_2'}_{\bfq_{2}\bfq'_{2}} \; \disc{q'_{2}}\left[ \psi_{ \bfk_1 \bfq_2' \bfq_1}^{\alpha_{1}\lambda_2'\lambda_1}  \right] \disc{q'_{1}} \left[ \psi_{\bfk_2 \bfq_2 \bfq_1'}^{\alpha_{2}\lambda_2\lambda'_1}\right] \; .  \label{eqn:CTTgrav}
\end{align}
where the superscript on each wavefunction coefficient indicates the intrinsic quantum numbers of each field (e.g. the spin and helicity). 

As a concrete example, consider the interaction, 
\begin{equation}
    \mathcal{L}_{\text{int}}=-\frac{ g (t)}{2}h^{ij}\partial_{i}\phi\partial_{j}\phi,
\end{equation}
between the scalar $\phi$ and a metric fluctuation $h_{ij}$ (in a transverse, traceless gauge). 
This produces a contact diagram:
\begin{equation}
    \fig{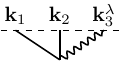} \; =\; \int_{t} i g (t) \, V^{\lambda}(\bfk_{1},\bfk_{2},\bfk_{3})K_{k_{1}}(t)K_{k_{2}}(t)K_{k_{3}}^{\lambda}(t)\label{eqn:gravcon}
\end{equation}
The vertex function $V^{\lambda}(\bfk_{1},\bfk_{2},\bfk_{3})$ is proportional to the momentum of the scalar lines and the polarisation tensor\footnote{
The polarisation tensors $\epsilon^\lambda (\bfk)_{ij}$ are transverse, symmetric and traceless. They are also conjugate under parity $(\epsilon^{\lambda}(\bfk)_{ij})^{*}=\epsilon(-\bfk)^{\lambda}_{ij}$. This leads to the vertex function  $V^{\lambda}(\bfk_{1},\bfk_{2},\bfk_{3})$ being hermitian analytic, $V^{\lambda}(\bfk_{1},\bfk_{2},\bfk_{3})=(V^{\lambda}(-\bfk_{1},-\bfk_{2},-\bfk_{3}))^{*}$. While not required for the tree theorem, which uses only $\disc{}$'s, this is crucial for other unitarity cutting rules that involve $\Disc{}$ operations.
} of the graviton with helicity $\lambda$,
\begin{equation}
    V^{\lambda}(\bfk_{1},\bfk_{2},\bfk_{3})\propto(\bfk_{1})_{i}(\bfk_{1})_{j}\epsilon^{\lambda}(\bfk_{3})_{ij}.
\end{equation}
This interaction leads to two different exchange diagrams---one with the exchange of a graviton,
\begin{equation}
\fig{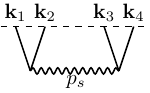}= \sum_{\lambda, \lambda'} \int_{t_{1}t_{2}} i g (t_1) g(t_2) \, V^{\lambda}(\bfk_{1},\bfk_{2},\bfp_{s})K_{k_{1}}(t_{1})K_{k_{2}}(t_{1})G^{\lambda\lambda'}_{p_{s}}(t_{1},t_{2})K_{k_{3}}(t_{2})K_{k_{4}}(t_{2})V^{\lambda'}(\bfk_{3},\bfk_{4},-\bfp_{s})\label{eqn:gravexchange}
\end{equation}
where the $\lambda, \lambda'$ indices are summed over, and one with the exchange of a scalar,
\begin{equation}
    \fig{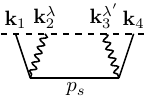}=\int_{t_{1}t_{2}}  i g(t_1) g(t_2) V^{\lambda}(\bfk_{1},\bfk_{2},\bfp_{s})K_{k_{1}}(t_{1})K^{\lambda}_{k_{2}}(t_{1})G_{p_{s}}(t_{1},t_{2})K^{\lambda'}_{k_{3}}(t_{2})K_{k_{4}}(t_{2})V^{\lambda'}(\bfk_{3},\bfk_{4},-\bfp_{s})\label{eqn:scalarexchange}
\end{equation}
where the indices are not summed (they are external kinematic data).
The loop correction to $\psi_2$ from this interaction contains one graviton and one scalar as internal lines,
\begin{align}
    \fig{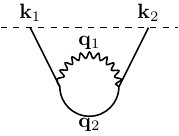}= \sum_{\lambda, \lambda'} \int_{t_{1}t_{2}} g (t_1) g(t_2) \int_{\bfq_{1}\bfq_{2}}&K_{k_{1}}(t_{1})G^{\lambda\lambda'}_{q_{1}}(t_{1},t_{2})G_{q_{2}}(t_{1},t_{2})K_{k_{2}}(t_{2})V^{\lambda}(\bfk_{1},\bfq_{1},\bfq_{2})V^{\lambda'}(\bfk_{2},-\bfq_{1},-\bfq_{2})\label{eqn:gravsunset}
\end{align}
Since both $G^{\lambda \lambda'}$ and $P^{\lambda \lambda'}$ are $\propto \delta^{\lambda \lambda'}$, we see that indeed \eqref{eqn:CTTgrav} is satisfied by the above four diagrams.

\paragraph{Comparison with OFPT.}
Finally, note that on a Minkowski background one can use old-fashioned perturbation (OFPT) of the Lippmann-Schwinger equation to similarly express every wavefunction coefficient without the need for explicit time integrals. 
Let us close this section by making some explicit comparisons between our Cosmological Tree Theorem and the use of OFPT. 
There are three key differences between the two approaches:

\begin{itemize}

\item[(i)] \emph{Different inputs.} OFPT uses only $\psi_1 = 1/\omega_k$ as input, and builds all higher-point coefficients by exploiting the time translation Ward identity of Minkowski. On the other hand, the Cosmological Tree Theorem makes use of all cut diagrams, which can be bootstrapped by unitarity and causality from all contact $\psi_n$.  

\item[(ii)] \emph{Different terms.} This is related to the different inputs: OFPT expands a diagram in terms of $\psi_1$, while the Cosmological Tree Theorem expands in a more varied basis of $\psi_n^{\rm contact}$ so often requires fewer terms overall. This is summarised in Table \ref{table:OFPTvsCTT}.
For OFPT, the number of terms required is $E!$, where $E$ being the number of edges\footnote{
This can be seen by induction: since the OFPT representation of a $E$ edges diagram has $E$ single-cut subdiagrams, that have $E-1$ cut subdiagrams each, that have $E-2$ cut subdiagrams each, and so on... we can follow this procedure until one reaches the single vertex diagram that is only $1/\Omega_k$.
}.
On the other hand, explicitly performing the nested time integrals in what is traditionally called time-orderd perturbation theory (TOPT) requires $3^{E}$ terms (since each internal propagator $G_p (t_1, t_2)$ contains the 3 terms shown in \eqref{eqn:G_to_K}). The Cosmological Tree Theorem is an improvement over both approaches when it comes to the number of terms, since it decomposes the $E$ edges loop into just $2^{E}-1$ terms that we can relate to the discontinuity of different tree-level wavefunction coefficients. 

\item[(iii)] \emph{Different applicability}. OFPT becomes difficult beyond Minkowski (although see \cite{Hillman:2021bnk} for recent progress), but the Cosmological Tree Theorem can be applied on any time-dependent spacetime background. This makes it well-suited for computing cosmological correlators: for instance we show in section~\ref{sec:power} below that it leads to a simple expression for the one-loop integrand for the inflationary power spectrum in the EFT of inflation. 

\end{itemize}

\begin{table}[t]
\centering
\begin{tabular}{|c | c | c | c|} 
 \hline
\multirow{2}{*}{} &  \multirow{2}{*}{OFPT} & \multirow{2}{*}{TOPT} &  Cosmological \\
 & & &  Tree Theorem \\ [0.5ex] 
 \hline
 $E=1$ & 1 & 1 & 1 \\ \hline
 $E=2$ & 2 & 9 & 3 \\ \hline
 $E=3$ & 6 & 27 & 7 \\ \hline
 $E=4$ & 24 & 81 & 15 \\ \hline
 $\vdots$ & $\vdots$ & $\vdots$ & $\vdots$ \\  \hline
 $E$ & $E!$ & $3^{E}$ & $2^{E}-1$ \\ \hline
\end{tabular}
\caption{Number of terms in the different representations of a one-loop diagram with $E$ internal edges.}
\label{table:OFPTvsCTT}
\end{table}

\section{Correlation functions}
\label{sec:corr}

Having explored the consequences of causality for the wavefunction coefficients and derived a number of useful identities---including the tree-level cutting rule~\eqref{eqn:caus_cut_0}, the colinear identity~\eqref{eqn:Ctype_2} and the Cosmological Tree Theorem~\eqref{eqn:CTT} for any loop diagram---we now turn to the question of how these can be used to constrain inflationary observables. 
In particular, we consider the \emph{equal-time correlation functions} of a weakly coupled scalar field $\phi$ on a generic time-dependent background. 
In the context of the EFT of inflation, this background is quasi-de Sitter and $\phi$ represents the Goldstone of broken time translation: its correlators are closed related to the comoving curvature perturbations which later re-enter the horizon and ultimately seed the initial conditions for the Cosmic Microwave Background and Large Scale Structure \cite{Cheung:2007st}. 

First, we briefly review the standard map from wavefunction to correlators via the Born rule.
Then in subsection~\ref{sec:KLN}, we show on general grounds (for any interactions and spacetime background) that the Cosmological Tree Theorem implies a delicate cancellation between different wavefunction contributions to cosmological correlators. 
This is the cosmological analogue of the KLN theorem for amplitudes and cross sections, and in particular it leads to a complete cancellation of certain total energy singularities. 
Then in subsection~\ref{sec:power}, we provide a simplified expression for the one- and two-point function in any theory of massless fields at one-loop, exploiting this cancellation to remove all loops and redundant terms which vanish in dimensional regularisation. 
Finally, we specialise to the EFT of inflation and use this expression to evaluate the inflationary power spectrum. 
We give analogous simplified expressions for the one-loop corrections to primordial non-gaussianity (the bispectrum) in appendix~\ref{app:bispectrum}.

\paragraph{From wavefunction to correlators.}
From the wavefunctional $\Psi [ \phi ]$, one can extract any desired equal-time correlation function using the (functional version of the) Born rule\footnote{
We have implicitly normalised the state / wavefunctional so that $\langle 1 \rangle = 1$. Without this condition, the well-defined observables in this field theory are the ratios $\langle \mathcal{O} \rangle / \langle 1 \rangle$, which introduces an explicit normalisation factor of $1/\int \mathcal{D} \phi | \Psi [ \phi ] |^2$ on the right-hand-side. Diagrammatically, the role of this normalisation is to cancel all vacuum bubble contributions.
}, 
\begin{equation}
 \left\langle  \Omega \left| \;  \mathcal{O} \left( \hat{\phi} , i \hat{\Pi}   \right) \; \right|   \Omega \right\rangle =  \int\mathcal{D}\phi \; \Psi^* [\phi] \; \mathcal{O} \left(  \phi  , \frac{\delta}{\delta \phi}  \right) \Psi [\phi]
 \label{eqn:Born}
 \end{equation}
which follows immediately from inserting a complete set of $\phi$ eigenstates to the left of the operator $\mathcal{O}$. 
The $\mathcal{D} \phi$ represents a functional integral over all $\phi_{\bfk}$ modes (this is not a path integral, but rather an integral over all field configurations at a fixed time). 

In practice, we evaluate \eqref{eqn:Born} by treating any non-Gaussianity in $\Psi [\phi ]$ as a small perturbation. Concretely, we expand the wavefunction as in \eqref{eqn:psi_def}, namely,
\begin{align}
  \Psi [\phi ]   = \exp \left( 
\sum_{n=0}^{\infty} \int_{\bfk_1 ... \bfk_n} \, \frac{ \psi_{\bfk_1 ... \bfk_n} }{n!} \, \phi_{\bfk_1} ... \phi_{\bfk_n}
\right)
\end{align}
where $\psi_0$ is simply an overall normalisation, and in perturbation theory we treat each $\psi_n \sim \mathcal{O} \left( g_*^{n-2} \right)$, except for the tadpole $\psi_1$ which $\sim \mathcal{O} \left( g_* \right)$, where $g_*$ is a small power counting parameter. 

For instance, suppose we define the power spectrum of the full interacting theory via,
\begin{align}
\langle \Omega |  \; \hat{\phi}_{\bfk} \hat{\phi}_{\bfk'} \; | \Omega \rangle =  \mathcal{P}_k \; \tilde{\delta}^3 \left( \bfk + \bfk' \right)
\label{eqn:P_def}
\end{align}
where we use a caligraphic $\mathcal{P}_k$ to distinguish this from the free-theory power spectrum $P_k$. 
A perturbative expansion of the Born rule~\eqref{eqn:Born} then fixes $\mathcal{P}_k$ in terms of the wavefunction coefficients of the previous section.  
As is often the case with quadratic correlators (e.g. the propagator), it is simpler to give the perturbative expansion for the inverse,
\begin{align}
- \frac{ \tilde{\delta}^3 \left( \bfk + \bfk' \right) }{ \mathcal{P}_k }  &=  2 \text{Re}\,  \psi_{\bfk \bfk'} + \int_{\bfq \bfq'} P_{\bfq \bfq'} \left(  \text{Re} \, \psi_{\bfk \bfk' \bfq \bfq'}  +  4 \text{Re} \, \psi_{\bfk \bfk' \bfq} \text{Re} \, \psi_{\bfq'}   \right)    \label{eqn:P_from_wvfn}\\
&+ 2 \int_{ \substack{ \bfq_1 \bfq_1' \\ \bfq_2 \bfq_2' } } P_{\bfq_1 \bfq_1'} P_{\bfq_2 \bfq_2'}  \left( \text{Re} \, \psi_{\bfk \bfq_1 \bfq_2'} \text{Re} \, \psi_{\bfk' \bfq_1' \bfq_2} 
+ \text{Re} \, \psi_{\bfk \bfk'  \bfq_1} \text{Re} \, \psi_{\bfq_1 \bfq_2 \bfq_2'}  \right)  + \mathcal{O} \left( g_*^3 \right)  \nonumber
\end{align}
where we have treated all $\hat{\phi}$ fields as indistinguishable (and hence $\psi_n$ is a symmetric function of its arguments).  

Similar expansions can be given for all other $n$-point functions. 
The other correlator we consider in the main text will be the one-point function,
\begin{align}
\frac{ \langle \hat{\phi}_{\bfk} \rangle }{ \mathcal{P}_k } = v \, \tilde{\delta} \left( \bfk \right)
\label{eqn:v_def}
\end{align}
where we have normalised by $\mathcal{P}_k$ in order to cancel various contributions which correspond diagrammatically to corrections that only affect the propagation of a single leg as it propagates to the boundary (i.e. diagrams of the form \eqref{eqn:trimming}). 
The Born rule \eqref{eqn:Born} can be used to determine $v$ perturbatively in the non-Gaussianity,
\begin{align}
& v \, \tilde{\delta}^3 \left( \bfk \right) = 2 \text{Re} \, \psi_{\bfk} +  \int_{\bfq \bfq'} P_{\bfq \bfq'} \,  \text{Re} \, \psi_{\bfk \bfq \bfq'}
   + \mathcal{O} \left( g_*^3 \right) \; .
   \label{eqn:v_from_wvfn}
\end{align}

\paragraph{Loop expansion.}
The rationale for the power counting $\psi_n \sim \mathcal{O} ( g_*^{n-2} )$ is that each wavefunction coefficient stems from a weakly coupled Lagrangian of the form, $g_*^{n-2} \mathcal{L} \left[ g_* \phi   \right]$, where $g_* \ll 4 \pi$ suppresses Feynman diagrams which contain loops. 
As a result, in the loop expansion~\eqref{eqn:loop_exp} of the wavefunction coefficients, we should treat,
\begin{align}
 \psi_n^{L\text{-loop}} \sim \mathcal{O} \left(  g_*^{2L+n-2} \right) \; .
\end{align}
Equations~\eqref{eqn:P_from_wvfn} and ~\eqref{eqn:v_from_wvfn} then have expansions in $g_*$, for instance,
\begin{align}
v &=  v^{(1)} + \mathcal{O} \left( g_*^3 \right) \; , \;\; & \mathcal{P}_k &= \mathcal{P}_k^{(0)}  +  \mathcal{P}_k^{(2)} + \mathcal{O} \left( g_*^4 \right)  
\end{align}
where $\mathcal{P}_k^{(0)} = P_k$ is the free-theory power spectrum and $\mathcal{P}_k^{(2)}$ is determined by $\psi_2^{\rm 1-loop}$, $\psi_4^{\rm tree}$ and $\psi_3^{\rm tree} \times \psi_3^{\rm tree}$. 
Note that we have defined the free theory such that $\psi_1^{\rm tree} = 0$ and hence the leading order $v^{(1)}$ is determined by $\psi_3^{\rm tree}$ and $\psi_1^{\rm 1-loop}$. 

Notice that each $n$-point correlator starts at $\mathcal{O} \left( g_*^{n-2} \right)$, so at leading order (i.e. $\mathcal{O} (g_*)$) the only non-zero correlators are the one-loop correction to the vev and the tree-level bispectrum. At next-to-leading order (i.e. $\mathcal{O} (g_*^2 )$), the non-zero correlators are the one-loop correction to the power spectrum and the tree-level trispectrum. 
These are the objects we focus on in the main text.
At next-to-next-to-leading order (i.e. $\mathcal{O} ( g_*^3 )$), the non-zero correlators are the two-loop vev, the one-loop bispectrum and the tree-level five-point function. We describe these in appendix~\ref{app:bispectrum}. 

Finally, although we focus predominantly on $\phi$ correlators, since these give the largest signal at the end of inflation, we have checked that similar conclusions can also be drawn for mixed correlators which also contain the conjugate momenta $\Pi$. 

\subsection{IR singularities and the cosmological KLN theorem}
\label{sec:KLN}

Before analysing any particular $n$-point function, in this subsection we give a general argument about how the Cosmological Tree Theorem will affect a generic correlator. 
In short, once our tree theorem replaces loop wavefunction diagrams with tree-level ones, it makes manifest the cancellations that can place between e.g. $\psi_2^{\text{1-loop}}$, $\psi_4^{\rm tree}$ and $\psi_3^{\rm tree} \times \psi_3^{\rm tree}$ in the power spectrum.

\paragraph{Analytic structure of wavefunction.}
Earlier in section~\ref{sec:tree}, we made use of the condition that tree-level Feynman-Witten diagrams for the Bunch-Davies initial state can only have singularities when the total energy flowing into one or more vertices vanishes. 
The idea is that, for a Bunch-Davies initial state, the bulk-to-boundary propagators appearing in any contact diagram have the asymptotic behaviour\footnote{
On Minkowski, a massive field would have $\Omega_k = \sqrt{k^2 + m^2}$ in place of $k$ in this expression. On a quasi-de Sitter background, however, since $k$ blueshifts in the far past any finite mass parameter becomes negligible and the propagators scale like \eqref{eqn:K_exp}. 
},
\begin{align}
 K_{k_1} (t ) ... K_{k_n} (t) \sim e^{i \omega_T t}
 \label{eqn:K_exp}
\end{align} 
at large $t$. 
Integrating this from $t= - \infty$ is what produces singularities when $\omega_T \to 0$. 
Physically, this divergence corresponds to the interaction becoming long-lived: when there is no longer any $e^{ i \omega_T t}$ to suppress the vertex in the far past, it gives an infinite contribution to the wavefunction. 
Since these contact diagrams can be used to construct any tree-level exchange diagram, we similarly conclude that those will have singularities when the energy at any vertex (or collection of vertices) vanishes\footnote{
The singularities when a collection of vertex energies vanish arise naturally in the bootstrap procedure of section~\ref{sec:tree} from the matching of $D_R$ onto the unphysical singularities of $D_C$: for instance the terms $1/E_1 ( \omega_{k_3} + \omega_{q})$ and $1/E_1 ( \omega_{k_3} + \omega_{k_4} + \omega_{k_5} )$ in \eqref{eqn:DR_eg_2} are producing from the single-vertex function $E_1$ poles which depends on the total energy of two or three vertices. 
}. 
This typically occur in regions of parameter space that are not observationally accessible: for instance when some of the $\omega_k$ are negative. 

Recently, \cite{Salcedo:2022aal} studied the analytic structure of the wavefunction on Minkowski, and used old-fashioned perturbation theory to show that loop integrands similarly can only have singularities when the total energy flowing into one or more vertices vanishes (the ``energy conservation condition''). As a result, a simple Landau analysis shows that $\psi_n$ has branch points at certain ``thresholds'': the minimum energy at which this energy conservation condition can be met. 

The Cosmological Tree Theorem can be used to extend this argument to other spacetime backgrounds. 
Thanks to \eqref{eqn:CTT}, a general loop integrand can be decomposed into tree-level diagrams, which have known singularities (when the total energy flowing into one or more vertices vanishes).
The subtlety is that \eqref{eqn:CTT} involves discontinuities that change the sign of some energies and produce unphysical singularities, which ultimately cancel one another in the final sum. 
Applying the energy conservation condition of \cite{Salcedo:2022aal} directly to \eqref{eqn:CTT} will therefore produce a list of \emph{possible} thresholds, but in practice we expect several of them to cancel out. 
We leave a systematic study of these cancellations for the future. 
For now, all we need is the idea that the singularities of a diagram arise when the total energy flowing into one or more vertex vanishes. 
 Together with the Cosmological Tree Theorem, this is sufficient to demonstrate the cancellation of certain singular points which takes place for equal-time correlators.

\paragraph{Tree-level singularities.}
At their first non-trivial order, equal-time correlators are determined simply by tree-level wavefunction coefficients. 
There are three kinds of singularities which can occur in these objects when a total or partial energy vanishes:
\begin{itemize}

\item[(i)] \emph{Poles}.  
As can be seen in the various examples of section~\ref{sec:wvfn}, for massless fields on both Minkowski and de Sitter there are generally poles in $\psi_n$ whenever the energy flowing into a connected set of vertices vanishes. On Minkowski, these poles are the finite-time avatar of the energy-conserving $\delta$-functions which appear in asymptotic observables such as the $S$-matrix (indeed, pushing the wavefunction to $t \to + \infty$ would replace each simple pole with such a $\delta$ function). 
On de Sitter, the order of the pole is fixed by dilation invariance. 

\item[(ii)] \emph{Late-time logs}. 
But poles are not the only kind of singularity that can appear at tree-level. Unlike for amplitudes, which are always finite at tree-level (thanks to the energy-conserving $\delta$ functions), when it comes to finite-time wavefunction coefficients particular Feynman-Witten diagrams can diverge and require renormalisation even at tree-level. 
This happens whenever the masses of the fields satisfy certain sum rules, which leads to them persisting at arbitrarily late times (by contrast, the simple poles above correspond to the early-time limit). For instance, for the cubic interaction $\tfrac{1}{3!} a^4 (t)\phi^3$ in which each field has mass $m^2 = 2H^2$ (so that the total conformal dimension $3 \Delta = 3$ matches the spacetime dimension), the $\psi_3^{\rm tree}$ coefficient
\begin{align}
\fig{contactkkk} = \int_{-\infty}^{t_0} \frac{d t}{t} e^{ i k_T t} \sim \log (  k_T t_0 ) 
\end{align} 
and diverges as we approach the $t_0 = 0$ conformal boundary. 
As discussed in \cite{Bzowski:2015pba, Cespedes:2020xqq}, these can be renormalised into the Boundary Operator Expansion (in the same way that UV divergences in ordinary QFT are absorbed into the Operator Product Expansion), which effectively replaces $t_0 \to 1/\mu$ with some sliding RG scale.
This is essentially the phenomenon of holographic renormalisation (see e.g. \cite{Skenderis:2002wp}), and in the context of de Sitter is perhaps most usefully viewed in terms of dynamical RG \cite{Green:2020txs, Cohen:2021fzf, Premkumar:2021mlz}.

\item[(iii)] \emph{Massive branch cuts.} 
The other way in which these simple poles become branch points is for fields with non-zero mass. 
Even on Minkowski, if we consider the complex $k$ plane for massive fields we would encounter a branch point at $ k = -m$ rather than a simple pole due to the square root in $\omega_k = \sqrt{k^2 + m^2}$. 
On de Sitter, we similarly find branch cuts in the complex $k$ plane whenever massive fields are involved.
For instance, consider the interaction $\tfrac{1}{2} a^4 (t) \phi^2 \sigma$ between two fields of mass $m^2 = 2H^2$ and one field of mass $m^2 = \frac{9}{4} + \nu^2$. 
The corresponding $\psi_3^{\rm tree}$ is\footnote{
While $\psi_n \sim t_0^n$ for such fields near the boundary, one can extract the finite $\psi_n t_0^{-n}$ by a suitable redefinition of the boundary operators in the theory~\cite{Cespedes:2020xqq}, and this is what is shown in \eqref{eqn:branch_eg}. 
},
\begin{align}
\fig{contactkkk} \propto \int_{-\infty}^0  d t \, e^{i k_{12} t}  \; \left( - k_3 t \right)^{1/2}  \; H_{i\nu}^{(2)} ( - k_3 t ) 
\propto \frac{  P^{-1}_{i \nu - \tfrac{1}{2}} \left( \frac{k_{12}}{k_3} \right)  }{ \sqrt{ k_3^2 - k_{12}^2 } }
\label{eqn:branch_eg}
\end{align}
This Legendre function has a \emph{branch point} at $z = k_{12}/ k_3  = -1$, which conventionally runs along the negative real axis to $z = -\infty$.  

\end{itemize}
Since we are primarily interested in corrections to inflationary correlators, which are dominated by an approximately massless Goldstone mode, we will mostly focus on the pole singularities (i). 
However, it is important to mention singularities (ii) and (iii) since these would play an important role in future applications to the cosmological collider signals produced by heavy states during inflation.

\paragraph{One-vertex loop.}
The first kind of cancellation which takes place due to our tree theorem is for loops containing a single vertex. These always appear together with a partner contribution,
\begin{align}\label{eqn:oneBlob}
\fig{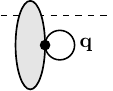}+\int_{\bfq \bfq'} P_{\bfq \bfq'} \fig{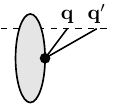} 
\end{align}
where the grey blob represents any particular set of other $\psi$ diagrams and their complex conjugates. The two terms in \eqref{eqn:oneBlob} are the in-in analogue of the \emph{virtual} and \emph{real} emission contributions to the inclusive cross-section in particle physics. 
By applying the tree theorem, the one-loop virtual emission diagram can be replaced by a $\text{disc}$, so that \eqref{eqn:oneBlob} becomes,
\begin{align}
 \int_{\bfq \bfq'} P_{\bfq \bfq'} \, \left\{ \fig{oneBlobOneCut}  - \discn{\bfq'} \left[ \fig{oneBlobOneCut} \right] \right\}=\int_{\bfq \bfq'} P_{\bfq \bfq'} \, \,  \fig{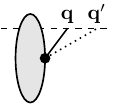}
\label{eqn:oneBlobCTT}
\end{align}
where the dotted line represents an analytic continuation of the corresponding energy. 
Crucially, since a diagram can only contain singularities when the total energy flowing into a vertex vanishes, we see that this final diagram cannot contain any singularities which depend on both $q$ and $\omega_{\rm blob}$, since the total energy at the shown vertex is $\omega_q - \omega_{q'} + \omega_{\rm blob} = \omega_{\rm blob}$ (where $\omega_{\rm blob}$ is the total energy flowing into the blob from the boundary).
As a result, for massless fields this integral takes the simple form $\int_q  P_q \; \text{Poly} (q)$ and such (``scale-free'') integrals \emph{vanish} in dimensional regularisation. This is perhaps the most striking consequence of the tree theorem: it guarantees that the combination of real and virtual emission shown in \eqref{eqn:oneBlobCTT} exactly cancel when evaluated in dim reg. In the analogous calculation of the inclusive cross-section, this cancellation is the so-called KLN theorem. The fact that in-in correlators enjoy a ``cosmological KLN theorem'', thanks to causality and the tree theorem~\eqref{eqn:CTT}, explains why several recent calculations of equal-time correlation functions have found a simpler analytic structure than the corresponding wavefunction coefficients \cite{Salcedo:2022aal, Lee:2023jby}. 

Note that for massive fields, the cancellation shown in~\eqref{eqn:oneBlobCTT} still takes place (leaving just a single analytically continued diagram with total energy $\omega_{\rm blob}$), however the resulting integral is no longer scale-free and need not vanish. 
For instance on Minkowski $2 P_{q} = 1/\sqrt{q^2 + m^2}$ and hence gives a finite integral (which $\sim m^{d-1}$). 
On de Sitter, contact diagrams with two heavy external legs typically $\sim P_{i \nu - \frac{1}{2}}^{j} \left( z \right) \left( 1 - z^2 \right)^{j/2}$ where $z = -1 + \left( ( q + q' )^2 - \omega_{\rm blob}^2 \right)/(2 q q')$ and again $z=-1$ corresponds to the branch point: so although $q + q' = 0$ in \eqref{eqn:oneBlobCTT} (and hence the branch point becomes simply $\omega_{\rm blob} = 0$), this function nonetheless produces a non-zero integral over $q$. 
However, note that in both of these cases the kind of singularity which emerges from performing the loop integral is \emph{the same} as the singularity in $\omega_{\rm blob}$ already present in the tree-level integrand. 
This is guaranteed by a simple Landau analysis: the only way to increase the order of the singularity (e.g. to produce dilogs) from a singularity in the integrand is if this integrand has a singular point that mixes $\omega_{\rm blob}$ and $q$. Since our tree theorem shows this cannot happen, in effect it has shown that the analytic structure of the equal-time correlator remains that of the tree-level wavefunction (even at loop-level).

\paragraph{Two-vertex loop.}
This KLN cancellation is not limited to loops with a single vertex. An analogous cancellation takes place for loops containing two vertices, which always appear together with three other contributions, 
\begin{align}
&\text{Re} \left[ \fig{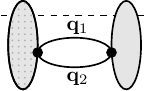} \right]+\int_{\bfq_1 \bfq'_1} P_{\bfq_1 \bfq'_1}  \text{Re} \left[  \fig{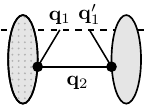}  \right]\label{eqn:twoBlob} \\ 
&+\int_{\bfq_1 \bfq'_1} P_{\bfq_2 \bfq'_2}  \text{Re} \left[  \fig{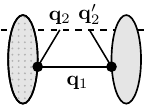}   \right]+\int_{ \substack{\bfq_1 \bfq_1' \\ \bfq_2 \bfq_2'} }  P_{\bfq_1 \bfq_1'} P_{\bfq_2 \bfq_2'}  2 \text{Re} \left[\fig{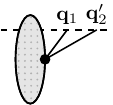}\right] \text{Re}\left[\fig{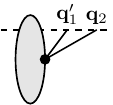}\right] \; .\nonumber 
\end{align}
As above, each gray blob represents a particular set of $\psi_n$ diagrams (or their conjugates) and we have added a hatched pattern to indicate that there could be different sets attached to the left and right vertices. The first of these terms represents the ``virtual'' emission, while the latter three terms represent the ``real'' emission. Again we notice the close parallel between extracting an in-in correlator from the wavefunction and extracting an inclusive cross-section from an amplitude. Applying our tree theorem replaces the virtual one-loop emission with $\text{disc}$'s of real tree-level emission, 
\begin{align}
&\int_{\bfq_1 \bfq'_1} P_{\bfq_1 \bfq'_1}    \, \text{Re} \left[ \fig{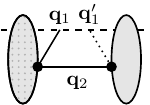}  \right] + \int_{\bfq_2 \bfq'_2} P_{\bfq_2 \bfq'_2}    \, \text{Re}  \left[ \fig{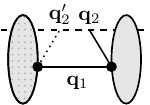}  \right]  \label{eqn:twoBlobCTT} \\
&+ \int_{ \substack{\bfq_1 \bfq_1' \\ \bfq_2 \bfq_2'} }  P_{\bfq_1 \bfq_1'} P_{\bfq_2 \bfq_2'}  \left\{ 2 \text{Re} \left[ \fig{oneBlobOneCutWithDotsq1q2}    \right] \text{Re} \left[\fig{oneBlobOneCutWithDotsq2q1}
\right]-\text{Re} \left[ \discn{\bfq_2'} \left[  \fig{oneBlobOneCutWithDotsq1q2}   \right]  \discn{\bfq_1'} \left[  \fig{oneBlobOneCutWithDotsq2q1} \right]\right]\right\}  \nonumber 
\end{align}
where again the dotted line represents an analytic continuation to negative energy of that particular field. The first line no longer vanishes for massless fields in dim reg, but it does have a simpler singularity structure than the original wavefunction coefficients. Explicitly, suppose that $\omega_L$ and $\omega_R$ are the total energies flowing into the left and right blobs from the boundary. This integrand can have singularities when the energy at either vertex vanishes, which would lead to singularities in $\omega_{L}$ or $\omega_{R}$ separately. But since the total energy flowing into both vertices is independent\footnote{
Explicitly, the total energy flowing into both vertices is $\omega_{q_1} - \omega_{q_1'} + \omega_{L} + \omega_{R} = \omega_{L} + \omega_{R} $.} of the loop momenta, we find that there is no thershold in $\omega_L + \omega_R$. As a result, the integral on the first line of \eqref{eqn:twoBlobCTT}  cannot produce any branch cut singularity in the total energy $\omega_L + \omega_R$ if the tree-level diagram contains only poles (and it cannot increase the order of the branch point if the tree-level diagram already contains a branch cut). 
In the second line there is a partial cancellation between the various disconnected diagrams, but the important observation is that these integrals also cannot produce any singularity in the total $\omega_L+\omega_R$, since each disconnected factor can only have singularities in $\omega_L$ or $\omega_R$ separately. So overall, we conclude that the virtual and real emissions shown in \eqref{eqn:twoBlob} combine in such a way that all branch cut singularities in $\omega_L + \omega_R$ exactly cancel for massless fields. This is a direct consequence of causality and the tree theorem.  It also further strengthens the analogy with the KLN cancellation of IR divergences in amplitudes: the KLN theorem often implies that the soft divergences $\sim 1/\omega$ in the amplitude will cancel out between the real and virtual contributions to the inclusive cross-section, rendering the latter an IR-safe observable. Here we similarly find that while the wavefunction may contain higher-order singularities as the total energy $\omega_L + \omega_R \to 0$, these are guaranteed to cancel out when computing an in-in correlator.

\paragraph{Three-vertex loop.}
This pattern of cancellations continues for loops with arbitrarily many vertices.
For instance, a loop with three vertices always comes with real emission diagrams corresponding to the different ways of cutting the loop, 
\begin{align}
&\fig{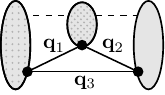}+\sum_{\rm perm.}^3 \int_{\bfq_1 \bfq_1'} P_{\bfq_1 \bfq_1'} \, \fig{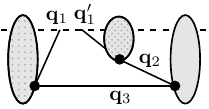}\label{eqn:threeBlob} \\
&+2 \sum_{\rm perm.}^3 \int_{ \substack{ \bfq_1 \bfq_1' \\  \bfq_2 \bfq_2'  } } P_{\bfq_1 \bfq_1'} P_{\bfq_2 \bfq_2'} \fig{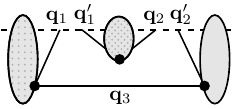}+ 2 \int_{ \substack{ \bfq_1 \bfq_1' \\  \bfq_2 \bfq_2'  \\ \bfq_3 \bfq_3' } } P_{\bfq_1 \bfq_1'} P_{\bfq_2 \bfq_2'} P_{\bfq_3 \bfq_3'} \fig{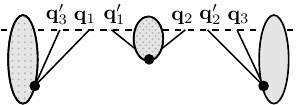}\nonumber
\end{align}
where the three gray blob represent any three particular sets of $\psi_n$ diagrams and their conjugates.
Applying the tree theorem, we again find that the connected terms on the first line combine to give,
\begin{align}
\sum_{\rm perm.}^3 \int_{\bfq_1 \bfq_1'} P_{\bfq_1 \bfq_1'} \, \fig{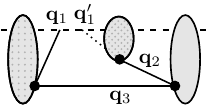}
\label{eqn:threeBlobCTT}
\end{align}
plus the $\text{disc}$ of disconnected diagrams like those on the second line. As before, if $\omega_L$, $\omega_C$ and $\omega_R$ denote the total energies flowing into the left-, central- and right-blob, we find that the analytic continuation of the energy in \eqref{eqn:threeBlobCTT} means that this integral cannot produce branch cut singularities in the total $\omega_L + \omega_C + \omega_R$ that were not already present in the tree-level wavefunction\footnote{
The argument is the same as before: only the combination of all three vertices could produce the required threshold, but the total energy flowing into all three vertices is independent of the loop momentum. 
} i.
It can, however, produce partial energy singularities in any single $\omega_i$ or pair $\omega_i + \omega_j$, as can the other disconnected diagrams.

\paragraph{Beyond one loop.}
Now we briefly discuss higher loop diagrams.
In all of the preceding diagrams, the grey blobs may contain further loops: this does not affect our argument and in such cases the total energy singularity still cancels.
The qualitative difference when going to higher loops is the presence of overlapping loops such as,
\begin{align}
\fig{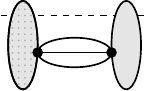}
\qquad \qquad \fig{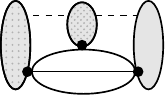}
\end{align}
We find that the same pattern of KLN cancellations continues for these higher loop corrections. 
For instance, the diagram on the left always appears with the following terms,
\begin{align}
    \frac{1}{3}\,\fig{twoBlobTwoLoops}\,+\frac{2}{3}\sum_{\text{perm.}}^{3}\int_{\bfq_{3}\bfq'_{3}}P_{\bfq_{3}\bfq'_{3}}\fig{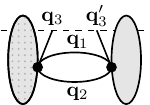}+\frac{1}{3}\sum_{\text{perm.}}^{3}\int_{\substack{\bfq_{1}\bfq'_{1}\\ \bfq_{2}\bfq'_{2}}}P_{\bfq_{1}\bfq'_{1}}P_{\bfq_{2}\bfq'_{2}}\fig{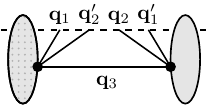}
    \label{eqn:twoBlobsTwoLoops}
\end{align}
Once we apply the tree theorem the remaining connected terms all take the same tree-level form but with different analytic continuations of the external legs. Once combined, we find that the only terms which survive have two pairs of identified external energies with opposite signs, so that again the total energy flowing into both vertices is independent of the loop momenta. 
An explicit example of this can be found in appendix~\ref{app:bispectrum} for the two-loop field vev $v^{(3)}$.

~\\
Our main conclusion from this general discussion may be stated as:
 \begin{tcolorbox}
 \textbf{Cosmological KLN theorem (Massive fields)}: \\ 
 \textit{Any total-energy branch point produced in the wavefunction by loop integration will not appear in the corresponding equal-time correlation functions.}
 \end{tcolorbox}
\noindent In particular, since for massless fields (with derivative interactions) the only tree-level singularities in the wavefunction are poles, we have a stronger corollary in that case,
 \begin{tcolorbox}
 \textbf{Cosmological KLN theorem (Massless fields)}: \\ 
 \textit{Equal-time correlation functions may only have poles in the total energy.}
 \end{tcolorbox}
\noindent Our argument above uses unitarity and causality to prove this theorem in perturbation theory for all one-loop diagrams with up to three vertices, and it also holds true in all of the other examples we have checked. 

Note although we have focussed on what happens in dimensional regularisation (where various scale-free integrals vanish), this KLN theorem also applies to other regularisation schemes. For instance, with a hard cut-off, integrals of the form $\int_q P_q \, \text{Poly} ( q , \Omega_T )$ will not vanish, but rather produce a polynomial $\text{Poly} ( \Lambda, \omega_T$), which is also free of branch cuts at $\omega_T = 0$ for massless fields.


\subsection{Some examples}
\label{sec:power}

Following our general discussion, we now focus on some specific examples to better illustrate the cancellations and their physical/computational significance.

\paragraph{One-loop vev.}
The simplest example of the above KLN cancellation is for the one-point function~\eqref{eqn:v_def}, i.e. the vacuum expectation value (vev) of the field, $v$. 
Note that we have defined the free theory such that $v = 0$ in the absence of any interactions (so that physically, $\phi$ represents the fluctuations around the classical trajectory). 
At next-to-leading order in $g_*$, a cubic interaction can produce corrections to $v$ via ``tadpole'' diagrams. In practice, these should be renormalised away by redefining $\phi \to \phi - v$, which introduces an extra step into the renormalisation process (which can mix e.g. the bispectrum into the power spectrum).  
Happily, the Cosmological Tree Theorem proves that this is not necessary: at least for massless fields at one loop order. 

To see this explicitly, consider the loop expansion of~\eqref{eqn:v_from_wvfn}, 
\begin{align}
v^{(1) } \, \tilde{\delta}^3 \left( \bfk \right) 
=  \text{Re} \left[ \fig{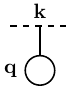}  + \int_{\bfq \bfq'} P_{\bfq \bfq'}   \fig{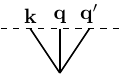} \right] \; . 
\end{align}
Applying the tree theorem produces simply,
\begin{align}
v^{(1) } \, \tilde{\delta}^3 \left( \bfk \right) 
= \text{Re} \left[ \int_{\bfq \bfq'} P_{\bfq \bfq'}   \fig{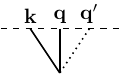}  \right] 
\; . 
\end{align} 
Finally, we invoke the ``energy conservation condition'' to determine the location of any singular points in this diagram. Since the total energy flowing into this vertex is independent of the loop momenta $q$, we conclude that the only singular points in the integrand come from $P_q$.
Since these singular points are independent of $k$, this loop integral does not introduce any kinematic branch points beyond those already present in $\psi_3$.
For instance, for a massless scalar on both Minkowski and de Sitter \eqref{eqn:G_eg_intro}, since $P_q$ is simply a power of $q$ this integral \emph{vanishes} in dim reg.

\paragraph{One-loop power spectrum.}
The next simplest example would be the one-loop correction to the power spectrum, given by expanding~\eqref{eqn:P_from_wvfn} to $\mathcal{O} \left( g_*^2 \right)$. 
Doing so produces terms built from the following 7 diagrams,
\begin{align}
 \fig{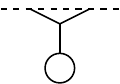}, \fig{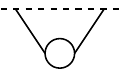} , \fig{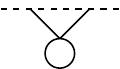} , \fig{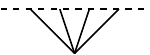}, \fig{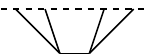} , \fig{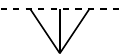} , \fig{P_f} 
 \label{eqn:P_diags}
\end{align}
These organise into the combinations~\eqref{eqn:oneBlob} and \eqref{eqn:twoBlob}.
Applying the tree theorem, we find that all of the \eqref{eqn:oneBlob} pairs of terms will vanish in dim reg for a massless scalar, and the \eqref{eqn:twoBlob} terms combine to give, 
\begin{align}
\frac{ \mathcal{P}_k^{(2)} }{ P_k P_k} \, \tilde{\delta}^3 \left( \bfk + \bfk' \right)=&\int_{\bfq_1 \bfq'_1} P_{\bfq_1 \bfq'_1}    \, \text{Re} \left[ \fig{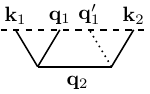}  \right] + \int_{\bfq_2 \bfq'_2} P_{\bfq_2 \bfq'_2}    \, \text{Re}  \left[ \fig{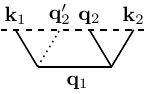}  \right]  \label{eqn:PowerspectrumCTTdimreg} \\
&+ \int_{ \substack{\bfq_1 \bfq_1' \\ \bfq_2 \bfq_2'} }  P_{\bfq_1 \bfq_1'} P_{\bfq_2 \bfq_2'}   2 \text{Re} \left[ \fig{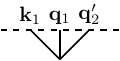}    \right] \text{Re} \left[\fig{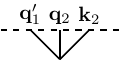}
\right]\nonumber \\
&-\int_{ \substack{\bfq_1 \bfq_1' \\ \bfq_2 \bfq_2'} }  P_{\bfq_1 \bfq_1'} P_{\bfq_2 \bfq_2'}\text{Re} \left[ \discn{\bfq_2'} \left[  \fig{Powerspectrumk1q1q2p}   \right]  \discn{\bfq_1'} \left[  \fig{Powerspectrumq1pq2pk2} \right]\right]  \nonumber 
\end{align}
The fact that the one-loop integrand from the power spectrum can be constructed from just two diagrams---the exchange four-point and the contact three-point---is a great simplification compared with the seven diagrams in~\eqref{eqn:P_diags}.
Furthermore, since we have argued that the exchange diagram can be effectively bootstrapped from the contact three-point diagram, in practice the only time integral one needs to compute is that of $\psi_3^{\rm tree}$: from that input alone, one can construct both the tree-level four-point exchange and the entire loop integrand above. 

For example, using
the results \eqref{eqn:psi3_Mink} and \eqref{eqn:exch_Mink_eg} for these tree-level diagrams on Minkowski, our simplified equation \eqref{eqn:PowerspectrumCTTdimreg} produces,
\begin{align}
\frac{ \mathcal{P}_{k}^{(2)} }{ P_{k}^2 } \tilde{\delta}^3 \left( \bfk + \bfk' \right) = \frac{ \tilde{\delta}^3 \left( \bfk + \bfk' \right) }{ \omega_k + \omega_{k'} }  \int_{\bfq} \frac{ \omega_k + \omega_{k'} + \omega_{q_1} + \omega_{q_2} }{  2 \omega_{q_1} \omega_{q_2}   ( \omega_k + \omega_{q_1} + \omega_{q_2} )( \omega_{k'} + \omega_{q_1} + \omega_{q_2} )}  \; , 
\end{align}
for a massless scalar, where $\bfq_2 - \bfq_1 = \bfk$ and $\bfq$ denotes the remaining loop momentum. 
We have checked that this agrees with performing a traditional in-in calculation (details can be found in appendix~\ref{app:symm}, cf. \eqref{eqn:ininsunet}).  
Notice that, as anticipated in section~\ref{sec:KLN}, there are no singularities in this integrand in $\omega_T = \omega_k + \omega_{k'}$ thanks to the analytic continuation of the external lines (beyond the $\omega_T$ pole already present in the tree-level wavefunction).  
However, note also that while the individual exchange terms in \eqref{eqn:PowerspectrumCTTdimreg} also introduce singularities at $\omega_{k'} + \omega_{q_2} - \omega_{q_1}$ and $\omega_k - \omega_{q_2} + \omega_{q_1}$, these singular points exactly cancel once combined with the $\disc{} [ \psi_3^{\rm tree}]$ terms---there are therefore further cancellations taking place and leading to an even simpler correlator than \eqref{eqn:PowerspectrumCTTdimreg} would suggest\footnote{
In fact, if this observation were promoted to an assumption about the loop integrands of the Bunch-Davis wavefunction, then one could proceed as in section~\ref{sec:tree} and fix the analytically continued exchange diagrams appearing in \eqref{eqn:PowerspectrumCTTdimreg} so that they cancelled the undesired singularities in the $\disc{} [ \psi_3^{\rm tree} ]$ terms. 
}. 
Finally, once momentum conservation is imposed, the two-point function is special in that $\omega_k = \omega_{k'} = \omega_T/2$ and therefore branch points in either partial energy become indistinguishable from branch points in the total energy. 
However, one virtue of writing the wavefunction/correlators in terms of general $\omega_k$ functions is that it is straightforward to consider the analogous contribution to the 3-point function, in which $\omega_k = \omega_{k_1} + \omega_{k_2}$ and $\omega_{k'} = \omega_{k_3}$ are no longer tied together by momentum conservation.
So while our KLN theorem above should be applied carefully when considering the on-shell two-point function, it applies unambiguously to all higher-point correlators. 

\paragraph{EFT of Inflation.}
As our last example, we consider the effective field theory of inflation. The leading correction to the power spectrum comes from a loop of two $\dot{\pi}^3$ interactions, and the tree-level wavefunction coefficients from this interaction on a quasi-de Sitter inflationary spacetime were given in \eqref{eqn:psi3_dS} and\eqref{eqn:dS_eg_1}.
From these, our simplified expression \eqref{eqn:PowerspectrumCTTdimreg} immediately gives the loop integrand,
\begin{align}
 \frac{ \mathcal{P}_k^{(2)} }{ P_k^2 } 
 =  \int_{\bfq} \frac{ k_1^2 k_2^2 }{
 k_T^5 } \left[  - 12 q_T  + 
    q_1 q_2 \left( \frac{ 2 k_T^2 (k_T + q_T) (k_T^2 + 2 k_T q_T + 4 q_T^2) }{e_L^3 e_R^3} +
      \frac{ 6 k_T q_T (k_T + 2 q_T) }{e_L^2 e_R^2}  + \frac{ 12 q_T  }{e_L e_R} \right) \right]  \nonumber 
\end{align}
where we have adopted the shorthands,
\begin{align}
 e_L &= k + q_1 + q_2 \; , &e_R &= k' + q_1 +q_2 \; , &k_T &= k+ k'  \; , &q_{T} &= q_1 + q_2 \; ,
\end{align}
Again we see that the only singular points depend on $k$ and $k'$ individually, and that the apparent singularities at $k' + q_2 - q_1$ and $k - q_2 + q_1$ in \eqref{eqn:PowerspectrumCTTdimreg} have cancelled out. 
Performing the remaining momentum integral (using e.g. the identities in \cite{Salcedo:2022aal}) reproduces the result of \cite{Senatore:2009cf} from a traditional in-in calculation.
In fact, we find that complete off-shell expression (treating $\omega_{k_1}, \omega_{k_2}$ and $k = |\bfk_1| = |\bfk_2|$ as independent) is given by,
\begin{align}
    \frac{ \mathcal{P}_k^{(2)} }{ P_k^2 } 
 = -\frac{\pi \text{Poly}_{1}(k, \omega_{k_{1}}, \omega_{k_{2}})}{40 ( \omega_{k_{1}} -\omega_{k_{2}} )^4 ( \omega_{k_{1}} + \omega_{k_{2}})^3}+\frac{\pi\text{Poly}_{2}(k, \omega_{k_{1}}, \omega_{k_{2}})\text{log}\left(\frac{ \omega_{k_{1}}}{\Lambda}\right)-( \omega_{k_{1}}\leftrightarrow \omega_{k_{2}})}{120 ( \omega_{k_{1}}^{2} - \omega_{k_{2}}^{2})^{5}}  
\end{align}
where:
\begin{align}
   \text{Poly}_{1}(k, \omega_{k_{1}}, \omega_{k_{2}})=&10 k^2 \omega_{k_1}^2 \omega_{k_2}^2 ( \omega_{k_1} + \omega_{k_2} )^4 - 5 \omega_{k_1}^3 \omega_{k_2}^3 (3 \omega_{k_1}^4 + 4 \omega_{k_1}^3 \omega_{k_2} +34 \omega_{k_1}^2 \omega_{k_2}^2 + 4 \omega_{k_1} \omega_{k_2}^3 + 3 \omega_{k_2}^4) \\
   &+  k^4 ( \omega_{k_1}^6 - \omega_{k_1}^5 \omega_{k_2} - 21 \omega_{k_1}^4 \omega_{k_2}^2 - 6 \omega_{k_1}^3 \omega_{k_2}^3 - 21 \omega_{k_1}^2 \omega_{k_2}^4 - \omega_{k_1} \omega_{k_2}^5 + \omega_{k_2}^6) \nonumber\\
   \text{Poly}_{2}(k, \omega_{k_{1}}, \omega_{k_{2}})=&4 \omega_{k_{1}}^2 \omega_{k_{2}}^3 (9 k^4 (5 \omega_{k_{1}}^4 + 10 \omega_{k_{1}}^2 \omega_{k_{2}}^2 + \omega_{k_{2}}^4) + 45 \omega_{k_{1}}^2 \omega_{k_{2}}^2 (5 \omega_{k_{1}}^4 + 10 \omega_{k_{1}}^2 \omega_{k_{2}}^2 + \omega_{k_{2}}^4) \\
   &- 5 k^2 (9 \omega_{k_{1}}^6 + 55 \omega_{k_{1}}^4 \omega_{k_{2}}^2 + 31 \omega_{k_{1}}^2 \omega_{k_{2}}^4 + \omega_{k_{2}}^6))\nonumber
\end{align}
Indeed, we find that the power spectrum contains only partial energy branch cuts in $\omega_{k_1}$ and $\omega_{k_2}$, unlike the wavefunction coefficients $\psi_2^{\rm 1-loop}$, which contains a dilogarithmic branch cut in $\omega_{k_1} + \omega_{k_2}$. 

~\\
Overall, the Cosmological Tree Theorem~\eqref{eqn:CTT} for wavefunction coefficients can be used to replace ``virtual'' (loop) contributions with additional ``real'' (tree) contributions when computing equal-time correlators, and this makes manifest the various cancellations which can take place in a scheme such as dim reg. 
This provides a simpler way to compute the loop corrections to observable cosmological correlators from the wavefunction of the Universe.

\section{Discussion}
\label{sec:disc}

To sum up, we have shown how non-relativistic causality---that free theory propagation can be described by a retarded Green's function---can be used to place perturbative constraints on both the cosmological wavefunction and cosmological correlators.
When combined with recent cutting rule from unitarity and an analytic structure mandated by the Bunch-Davies initial condition, this provides a new way to bootstrap tree-level exchange diagrams from their simpler contact building blocks. 
At loop-level, causality and unitarity naturally lead to a cosmological analogue of Feynman’s tree theorem which can replace any closed loop in a Feynman-Witten diagram with a sum over cut diagrams. 
This Cosmological Tree Theorem fixes the whole loop integrand of wavefunction coefficient, complementing the recent unitarity cutting rules which fix only the discontinuity, and can be applied to fields of any mass or spin with any unitarity interaction on any time-dependent spacetime background. 
We have given several explicit examples of these constraints at both tree- and loop-level for both Minkowski and de Sitter backgrounds.
Applying the Cosmological Tree Theorem to cosmological correlators leads to various cancellations between real and virtual contributions which closely parallels the KLN theorem from scattering amplitudes. 
We have therefore named this phenomenon the Cosmological KLN theorem, and have shown that it leads to a greatly simplified expression for the one-loop power spectrum in terms of just two tree-level Feynman-Witten diagrams. 
In particular, the loop integration may not introduce any additional total energy singularities, and so the loop-level correlators have the same analytic structure in $\omega_T$ as the tree-level wavefunction. 

Thanks to these results, \emph{any} Feynman diagram (with an arbitrary number of edges and loops) can now be expressed in terms of the tree-level single-vertex Feynman diagrams of the theory.
Once these have been determined (e.g. by performing a single time integral), there is no need for any further time integration. 
For Minkowski scattering amplitudes, energy conservation removes the need to do any time integration---here, we are showing that for the Bunch-Davies wavefunction on an arbitrary time-dependent background, unitarity and causality remove the need to do all but one time integration. 
It would be interesting to combine these results with other bootstrap techniques for the single-vertex diagrams, which would remove the need for any time integration whatsoever. 

There are a number of interesting directions to be explored in the future:

\begin{itemize}

\item[(i)] \emph{Perturbative vs. non-perturbative.}
Wavefunction identities can be organised according to whether they hold:
\begin{itemize}

\item[(a)] for the full $\psi_n$ coefficients of the interacting theory (i.e. sum over all Feynman diagrams with $n$ external legs),

\item[(b)] for the $\psi_n^{L\text{-loop}}$ coefficients at a fixed order in the interactions (i.e. sum over all Feynman diagrams with $n$ external legs and $L$ loops) 

\item[(c)] for individual Feynman diagrams. 

\end{itemize}
At present, very few relations of type (a) are known. 
The tree theorem presented in this work, as well as the earlier cutting rules from perturbative unitarity in \cite{Melville:2021lst}, are of type (b). 
Further cutting rules, which involve any analytic continuation of the internal line energies, are of type (c). It is always natural to ask, therefore, which identities might be promoted to a more general (non-perturbative) type. 
One way in which the Cosmological Tree Theorem might be extended to type (a) would be to consider the causality constraints on the propagator of the full interacting theory: for instance using an expansion like~\eqref{eqn:DeltaHS_def} developed in the appendix.

%
%
%

\item[(ii)] \emph{Landau analysis.}
As mentioned briefly in section~\ref{sec:KLN}, the Cosmological Tree Theorem can provide new insights into the analytic structure of wavefunction coefficients in curved spacetime. In essence, the analytic structure of loop diagrams should now be determinable by the analytic structure of the tree-level diagrams (which is comparatively much easier to determine). In particular, this will mean that the branch points of loop diagram are determined by the poles of its tree-level cut diagrams, analogous to what was recently found for Minkowski \cite{Salcedo:2022aal}. 

\item[(iii)] \emph{UV/IR sum rules.}
This tree theorem is therefore an important step towards UV/IR sum rules for cosmological spacetimes. 
At present, such relations are mostly limited to subhorizon scattering amplitudes \cite{Baumann:2015nta, Grall:2020tqc, Grall:2021xxm}, which share many properties of the Minkowski amplitude but are fundamentally disconnected from the horizon-scale physics that we ultimately observe in the CMB (although see \cite{Creminelli:2022onn} for important recent progress towards positivity bounds directly on correlation functions). 
Building on these recent subhorizon applications of unitarity and causality \cite{Melville:2019wyy, Kim:2019wjo, Ye:2019oxx, Davis:2021oce, Melville:2022ykg, Freytsis:2022aho, Aoki:2021ffc}, it would be interesting to see the wavefunction emerge as a new object which shares enough similarly with the Minkowski amplitude that it admits usable UV/IR relations and yet remains firmly connected to the cosmological correlators that we actually measure.

\item[(iv)] \emph{Stronger causality conditions.}
On Minkowksi, the free propagators are also constrained by the future light-cone, i.e. $t_1 - t_2 > | \bfx_1 - \bfx_2 |$, which is stronger than the $t_1 - t_2 > 0$ condition used in this work. 
For amplitudes, this ultimately corresponds to analyticity in the Lorentz-invariant $p^2$ rather than the energy, and this is what underpins the Kallen-Lehmann spectral representation and other relativistic dispersion relations. 
It would be interesting to explore whether such a stronger condition, and corresponding spectral representation, can exists for wavefunction coefficients on a cosmological background. 

\end{itemize}

\paragraph{Acknowledgements.}
We thank Harry Goodhew, Mang Hei Gordon Lee, Prahar Mitra,  Enrico Pajer and David Stefanyszyn for useful discussions. 
S.M. is supported by a UKRI Stephen Hawking Fellowship (EP/T017481/1). 
S.A.S. is supported by a Harding Distinguished Postgraduate Scholarship. 
This work has been partially supported by STFC consolidated grant ST/T000694/1.

\paragraph{Note added.}
In the final stages of preparing this manuscript, \cite{Albayrak:2023hie} appeared on the arXiv, which discusses cutting rules from a different (polytope) perspective. It would be interesting to investigate how the causality conditions and loop-level cutting rules presented here could be recovered from the optical polytope of \cite{Albayrak:2023hie}.

\appendix
\section{Comparison with previous cutting rules}
\label{sec:overview}

In this appendix, we give an short overview of the various cutting rules which follow from unitarity (along the line of \cite{Melville:2021lst, Goodhew:2021oqg}) and from causality (this work).

\paragraph{General procedure.}
Throughout this work, we have introduced a number of different ``cutting rules'' for the wavefunction coefficients: equations which relate diagrams with different internal and external lines.
All of these rules can be derived following the same three-step procedure, 
\begin{itemize}
	
	\item[(i)] Identify a combination of the propagators $G_p$ and $K_k$ (and their complex conjugates) which vanishes thanks to unitarity or causality properties of the free theory,   
	
	\item[(ii)] Multiply this combination by any further function of $G_p, K_k$ and vertex factors, and then integrate over all time arguments, 
	
	\item[(iii)] Relate each term in the resulting identity to a Feynman-Witten diagram, invoking the unitarity of the interacting theory (hermiticity of $\mathcal{H}_{\rm int}$) so that the discontinuity operations may be used to convert between $G_p, K_k$ and $G_p^*, K_k^*$ thanks to \eqref{eqn:hermitian_analyticity}. 
	
\end{itemize}

\paragraph{Free propagation.}
Step (i) of this procedure requires finding combinations of the propagators that vanish identically. 
To achieve this, we introduce a useful representation for $G_p (t_1, t_2)$, 
\begin{align}
	G_p (t_1, t_2) = \frac{i}{2} \left( \Delta^H_p (t_1, t_2) + i \Delta^S_p (t_1, t_2) \text{sign} ( t_1 - t_2 ) \right) 
	\label{eqn:G_def}
\end{align}
where we have split the two-point function into its symmetric/antisymmetric parts,
\begin{align}
	\Delta_k^H (t_1, t_2) &= \langle \phi = 0 | \, \{ \phi_{\bfk} (t_1) , \phi_{\bfk'} (t_2 ) \} | \Omega \rangle' 
	\; , \;\; &	\Delta_k^S (t_1, t_2) &= \langle \phi = 0 | \, [ \phi_{\bfk} (t_1) , \phi_{\bfk'} (t_2 ) ] | \Omega \rangle'   \; . 
	\label{eqn:DeltaHS_def}
\end{align}
where the prime denotes that a momentum-conserving $\tilde{\delta}$ function has been removed.
These are the wavefunction analogues of the Hadamard and Schwinger functions.
In particular, the Schwinger function $\Delta^S_k$ is closely related to the classical response of the field to applied sources (more on this below), and in fact for the free theory it is insensitive to the choice of in- and out- state thanks to the canonical commutation relations\footnote{
This $\Delta^S_k$ is therefore equal to the usual Schwinger function for the vacuum state, $\tilde{\Delta}_k^S (t_1, t_2 ) = \langle \Omega | \, [ \phi_{\bfk} (t_1) , \phi_{\bfk'} (t_2 ) ] | \Omega \rangle'$.
}. 
The Hadamard function $\Delta^H_k$, on the other hand, characterises quantum aspects of the propagation and depends on the in- and out-state\footnote{
For instance, the usual Hadamard function for the vacuum state,
\begin{align}
	\tilde{\Delta}_k^H (t_1, t_2 ) = \langle \Omega | \, \{ \phi_{\bfk} (t_1) , \phi_{\bfk'} (t_2 ) \} | \Omega \rangle' = \Delta^H_k (t_1, t_2 )  + 2 P_k K_k (t_1) K_k (t_2) \; , 
\end{align}
differs from \eqref{eqn:DeltaHS_def} by a boundary term, and would give the usual Feynman propagator if used in \eqref{eqn:G_def} in place of $\Delta^H_p$. 
}.
Unitarity and causality of the free theory then imply relations between $\Delta^H_k, \Delta^S_k$ and the bulk-to-boundary propagator $K_k$, which lead to various cutting rules.

\paragraph{Tree-level cutting rules.}
For example, the tree-level cutting rules of \cite{Melville:2021lst, Goodhew:2021oqg} follow from two particular properties of these Hadamard/Schwinger functions,
\begin{align}
	\text{Im} \, \Delta^S_k (t_1, t_2) &= 0, \tag{$U_1$} \label{eqn:U1} \\  
	\text{Re} \, \Delta^H_k (t_1, t_2) &= 4 P_k \text{Im} K_k (t_1) \text{Im} K_k (t_2)  \tag{$U_2$} \label{eqn:U2}
\end{align}
\eqref{eqn:U1} guarantees that the imaginary part of any product of $N$ bulk-to-bulk propagators will contain at most $N-1$ factors of $\text{sign} (t_1 - t_2)$, and can therefore be re-expressed in terms of products of at most $N-1$ bulk-to-bulk propagators.
\eqref{eqn:U2} then allows us to rewrite these reduced products in terms of $G_p$ and $K_k$ only.
For instance, the combination,
\begin{align}
		\mathcal{I}_p (t_1, t_2) \equiv \text{Im} \, G_p (t_1, t_2 ) - 2 P_p  \text{Im} \, K_p (t_1) \text{Im} \, K_p (t_2) 
\end{align} 
vanishes identically once \eqref{eqn:U1} and \eqref{eqn:U2} are imposed.
Integrating this over time leads to a cutting rule for any diagram containing at least one internal line: for instance, 
\begin{align}
	\int_{t_L, t_R} \, K_{k_1} (t_L) K_{k_2} (t_L) K_{k_3} (t_R) K_{k_4} (t_R) \mathcal{I}_{p_s} (t_L, t_R) = 0 
	\label{eqn:int_I1}
\end{align}
corresponds to the diagrammatic identity\footnote{
Note that the momentum integral is trivially performed using the $\delta$ function in $P_{\bfq \bfq'}$. Introducing separate $\bfq$ and $\bfq'$ labels for the cut line is simply a useful book-keeping device: particularly for later identities in which we will perform further $\Disc{}$ operations on either the left- or right-hand side of the cut.  
}, 
\begin{align}
\fig{exch_s_C}
	=
	- \int_{\bfq \bfq'} P_{\bfq \bfq'} \left(\fig{con3_L_C}\right) \left(\fig{con3_R_C}\right) \label{eqn:COTexamplepsi4}
\end{align}
As a second example, the combination
\begin{align}
	\mathcal{I}_{p_1 p_2} (t_1, t_2, t_3) &\equiv 
	\text{Im} \left( G_{p_{1}} (t_1, t_2 ) G_{p_{2}} (t_2, t_3 ) \right) +4P_{p_{1}} P_{p_{2}} \Im (K_{p_{1}} (t_{1})) \Im (K_{p_{1}} (t_{2})K_{p_{2}} (t_{2})) \Im(K_{p_{2}} (t_{3})) \nonumber\\
	& - 2P_{p_{1}}\Im(K_{p_{1}}(t_{1}))\Im(K_{p_{1}}(t_{2})G_{p_{2}}(t_{2},t_{3}))  
	- 2P_{p_{2}} \Im( G_{p_{1}}(t_{1},t_{2}) K_{p_{2}}(t_{2})) \Im(K_{p_{2}}(t_{3})) \nonumber 
\end{align} 
also vanishes thanks to \eqref{eqn:U1} and \eqref{eqn:U2}, and leads to cutting rules for any diagram containing at least two internal lines.
Similar cutting rules for an arbitrary tree-level diagram were systematically developed in~\cite{Melville:2021lst, Goodhew:2021oqg}, and also follow from \eqref{eqn:U1} and \eqref{eqn:U2}.

\paragraph{Loop-level cutting rules.}
By contrast, the loop-level cutting rules of \cite{Melville:2021lst} follow instead from exploiting the full two-point function,
\begin{align}
	\langle \phi = 0 | \, \hat{\phi}_{\bfk} (t_1) \hat{\phi}_{\bfk'} (t_2) | \Omega \rangle' = 2 P_k K_k (t_2) \text{Im} K_k (t_1)   \; \tag{$U$} \label{eqn:U}
\end{align}
as well as the identity,
\begin{align}
	\sum_{ \{ n_j \} } \prod_{j=1}^N \text{sign} (t_j - t_{j+1} )^{n_j} = 0 \qquad \quad \quad \hfill \left(\text{sum over all} \; n_j = 0 \; \text{or} \; 1 \; \text{such that} \; N - \sum_j n_j \; \text{is even} \right) \; ,  \tag{$C_1$} \label{eqn:C1} 
\end{align}
which holds whenever $t_{N+1} = t_1$. 
Note that \eqref{eqn:U} fully specifies both $\Delta^H_k$ and $\Delta^S_k$ (and hence implies both \eqref{eqn:U1} and \eqref{eqn:U2}). 
The identity \eqref{eqn:C1} is crucial because it guarantees that a product of $N$ bulk-to-bulk propagators whose arguments form a closed loop will contain at most $N-1$ factors of $\text{sign} (t_1 - t_2 )$, and hence can be re-expressed in terms of products of at most $N-1$ bulk-to-bulk propagators. 
For instance, the combination,
\begin{align}
		\mathcal{R}_{p_1 p_2} (t_1, t_2) &\equiv 
2\text{Re} \left( G_p (t_1, t_2) G_{p_2} (t_2, t_1) \right)  
+4P_{p_{1}} P_{p_{2}} \Im(K_{p_{1}}(t_{1})K_{p_{2}}(t_{1}))\Im(K_{p_{2}}(t_{2})K_{p_{1}}(t_{2}))  \nonumber \\ 
&-  2P_{p_{1}} \Im(K_{p_{1}}(t_{2})G_{p_{2}}(t_{2},t_{1})K_{p_{1}}(t_{1})) 
-  2P_{p_{2}} \Im(K_{p_{2}}(t_{1})G_{p_{1}}(t_{1},t_{2})K_{p_{2}}(t_{2}))
\end{align} 
vanishes identically once \eqref{eqn:U} and \eqref{eqn:C1} are imposed (the latter in this case is $\text{sign} (t_1 - t_2 ) \text{sign} (t_2 - t_1 ) = -1$).
Integrating this over time leads to a cutting rule for any diagram containing a loop with two internal lines: for instance, 
\begin{align}
	\int_{t_1, t_2} \, K_{k_1} (t_1) K_{k_2} (t_2)	\mathcal{R}_{p_1 p_2} (t_1, t_2) &= 0
\end{align}
corresponds to the diagrammatic identity, 
\begin{align}
	-  \fig{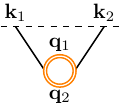}  &= \int_{\bfq_{1},\bfq'_{1}}P_{\bfq_{1}\bfq'_{1}}\left(\fig{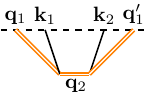}\right)  +\int_{\bfq_{2},\bfq'_{2}}P_{\bfq_{2}\bfq'_{2}}\left(\fig{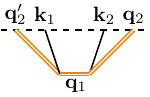}\right) \nonumber \\
	&+\int_{\substack{\bfq_{1},\bfq'_{1} \\ \bfq_{2},\bfq'_{2}}}P_{\bfq_{1}\bfq'_{1}}P_{\bfq_{2}\bfq'_{2}}\left(\fig{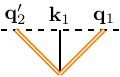}\right)\left(\fig{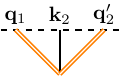}\right)  \label{eqn:CCRexample}
\end{align}
Analogous cutting rules for the $\Disc{}$ of an arbitrary loop diagram were developed in~\cite{Melville:2021lst}, and all follow from \eqref{eqn:U} and \eqref{eqn:C1}\footnote{
Stated this way, it is clear why additional cutting rules can exist for loop-diagrams but not for trees: because the identity \eqref{eqn:C1} only holds when the time arguments form a closed loop.  
}.

\paragraph{New cutting rules.}
Since both the $\text{Im}$ and $\text{Re}$ part of the bulk-to-bulk propagators appearing in a loop can be reduced to terms with fewer bulk-to-bulk propagators, the entire complex loop diagram (without any discontinuity) must be expressible in terms of cut tree-level diagrams. 
This could be achieved by straightforwardly applying the above identities: for instance $\mathcal{R}_{p_1 p_2} (t_1, t_2) + \mathcal{I}_{p_1 p_2} (t_1, t_2, t_1) = 0$ could be used to cut any $G_{p_1} (t_1, t_2) G_{p_2} (t_2, t_1)$ loop.
However, we will show that a more elegant formulation of the loop-level cutting rules follows from the identity, 
\begin{align}
	\prod_{j=1}^N \Theta ( t_j - t_{j+1} ) = 0 \; ,  	\tag{$C$} \label{eqn:C}
\end{align}
which holds whenever $t_{N+1} = t_1$. 
For instance, the combination, 
\begin{align}
		\mathcal{L}_{p_1 p_2} (t_1, t_2) &\equiv 
		G_{p_1} (t_1, t_2) G_{p_2} (t_2, t_1) 
		+4P_{p_{1}}P_{p_{2}} K_{p_{1}}(t_{1})  \Im(K_{p_{2}}(t_{1})) K_{p_{2}}(t_{2}) \Im(K_{p_{1}}(t_{2}))  \nonumber \\ 
		&- 2 P_{p_{1}}  \Im(K_{p_{1}}(t_{2})) G_{p_{2}}(t_{2},t_{1}) K_{p_{1}}(t_{1}) 
		- 2 P_{p_{2}}  \Im(K_{p_{2}}(t_{1})) G_{p_{1}}(t_{1},t_{2}) K_{p_{2}}(t_{2}) \; ,
\end{align}
vanishes identically thanks to \eqref{eqn:U} and \eqref{eqn:C}. 
This leads to a diagrammatic cutting rule, 
\begin{align}
	- \fig{loopkk_qq} &= \int_{\bfq_1 \bfq_1'} P_{\bfq_1 \bfq_1'} \left( \fig{exchkqqCk_1} \right) + \int_{\bfq_2 \bfq_2'} P_{\bfq_2 \bfq_2'} \left( \fig{exchkqCqk_2} \right)   \nonumber \\ 
	&+ \int_{\substack{\bfq_1 ,\bfq_2 \\ \bfq_1' ,\bfq_2'}} \, P_{\bfq_1 \bfq_1'} P_{\bfq_2 \bfq_2'}  \left(  \fig{contactkqqC_1}  \right)   \left(  \fig{contactkqqC_2}  \right) 
\end{align}
which goes beyond \eqref{eqn:CCRexample} because it captures both the real and imaginary part of the loop diagram.

Note that \eqref{eqn:C} has assumed a particular orientation for the loop. 
The other orientation (again with $t_{N+1} = t_1$),
\begin{align}
	\prod_{j=1}^N \Theta ( t_{j+1} - t_{j} ) = 0 \; ,  	\tag{$C'$} \label{eqn:C'}
\end{align}
gives rise to a second identity ($\mathcal{L}_{p_1 p_2} (t_2, t_1) = 0$ in the preceding example), and hence a second cutting rule, which differs from the first only in which momenta are held fixed in the $\disc{}$'s.

In fact, since $\Theta ( t_1 - t_2) = \frac{1}{2} \left( 1 + \text{sign} ( t_1 - t_2 ) \right)$, the identities \eqref{eqn:C} and \eqref{eqn:C'} are equivalent to \eqref{eqn:C1} and a further identity,
\begin{align}
		\sum_{ \{ n_j \} } \prod_{j=1}^N \text{sign} (t_j - t_{j+1} )^{n_j} = 0  \qquad \quad \quad \hfill  \left( \text{sum over all} \; n_j = 0 \; \text{or} \; 1 \; \text{such that} \; N - \sum_j n_j \; \text{is odd} \right) \; .  \tag{$C_2$} \label{eqn:C_2} 
\end{align}
This leads to one further set of cutting rules, which are quite distinct from all previous examples in that they combine the same diagram with different collinear kinematics in order to achieve a cut. 
For instance, the combination,
\begin{align}
	&\mathcal{C}_{p_1 p_2} (t_1, t_2)  \\ 
	&\equiv  \Re G_{p_{1}} (t_1, t_2) \text{Im} \left( K_{p_2} (t_2) K_{p_2}^* (t_1)  \right) - 2 \Im K_{p_1} (t_1)  \Im  K_{p_2} (t_2)  \Re  K_{p_1} (t_2) \Re K_{p_2} (t_1)  - \left( p_1 \leftrightarrow p_2 \right)  . \nonumber 
\end{align}
vanishes once \eqref{eqn:U} and \eqref{eqn:C_2} are imposed (the latter in this case is $\text{sign} (t_1 - t_2) + \text{sign} (t_2 - t_1 ) = 0$). When integrated against two bulk-to-boundary propagators as in \eqref{eqn:int_I1}, this leads to the diagrammatic relation shown in \eqref{eqn:Ctype_2}.
So while unitarity can fix only the $\Disc{}$ of tree-level diagrams, causality imposes additional constraints on their $\Disc{}$-less part.

\paragraph{Unitarity and Causality.}
Property \eqref{eqn:U} (and hence also \eqref{eqn:U1} and \eqref{eqn:U2}) can be viewed as \emph{unitarity} of the free theory, since Hermiticity of the free Hamiltonian allows us to resolve the identity using a complete basis of $n$-particle states, 
\begin{align}
\sum_n \langle \phi = 0 | \hat{\phi}_{\bfk} (t) | n \rangle \langle n |  \hat{\phi}_{\bfk'} (t') | \Omega \rangle = \int_{\bfp} \langle \phi = 0 | \hat{\phi}_{\bfk} (t) \hat{a}^\dagger_{\bfp} | \Omega \rangle \langle \Omega | \hat{a}_{\bfp}  \hat{\phi}_{\bfk'} (t') | \Omega \rangle \; ,
\end{align}
since in the free theory $\hat{\phi}_\bfk | \Omega \rangle$ overlaps only with the 1-particle states.
The matrix elements on the right-hand-side evaluate to $\sqrt{P_k} \Im K_k (t)$ and $\sqrt{P_{k'} } K_{k'} (t')$ (up to an overall unimportant phase), hence producing~\eqref{eqn:U}\footnote{
There is another sense in which \eqref{eqn:U1} is related to unitarity: it follows from the canonical commutation relation for $[ \phi_{\bfk} , \hat{\Pi}_{\bfk} ]$, which (by the Stone-von Neumann theorem) is what guarantees that the Hilbert spaces at different times are unitarily related, i.e. that time evolution is implemented by a unitary operator.
}.  

The identities \eqref{eqn:C} and \eqref{eqn:C'}, on the other hand, are related to \emph{causality}. Causality in the classical sense: signals may not precede their sources. This amounts to the existence of a retarded Green's function,
\begin{align}
	G^R_p (t_1, t_2) = G_p (t_1, t_2) - \langle \phi = 0 | \hat{\phi}_{\bfp} (t_1) \hat{\phi}_{\bfp_2} (t_2) | \Omega \rangle' =  \Delta^S_p (t_1, t_2) \Theta ( t_1 - t_2 ) \; .
\end{align}
which would describe the classical response of the field to an applied source, and for which identity~\eqref{eqn:C} implies,
\begin{align}
	\prod_{j=1}^N G^R_p (t_j, t_{j+1} ) = 0 \; ,
\end{align}
whenever $t_{N+1} = t_1$ and the arguments form a closed loop. 
Put another way: a closed loop is forbidden by classical causality, since it requires the signal to propagate backwards in time. This is ultimately the reason why loop-level Feynman diagrams are so tightly constrained, and can be expressed in terms of tree-level diagrams.  
The identity~\eqref{eqn:C'} similarly implies that the product of advanced propagators is zero for a closed loop. Since the identities \eqref{eqn:C1} and \eqref{eqn:C_2} are equivalent to \eqref{eqn:C} and \eqref{eqn:C'}, they too can be viewed as causality conditions.

Finally, beyond unitarity of the free theory, all of the above cutting rules make use of unitarity of the interactions. 
This is ultimately the condition that all vertex factors commute with the discontinuity operations, so that we can extract the $\text{Im}$ part of propagators inside the diagram without changing the structure of any interaction vertex.

\paragraph{Cosmological Tree Theorem.}
Altogether, the different propagator identities, their corresponding cutting rules and their underlying assumptions are
listed in Table~\ref{table:summary}.
The cutting rules stemming from \eqref{eqn:C}/\eqref{eqn:C'} will turn out to be particularly useful, and so we have given them a name: the \emph{Cosmological Tree Theorem}.
In particular, when applied to equal-time correlation functions, these ``tree theorem'' identities immediately explain the cancellation of certain singularities. 
This is reminiscent of the cancellation of IR divergences when passing from the amplitude to the cross-section (the KLN theorem).
Furthermore, singling out the Cosmological Tree Theorem seems natural because it is a complex relation (whereas the other cutting rules reviewed here are purely real), and so it is a more efficient way of encoding the cutting rules.
Of course, since all of these rules follow in some way from the same \eqref{eqn:U} identity, they are ultimately related to one another. 
For instance, in the examples given above, the two complex relations $\mathcal{L}_{p_1 p_2} (t_1, t_2) = 0$ and $\mathcal{L}_{p_1 p_2} (t_2, t_1) = 0$ are actually related to the four real relations $\{ \mathcal{R}_{p_1 p_2} (t_1, t_2), \mathcal{I}_{p_1 p_2} (t_1,t_2,t_1) , \mathcal{C}_{p_1 p_2} (t_1, t_2) , \mathcal{I}_p (t)  \}$, since,
\begin{align}
	\text{Re} \left[  \mathcal{L}_{p_1 p_2} (t_1, t_2) +        \mathcal{L}_{p_1 p_2} (t_2, t_1) \right] =& \mathcal{R}_{p_{1}p_{2}}(t_{1},t_{2})
	+ 2 \Re \left( K_{p_1} (t_1) K_{p_1}^* (t_2) \right) \mathcal{I}_{p_{2}}(t_{2},t_{1}) 
	+ 2 \Re \left( K_{p_2} (t_2) K_{p_2}^* (t_1) \right) \mathcal{I}_{p_{1}}(t_{1},t_{2}) \nonumber \\
	\text{Im} \left[  \mathcal{L}_{p_1 p_2} (t_1, t_2) + \mathcal{L}_{p_1 p_2} (t_2, t_1) \right] =&\mathcal{I}_{p_{1}p_{2}}(t_{1},t_{2},t_{1}) + \Re G^{R}_{p_{1}}(t_{1},t_{2}) \mathcal{I}_{p_{2}}(t_{2},t_{1}) + \Re G^{R}_{p_{2}}(t_{1},t_{2}) \mathcal{I}_{p_{1}}(t_{1},t_{2}) \nonumber\\ 
	\text{Re} \left[  \mathcal{L}_{p_1 p_2} (t_1, t_2) - \mathcal{L}_{p_1 p_2} (t_2, t_1) \right] =&  2 \mathcal{C}_{p_1 p_2} (t_1, t_2)   \label{eqn:LtoIRC}  \\
	\text{Im} \left[  \mathcal{L}_{p_1 p_2} (t_1, t_2) - \mathcal{L}_{p_1 p_2} (t_2, t_1) \right] =&  2 \Im \left( K_{p_1} (t_1) K_{p_1}^* (t_2) \right) \mathcal{I}_{p_{2}}(t_{2},t_{1}) 
	+ 2 \Im \left( K_{p_2} (t_2) K_{p_2}^* (t_1) \right) \mathcal{I}_{p_{1}}(t_{1},t_{2})  \nonumber 
\end{align}
irrespective of any connection\footnote{
in this equation we have used the shorthand, $G^R_p (t_1, t_2) = G_p (t_1, t_2) -2 K_{p} (t_1) \Im  K_{p_1} (t_2)$.
} between $G_p$ and $K_k$.

\begin{table}
	\renewcommand{\arraystretch}{1.5}
	\centering
	\begin{tabular}{| c | c | c | c | c | } 
		\hline
		Identity & follows from & used to fix & leads to cutting rules for \\ [0.5ex] 
		\hline
		$\mathcal{I} = 0$ & \eqref{eqn:U1} + \eqref{eqn:U2}  & $\text{Im} \left( G_1 ... G_N \right)$& $\Disc{}$ of any tree diagram, see \cite{Melville:2021lst, Goodhew:2021oqg}  \\ \hline
		
		$\mathcal{R} = 0$ & \eqref{eqn:U} + \eqref{eqn:C1} & $\text{Re} \left(  G_1 ... G_N \right)_{\rm loop}$ &  $\Disc{}$ of any loop diagram, see \cite{Melville:2021lst}  \\ \hline
		
		$\mathcal{L} = 0$ & \eqref{eqn:U} + \eqref{eqn:C}  & $( G_1 ... G_N )_{\rm loop}$ & any loop diagram, see \eqref{eqn:CTT}  \\ \hline
		
		$\mathcal{L}' = 0$ & \eqref{eqn:U} + \eqref{eqn:C'} & $( G_1 ... G_N )_{\rm loop}$ & any loop diagram, see \eqref{eqn:CTT_2}  \\ \hline		
		
		$\mathcal{C} = 0$ & \eqref{eqn:U} + \eqref{eqn:C_2} & $ \sum_{\rm perm.}^N \Delta_1^S G_2 ... G_N$ & $\Disc{}$ of collinear tree diagrams, see \eqref{eqn:CTT_3}  \\ \hline
	\end{tabular}
	\caption{Summary of the various cutting rules considered in this section, and the different unitarity (\eqref{eqn:U} $\Rightarrow$ (\ref{eqn:U1}, \ref{eqn:U2})) and causality ((\ref{eqn:C1}, \ref{eqn:C_2}) $\Leftrightarrow$ (\ref{eqn:C}, \ref{eqn:C'})) properties from which they folow. 
    }
	\label{table:summary}
\end{table}

\section{Non-Gaussianity at next-to-next-to-leading order}
\label{app:bispectrum}

In this appendix we briefly describe primordial non-Gaussianity at next-to-next-to-leading order (NNLO). 
By ``order'', we are referring to the power of the small coupling $g_*$ which suppresses field insertions in the Lagrangian. 
For instance, the ``leading-order'' non-Gaussianity is the tree-level bispectrum, which is $\mathcal{O} ( g_* )$. 
In the main text we discuss the next-to-leading order (NLO) effects at $\mathcal{O} ( g_*^2)$, which includes the one-loop correction to the power spectrum. 
Here, we discuss NNLO effects at $\mathcal{O} ( g_*^3 )$, namely the one-loop bispectrum and the two-loop vev.

\subsection{Bispectrum at one loop}

Beyond the power spectrum \eqref{eqn:P_def}, we can analogously extract higher-point correlation functions from the wavefunction. 
The first of these is the bispectrum\footnote{
Strictly speaking, the full $\langle \hat{\phi}_{\bfk_1} \hat{\phi}_{\bfk_2} \hat{\phi}_{\bfk_3} \rangle$ also contains disconnected contributions which $\sim \tilde{\delta} ( \bfk_1 + \bfk_2 ) \tilde{\delta}^3 ( \bfk_3 )$ and its permutations. One should therefore interpret the bispectrum \eqref{eqn:B_def} as the \emph{connected} part of the three-point function.
},
\begin{align}
\frac{  \langle \hat{\phi}_{\bfk_1} \hat{\phi}_{\bfk_2} \hat{\phi}_{\bfk_3} \rangle }{ \mathcal{P}_{k_1} \mathcal{P}_{k_2} \mathcal{P}_{k_3}}  
= \mathcal{B}_{k_1 k_2 k_3} \tilde{\delta}^3 \left( \bfk_1 + \bfk_2 + \bfk_3 \right) \; . 
\label{eqn:B_def}
\end{align}
Note that we normalise the bispectrum using the power spectrum $\mathcal{P}_q$ of the full interacting theory.
This normalisation leads to a convenient cancellation of all diagrams in which the propagation of a single external leg to the boundary is modified, i.e. it removes all terms of the form,
\begin{align}
\int_{\bfq \bfq'} P_{\bfq \bfq'} \, \fig{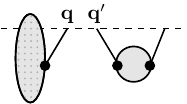} 
\label{eqn:trimming}
\end{align}
where the left blob can have arbitrarily many external legs but the right blob may only have the two external legs shown. 

The bispectrum is related to the non-Gaussian wavefunction coefficients by the Born rule~\eqref{eqn:Born}. Expanding this perturbatively in $g_*$ gives, 
\begin{align}
\mathcal{B}_{k_1 k_2 k_3} \tilde{\delta}^3 \left( \sum_{a=1}^3 \bfk_a  \right)  &= 2 \text{Re} \, \psi_{\bfk_1 \bfk_2 \bfk_3} 
+ \int_{\bfq \bfq'} P_{\bfq \bfq'} \, \left(
 \text{Re} \, \psi_{\bfk_1 \bfk_2 \bfk_3 \bfq \bfq'}
+
 4 \,  \text{Re} \, \psi_{\bfq'} \text{Re} \, \psi_{\bfk_1 \bfk_2 \bfk_3 \bfq}  \right)
  \nonumber \\
&+ 2 \int_{ \substack{ \bfq_1 \bfq_1' \\ \bfq_2 \bfq_2' } } P_{\bfq_1 \bfq_1'} P_{\bfq_2 \bfq_2'}   \left( 
\text{Re}\, \psi_{\bfq_1' \bfq_2 \bfq_2'} \text{Re} \, \psi_{\bfk_1 \bfk_2 \bfk_3 \bfq_1}   
+ \sum_{\rm perm.}^3   \text{Re} \, \psi_{\bfk_1 \bfq_1 \bfq_2'} \text{Re} \, \psi_{\bfk_2 \bfk_3 \bfq_2 \bfq_1'} \right) 
 \nonumber \\
&\;\;+ 4  \int_{ \substack{ \bfq_1 \bfq_1' \\ \bfq_2 \bfq_2' \\ \bfq_3 \bfq_3'} } P_{\bfq_1 \bfq_1'} P_{\bfq_2 \bfq_2'} P_{\bfq_3 \bfq_3'} \,  \text{Re} \, \psi_{\bfk_1 \bfq_1 \bfq_2'} \text{Re} \, \psi_{\bfk_2 \bfq_2 \bfq_3'} \text{Re} \, \psi_{\bfk_3 \bfq_3 \bfq_1'}  + \mathcal{O} \left( g_*^4 \right)  \; . 
\label{eqn:B_from_psi}
\end{align}
Each wavefunction coefficient can then be expanded in terms of Feynman-Witten diagrams, and as a result one arrives at the expansion,
\begin{align}
 \mathcal{B}_{k_1 k_2 k_3} =  \mathcal{B}^{(1)}_{k_1 k_2 k_3} + \mathcal{B}^{(3)}_{k_1k_2k_3} + \mathcal{O} \left( g_*^5 \right)
\end{align}
where $\mathcal{B}^{(1)} = 2 \text{Re} \, \psi_3^{\rm tree}$ is the tree-level part, and $\mathcal{B}^{(3)}$ is determined by $\psi_3^{\rm 1-loop}, \psi_5^{\rm tree}, \psi_4^{\rm tree}$ and $\psi_3^{\rm tree}$. 
The tree-level diagrams for $\psi_3$ and $\psi_4$ are given in \eqref{eqn:psi_intro_diagrams}, while for $\psi_5$ they are,
\begin{equation}
\psi_{\bfk_{1}\bfk_{2}\bfk_{3}\bfk_{4}\bfk_{5}}^{\rm tree} = \sum_{\rm perm.}^{15}  \fig{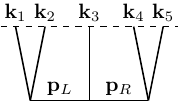} 
+
\sum_{\rm perm.}^{10} \fig{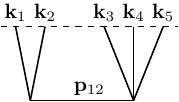} 
+\fig{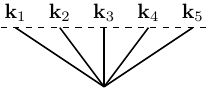}
\label{eqn:psi5_diagrams}
\end{equation}
and for the 1-loop cubic coefficient they are,
\begin{align}
\psi_{\bfk_1 \bfk_2 \bfk_3}^{\rm 1-loop} =& \frac{1}{2}\;\fig{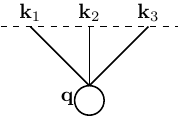}\;  \label{eqn:psi3_diagrams} \\
 &+\frac{1}{2}\;\fig{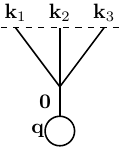}\;+\frac{1}{2} \sum_{\rm perm.}^3 \fig{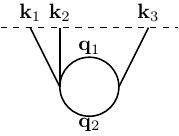} +\frac{1}{2}\sum_{\rm perm.}^3 \fig{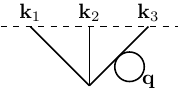} \nonumber\\
 &+\fig{TriangleGraph}+\frac{1}{2} \sum_{\rm perm.}^3 \fig{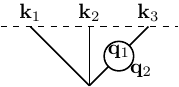}  +\frac{1}{2}\sum_{\rm perm.}^3 \fig{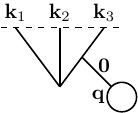} \; . \nonumber
\end{align}
Notice that the one- and two-vertex loops come with symmetry factors of $1/2$ (more on this in appendix~\ref{app:symm}), while the three-vertex loop has a symmetry factor of $1$.

Once the $\mathcal{B}^{(3)}$ correction is written in terms of Feynman-Witten diagrams, we find that they collect precisely into the groups identified in \eqref{eqn:oneBlob}, \eqref{eqn:twoBlob} and \eqref{eqn:threeBlob}. 
To be precise, the diagrams can be organised according to which vertices they contain and whether the external lines end on the same or different vertices.
This leads to,
\begin{align}
\mathcal{B}_{k_1 k_2 k_3}^{(3)} \, \tilde{\delta}^3 \left( \bfk_1 + \bfk_2 + \bfk_3 \right) 
= \mathcal{I}_{123} + \mathcal{I}_{12|3} +\mathcal{I}_{13|2} + \mathcal{I}_{23|1}  + \mathcal{I}_{1|2|3}
\label{eqn:B_to_I}
\end{align}
where the diagrams in $\mathcal{I}_{123}$ are all of the form~\eqref{eqn:oneBlob}, those in $\mathcal{I}_{12|3}$ and its permutations are of the form~\eqref{eqn:twoBlob}, and those in $\mathcal{I}_{1|2|3}$ are of the form~\eqref{eqn:threeBlob}. 
Using our tree theorem to replace all loop diagrams with tree-level diagrams then leads to the precise KLN cancellations between virtual and real emission described in section~\ref{sec:KLN}, and as a result there are no new total energy branch points introduced by the integration over loop momenta. 

In spite of these cancellations, there are still many tree-level diagrams required to compute the bispectrum for general massive fields. 
However, if we focus on massless internal lines, then many of these diagrams become scale-free integrals and hence vanish in dimensional regularisation. 
In that case, the only terms in \eqref{eqn:B_to_I} which give a non-zero contribution are,
\begin{align}
\mathcal{I}_{12|3}  &= \fig{Psi3sunset1PI}+\sum_{\rm perm.}^2 \fig{K3q1-q1k1k2}
+ 2 \; \; \fig{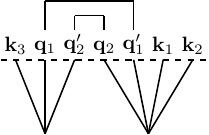}  \nonumber\\ 
&+ \fig{psi3sunsetred}+\sum_{\rm perm.}^2 \fig{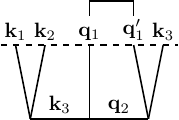} +2 \;\; \fig{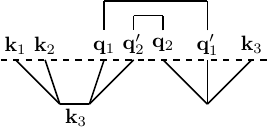} \nonumber \\ 
\mathcal{I}_{1|2|3} &= 2 \fig{TriangleGraph} + 2 \sum_{\rm perm.}^3 \fig{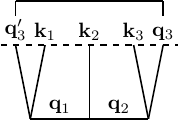}  \label{eqn:B3psi3triangle}\\
     &+4 \sum_{\rm perm.}^3  \fig{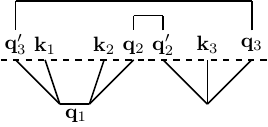}
     +4 \;\; \fig{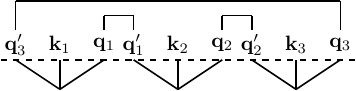}
     \nonumber
\end{align}
where we have omitted the $\text{Re}$ which should be taken of every connected component, and replaced each $\int_{\bfq \bfq'} P_{\bfq \bfq'}$ factor by a ``contraction'' of the corresponding $\bfq \bfq'$ pair in the diagram. 

Applying the tree theorem, we find that the one-loop correction to the bispectrum from massless fields can be written in terms of just 5 tree-level diagrams,
\begin{align}
 \fig{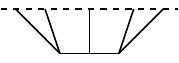} \; , \;\;  \fig{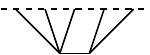} \; , \;\; \fig{P_g} \; ,  \;\; \fig{P_d} \; , \;\; \fig{P_e} 
 \label{eqn:B_diags}
\end{align}
which is just 2 more than the power spectrum. 
By contrast, the original \eqref{eqn:B_from_psi} requires 13 separate Feynman-Witten diagrams (6 more than the power spectrum). 
In fact, applying a bootstrap argument to fix the exchange diagrams in terms of contact diagrams, we can actually construct the integrand for $\mathcal{P}^{(2)}$ entirely from the single $\psi_3^{\rm tree}$ contact diagram and $\mathcal{B}^{(3)}$ from the two contact diagrams shown in \eqref{eqn:B_diags}.

The explicit relation between \eqref{eqn:B_diags} and the one-loop bispectrum \eqref{eqn:B_to_I} is,
\begin{align}
\mathcal{I}_{12|3}  &= \int_{\bfq\bfq'}P_{\bfq\bfq'}\left( \Re\left[ \fig{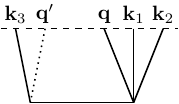}\right]+\Re\left[ \fig{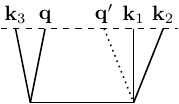}\right] \right)   \label{eqn:B_CTT_1} \\
    &+2 \int_{\substack{\bfq_{1}\bfq'_{1}\\ \bfq_{2}\bfq'_{2}}}P_{\bfq_{1}\bfq'_{1}}P_{\bfq_{2}\bfq'_{2}} \,  \text{Re} \left[  \fig{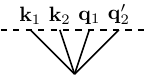}\right] \text{Re} \left[ \fig{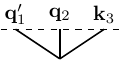}\right]   \nonumber \\
&    - \int_{\substack{\bfq_{1}\bfq'_{1}\\ \bfq_{2}\bfq'_{2}}} P_{\bfq_{1}\bfq'_{1}} P_{\bfq_{2}\bfq'_{2}}\; \text{Re} \left[ \disc{q'_{1}}\left[ \fig{3qqk}\right] \disc{q'_{2}}\left[\fig{4kkqq}\right]\right] 
      \nonumber\\ 
&+ \int_{\bfq\bfq'}P_{\bfq\bfq'} \left( \Re\left[ \fig{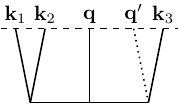}\right] +\Re\left[ \fig{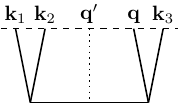}\right]  \right) \nonumber  \\
    &+2 \int_{\substack{\bfq_{1}\bfq'_{1}\\ \bfq_{2}\bfq'_{2}}}P_{\bfq_{1}\bfq'_{1}}P_{\bfq_{2}\bfq'_{2}} \Re\left[ \fig{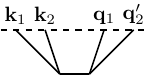}\right] \Re\left[ \fig{3qqk}\right] \nonumber\\
    &-\int_{\substack{\bfq_{1}\bfq'_{1}\\ \bfq_{2}\bfq'_{2}}}P_{\bfq_{1}\bfq'_{1}}P_{\bfq_{2}\bfq'_{2}} \Re\left[ \disc{q'_{1}}\left[ \fig{3qqk}\right]\disc{q'_{2}}\left[ \fig{4kk_qq}\right] \right] \nonumber 
  \end{align}
and its permutations, together with,
\begin{align}
\mathcal{I}_{1|2|3} &= 
   \sum_{\rm perm.}^3  \int_{\bfq\bfq'}P_{\bfq\bfq'}  
    \Re\left[ \fig{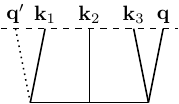}\right]  \label{eqn:B_CTT_2} \\
    &+ \sum_{\rm perm.}^3  2 \int_{\substack{\bfq_{2}\bfq'_{2}\\ \bfq_{3}\bfq'_{3}}}P_{\bfq_{2}\bfq'_{2}}P_{\bfq_{3}\bfq'_{3}} \Re\left[ \fig{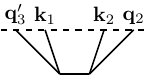}\right] \Re\left[ \fig{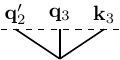}\right]  \nonumber \\ 
    &- \sum_{\rm perm.}^3 \int_{\substack{\bfq_{2}\bfq'_{2}\\ \bfq_{3}\bfq'_{3}}}P_{\bfq_{2}\bfq'_{2}}P_{\bfq_{3}\bfq'_{3}} \Re\left[ \disc{q'_{2}}\left[\fig{3qqk1}\right] \disc{q'_{3}}\left[\fig{4qk_kq}\right]\right]  \nonumber \\
    &+4 \int_{\substack{\bfq_{1}\bfq'_{1}\\\bfq_{2}\bfq'_{2}\\\bfq_{3}\bfq'_{3}}}P_{\bfq_{1}\bfq'_{1}}P_{\bfq_{2}\bfq'_{2}}P_{\bfq_{3}\bfq'_{3}} \, \Re\left[\fig{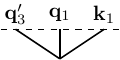}\right] \Re\left[\fig{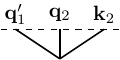}\right]\Re\left[ \fig{3qqk1}\right] \nonumber\\
    &-\int_{\substack{\bfq_{1}\bfq'_{1}\\\bfq_{2}\bfq'_{2}\\\bfq_{3}\bfq'_{3}}}P_{\bfq_{1}\bfq'_{1}}P_{\bfq_{2}\bfq'_{2}}P_{\bfq_{3}\bfq'_{3}}\Re\left[\disc{q'_{3}}\left[ \fig{3qqk3}\right]\disc{q'_{1}}\left[ \fig{3qqk2}\right]\disc{q'_{2}}\left[ \fig{3qqk1}\right] \right] \, . \nonumber
\end{align}

\subsection{Field vev at two loops}

To compute the vev~\eqref{eqn:v_def} at next-to-next-to-leading order, we first expand the Born rule up to $\mathcal{O} (g_*^3)$, 
\begin{align}
 v \, \tilde{\delta}^3 \left( \bfk \right) &= 2 \text{Re} \, \psi_{\bfk} +  \int_{\bfq \bfq'} \mathcal{P}_{\bfq \bfq'} \,  \text{Re} \, \psi_{\bfk \bfq \bfq'}
+  \int_{ \substack{ \bfq_1 \bfq_1' \\ \bfq_2 \bfq_2'} } P_{\bfq_1 \bfq_1'} P_{\bfq_2 \bfq_2'}  \, \tfrac{1}{4} \text{Re} \, \psi_{\bfk \bfq_1 \bfq_1' \bfq_2 \bfq_2'} \nonumber \\
&+  \int_{ \substack{ \bfq_1 \bfq_1' \\ \bfq_2 \bfq_2' \\ \bfq_3 \bfq_3'} } P_{\bfq_1 \bfq_1'} P_{\bfq_2 \bfq_2'}  P_{\bfq_3 \bfq_3'} \,   
\tfrac{2}{3} \text{Re} \, \psi_{\bfk \bfq_1 \bfq_2 \bfq_3} \text{Re} \, \psi_{\bfq_1' \bfq_2'  \bfq_3'} 
   + \mathcal{O} \left( g_*^4 \right)
\end{align}
where note that the $\mathcal{P}_{\bfq \bfq'}$ appearing in the next-to-leading order term is the full power spectrum and should be expanded as in \eqref{eqn:P_from_wvfn}.
Next, we expand each wavefunction coefficient as a series in the number of loops. 
This gives an expression for $v^{(3)}$ which depends on $\psi_1^{\rm 2-loop}$, $\psi_2^{\rm 1-loop}$, $\psi_3^{\rm 1-loop}$, $\psi_5^{\rm tree}$, $\psi_4^{\rm tree}$ and $\psi_3^{\rm tree}$.
The corresponding tree-level diagrams are given in \eqref{eqn:psi_intro_diagrams} and \eqref{eqn:psi5_diagrams}, while the 1-loop diagrams are given in \eqref{eqn:psi3_diagrams}.
The necessary two-loop diagrams are,
\begin{align}
 \psi_{\bfk}^{\text{2-loop}} = & \frac{1}{8} \; \fig{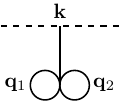} \;+ \frac{1}{4}  \;\fig{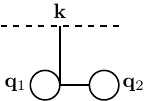}\;+\frac{1}{4}\;\fig{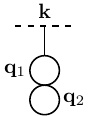}\;+\frac{1}{6}\;\fig{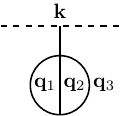}\; \\
     &+\frac{1}{4}\;\fig{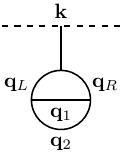}\;+\frac{1}{8}\;\fig{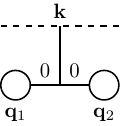}\;+\frac{1}{4}\;\fig{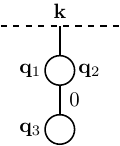}\; . \nonumber
\end{align}
We find that the terms in $v^{(3)}$ organise into the combinations~\eqref{eqn:oneBlob}, \eqref{eqn:twoBlob} and \eqref{eqn:threeBlob} for the one-loop diagrams, as well as \eqref{eqn:twoBlobsTwoLoops} for the two-loop diagrams.
Expanding each loop using the Cosmological Tree Theorem therefore leads to very many cancellations, and in particular the only connected terms which remain are found to have pairs of energies which are analytically continued so that there is no set of vertices whose total energy depends on both the total external energy and the loop momenta.
As a result, we find that while $v^{(3)}$ could be non-zero even for a massless field, it does not contain any additional branch cuts in $k$. 
Of course, once momentum conservation is imposed, in this example $k$ is fixed to be zero regardless. 
But when computing higher-point correlators, e.g. the $\mathcal{P}^{(4)}_{k}$ correction to the power spectrum, there will be a series of diagrams which resemble those above but with the $\bfk$ leg split into a pair of legs carrying momentum $\bfk$ and $\bfk'$. Since the KLN cancellation takes place at finite $k$, it means that this subset of diagrams in such higher-point correlators (for which the energy $k+k'$ is no longer set to zero) also do not introduce any additional branch points. 


\section{Wavefunction symmetry factors}
\label{app:symm}

In the literature, often single diagrams are considered so the overall normalisation is arbitrary. 
Here, when we compute in-in correlators in Section \ref{sec:corr}, the relative coefficient between loop and tree diagrams matters (in particular, the cancellation of singularities in subsection~\ref{sec:KLN} depends crucially on getting the relative factors right).

In this short appendix, we give three separate proofs that the correct symmetry factor for the sunset diagram in \eqref{eqn:psi_intro_diagrams} is $1/2$:  computing the Wick contractions directly, solving the Schrödinger equation and comparing the Born rule with an explicit in-in calculation. 
In each case, it is clear how the counting of symmetry factors should work for more general diagrams.

\paragraph{From Wick contractions.}
The way symmetry factors are often introduced in the context of amplitudes is through the Dyson series expansion of $\hat{U}$. 
The sunset diagram appears naturally in the self energy for a scalar field at one loop with $\frac{g}{3!}\phi^{3}$ interaction:
\begin{equation}
    \frac{1}{2}\frac{g^{2}}{3!^{2}}\langle\hat{\phi}(x)\hat{\phi}(y)\phi^{3}(z_{1})\phi^{3}(z_{2})\rangle=\frac{g^{2}}{2}\int_{z_{1}z_{2}}\Delta(x-z_{1})\Delta^{2}(z_{1}-z_{2})\Delta(z_{2}-y),
\end{equation}
where $\Delta(x_{1}-x_{2})$ is the Feynman propagator. The symmetry factor of $1/2$ arises from there being three different ways to contract the internal line with each vertex. There are then two ways to contract the remaining internal lines. There is an overall factor of 2 to account for the vertex permutation:
\begin{equation}
    \frac{3^2\times 2\times 2}{ 3!^{2}2}=\frac{1}{2}
\end{equation}
In the case of the wavefunction, the analogous series expansion of $\hat{U}$ inside \eqref{eqn:Psi_matrix} leads to the same set of Wick contractions. 
Consequently, the symmetry factors appearing in the wavefunctions coefficients are identical to those appearing in the usual amplitude calculation.

\paragraph{From Schrodinger equation.}
We now turn to solving the Schrödinguer equation,
\begin{equation}  
i\partial_{t}\Psi[\phi]=\hat{\mathcal{H}}\Psi[\phi],
\end{equation}
where $t$ is the time to which we have evolved the Bunch-Davies state. 
For the free Hamiltonian~\eqref{eqn:Lfree}\footnote{
Including the interaction Hamiltonian $\mathcal{H}_{\rm int}$ will only source the contact Feynman-Witten diagrams, but does not affect the exchange or loop-type diagrams for which we wish to count the symmetry factor.
}, the Schrodinger equation in momentum space becomes \cite{Cespedes:2020xqq},
\begin{equation}\label{eqn:Schq}
    -\partial_{t}\Gamma[\phi]=\frac{1}{2 a^{d-1}(t)}\int_{\textbf{p}_{1}\textbf{p}_{2}}\delta(\textbf{p}_{1}+\textbf{p}_{2})\left(\frac{\delta\Gamma[\phi]}{\delta\phi_{\textbf{p}_{1}}}\frac{\delta\Gamma[\phi]}{\delta\phi_{\textbf{p}_{2}}}-i\frac{\delta^{2}\Gamma[\phi]}{\delta\phi_{\textbf{p}_{1}}\delta\phi_{\textbf{p}_{2}}}\right)
\end{equation}
In order to discuss the symmetry factor of the sunset diagram we expand the phase (the on-shell action) to quadratic order in the coupling $g$:
\begin{align}
    \Gamma[\phi]=&\frac{1}{2}\int_{\textbf{k}_{1}\textbf{k}_{2}}\psi_{2}(\textbf{k}_{1},\textbf{k}_{2})\phi_{\textbf{k}_{1}}\phi_{\textbf{k}_{2}}\tilde{\delta}(\textbf{k}_{T})+\frac{1}{6}\int_{\textbf{k}_{1}\textbf{k}_{2}\textbf{k}_{2}}\psi_{3}(\textbf{k}_{1},\textbf{k}_{2},\textbf{k}_{3})\phi_{\textbf{k}_{1}}\phi_{\textbf{k}_{2}}\phi_{\textbf{k}_{3}}\tilde{\delta}(\textbf{k}_{T})\nonumber \\
    &+\frac{1}{24}\int_{\textbf{k}_{1}\textbf{k}_{2}\textbf{k}_{2}\textbf{k}_{4}}\psi_{4}(\textbf{k}_{1},\textbf{k}_{2},\textbf{k}_{3},\textbf{k}_{4})\phi_{\textbf{k}_{1}}\phi_{\textbf{k}_{2}}\phi_{\textbf{k}_{3}}\phi_{\textbf{k}_{4}}\tilde{\delta}(\textbf{k}_{T})
\end{align}
Doing an expansion $\psi_{2}(\textbf{k}_{1},\textbf{k}_{2})=\psi^{\text{free}}_{2}(\textbf{k}_{1},\textbf{k}_{2})+\psi_{2}^{1-\text{loop}}(\textbf{k}_{1},\textbf{k}_{2})+...$  leaves an Schrödinger equation of the form\footnote{
This agrees with \cite{Melville:2021lst} once we account for the factor of $1/2$ which is missing from the right-hand-side of their equation (A.15).
}:
\begin{equation}
   \frac{\partial_{t}(i\psi^{\rm 1-\text{loop}}_{2}(\bfk_{1},\bfk_{2})f^{*}_{k_{1}}(t)f^{*}_{k_{2}}(t))}{f^{*}_{k_{1}}(t)f^{*}_{k_{2}} (t)}=-\frac{1}{2a^{d-1}(t)}\int_{\bfq}\psi_{4}(\bfk_{1},\bfk_{2},\bfq,-\bfq).
\end{equation}
where we have also used $\psi'^{\rm free}_{2}(\textbf{k}_{1},\textbf{k}_{2})=a^{d-1}\partial_{t}\text{log} (f_{k_{1}}(t))$. 
 This equation is solved by:
\begin{align}
    \psi_{2}^{1-\text{loop}}(\bfk_{1},\bfk_{2})&=\frac{g^{2}}{2}\int_{\bfq_{1}\bfq_{2}}\int_{t_{1}t_{2}}\frac{\tilde{\delta}(\bfq_{1}+\bfq_{2}+\textbf{k}_{1})}{a^{d+1}(t_{1})a^{d+1}(t_{2})}K_{k_{1}}(t_{1})K_{k_{2}}(t_{2})G_{q_{1}}(t_{1},t_{2})G_{q_{2}}(t_{1},t_{2}) \label{eqn:psi2_Sch}\\
    \psi_{4}(\bfk_{1},\bfk_{2},\bfk_{3},\bfk_{4})&=ig^{2}\int_{t_{1}t_{2}}\frac{1}{a^{d+1}(t_{1})a^{d+1}(t_{2})}K_{k_{1}}(t_{1})K_{k_{2}}(t_{1})G_{p_{s}}(t_{1},t_{2})K_{k_{3}}(t_{2})K_{k_{4}}(t_{2})+\text{2 perm.}\nonumber
\end{align}
since we have the useful relation \cite{Anninos:2014lwa, Cespedes:2020xqq},
\begin{equation}
    \partial_{t} G_{q}( t_{1}, t_{2} )=-a^{1-d} (t) K_{q}(t_{1} )K_{q}( t_{2} )
\end{equation}
for the derivative of the bulk-to-bulk propagator with respect to the boundary time.
The relative factor of $1/2$ in \eqref{eqn:psi2_Sch} is the same symmetry factor we find in amplitudes for similar diagrams.

\paragraph{From canonical quantisation.}
Finally, in the in-in calculation of cosmological correlators one finds that the second order correction towards the expectation value of an observable is:
\begin{equation}
    \langle\hat{\mathcal{O}}\rangle=-\int_{-\infty}^{\tau}d\tau_{2}\int_{-\infty}^{\tau_{2}}d\tau_{1}\langle[\hat{\mathcal{H}}(\tau_{1}),[\hat{\mathcal{H}}(\tau_{2}),\hat{\mathcal{O}}]]\rangle,
\end{equation}
which can be rewritten as the sum of two terms:
\begin{equation}\label{eqn:inin2nd}
    \langle\hat{\mathcal{O}}\rangle=-2\int_{-\infty}^{\tau}d\tau_{2}\int_{-\infty}^{\tau_{2}}d\tau_{1}\left(\text{Re}\left(\langle\hat{\mathcal{H}}(\tau_{1})\hat{\mathcal{H}}(\tau_{2})\hat{\mathcal{O}}\rangle\right)-\text{Re}\left(\langle\hat{\mathcal{H}}(\tau_{1})\hat{\mathcal{O}}\hat{\mathcal{H}}(\tau_{2})\rangle\right)\right).
\end{equation}
In order to fix the symmetry factor for the sunset diagram, we will focus on the two point function $\langle\hat{\phi}_{\bfk_{1}}\hat{\phi}_{\bfk_{2}}\rangle$ and the interaction $\frac{g}{3!}\phi^{3}$. We do the calculation in Minkowski spacetime, as the symmetry factor does not depend on the spacetime. From equation (\ref{eqn:inin2nd}) we find two terms contributing towards $\langle\hat{\phi}_{\bfk_{1}}\hat{\phi}_{\bfk_{2}}\rangle$:
\begin{align}
&\langle\hat{\phi}_{\bfk_{1}}\hat{\phi}_{\bfk_{2}}\rangle=\delta(\textbf{k}_{1}+\textbf{k}_{2})g^{2}\int_{\textbf{q}_{1}\textbf{q}_{2}}\delta(\textbf{q}_{1}+\textbf{q}_{2}+\textbf{k}_{1})\left(\mathcal{I}_{1} - \mathcal{I}_{2}\right)\\
&\mathcal{I}_{1}=\frac{k_1 + k_2 + 2(q_1 + q_2)}{16k_1 k_2 q_1 q_2 (k_1 + k_2)(k_1 + q_1 + q_2)(k_2 + q_1 + q_2)} \nonumber \\
&\mathcal{I}_{2}=\frac{k_2 - k_1}{16 k_1 k_2 q_1 q_2 (k_1 - k_2)(k_1 + q_1 + q_2)(k_2 + q_1 + q_2)} ,\nonumber
\end{align}
which leaves:
\begin{equation}\label{eqn:ininsunet}
\langle\hat{\phi}_{\textbf{k}_{1}}\hat{\phi}_{\textbf{k}_{2}}\rangle=g^{2}\delta(-\textbf{k}_{1}-\textbf{k}_{2})\int_{\textbf{q}_{1}\textbf{q}_{2}}\frac{2 k+q_{1}+q_{2}}{16 k^3 q_{1} q_{2} (k+q_{1}+q_{2})^2}.
\end{equation}
This coincides with the calculation done in section \ref{sec:KLN} using the $1/2$ symmetry factor for the sunset diagram.

\section{Cosmological Loop-Tree Duality}
\label{sec:LTD}

Finally, we establish an explicit connection with the ``loop-tree duality'' (LTD), a modern descendent of Feynman's tree theorem.  
The main idea, due to Catani et al \cite{Catani:2008xa}, is that any one-loop diagram can be written as a sum over single-cut (tree) diagrams, where a diagram in which the line $i$ is cut uses the modified propagator,
\begin{align}
 \tilde{G}_j^i = G_j + \omega_j^i H_j  \; .
\end{align}
For amplitudes, $H_j \sim K_j K_j^*$, and for wavefunction coefficients $H_j \sim K_j \text{Im} \, K_j$ (where the time arguments are distinguished by going either clockwise or counterclockwise around the loop). 

The key to this construction is that the $\omega_j^i$ coefficients obey the relations,
\begin{align}
 \sum_{i}^N \prod_{j \neq i} \omega^i_j = 1 \; ,
 \label{eqn:w_condition}
\end{align}
where $N$ is any subset of the edges in the loop.
For instance, \cite{Catani:2008xa} use $\omega_j^i = \Theta ( q_j^0 - q_i^0 )$, where $q_j$ is the 4-momentum of the $j^{\rm th}$ internal line, since it is not hard to show that energy conservation then implies 
\eqref{eqn:w_condition}. 
For instance when $N=2$, 
\begin{align}
 \omega_1^2 + \omega_2^1 =  \Theta ( k^0 ) + \Theta (  - k^0 ) = 1 
\end{align}
where $\{ +k ,-k \}$ are the momenta of the external lines. 
When $N=3$,
\begin{align}
 \omega_1^3 \omega_2^3 + \omega_1^2 \omega_3^2 + \omega_2^1 \omega_3^1 = 
 \Theta (k_1^0 ) \Theta ( -k_3^0 ) + \Theta ( - k_2^0 ) \Theta ( k_3^0 ) + \Theta (k_2^0 ) \Theta ( - k_1^0 )  =  1  
\end{align}
since $k_1^0 + k_2^0 + k_3^0 = 0$. 

For instance, the LTD for the triangle graph follows from expanding, 
\begin{align}
 \tilde{G}_1^3 \tilde{G}_2^3  H_3 + \text{2 perm.} &= 
 G_1 H_2 H_3  \left( \omega_2^3 + \omega_3^2 \right) + \text{2 perm.}  \nonumber \\
 &+ H_1 H_2 H_3  \left( \omega_1^3 \omega_2^3 + \omega_1^2 \omega_3^2 + \omega_2^1 \omega_3^1  \right)  
\end{align}
and then using \eqref{eqn:w_condition} to set each $\omega_j^i$ sum to unity. 

So in order to use the LTD for the wavefunction, one needs to introduce a set of $\omega_j^i$ which obey \eqref{eqn:w_condition}. Since energy is no longer conserved, the Catani choice of $\Theta ( q_j^0 - q_i^0 )$ will not satisfy \eqref{eqn:w_condition}. However, the spatial momentum in any fixed direction is conserved, so an alternative modification of the propagator would be to introduce a fixed reference vector $\hat{n}$ and use $\Theta \left( \hat{n} \cdot ( \bfq_j - \bfq_i ) \right)$ instead. 
We have not explored this particularly thoroughly, and it would be interesting to develop this direction further in future, particularly in light of the recent progress which has been made in implementing the LTD both numerically \cite{Buchta:2015wna,Hernandez-Pinto:2015ysa,Sborlini:2016gbr,Sborlini:2016hat,Driencourt-Mangin:2017gop} and analytically for diagrams with up to five loops \cite{Ramirez-Uribe:2022sja}.

\bibliographystyle{JHEP}
\bibliography{refs}

\end{document}